\documentclass[a4paper,fleqn,usenatbib]{mnras}

\usepackage{newtxtext,newtxmath}

\usepackage[T1]{fontenc}
\usepackage{ae,aecompl}

\usepackage{graphicx}	
\usepackage{amssymb}	
\usepackage{lscape}	

\title[Variability power spectra]{Blazar variability power spectra from radio up to T\MakeLowercase{e}V photon energies: Mrk\,421 and PKS\,2155$-$304}

\author[A. Goyal]{
Arti Goyal,$^{1}$\thanks{E-mail: arti@oa.uj.edu.pl (AG)}
\\
$^{1}$Astronomical Observatory of the Jagiellonian University, ul. Orla 171, 30-244 Krak{\'o}w, Poland}

\date{Accepted XXX. Received YYY; in original form ZZZ}

\pubyear{2019}

\begin{document}
\label{firstpage}
\pagerange{\pageref{firstpage}--\pageref{lastpage}}
\maketitle

\begin{abstract}

We present the results of the power spectral density (PSD) analysis for the blazars Mrk\,421 and PKS\,2155$-$304, using good-quality, densely sampled light curves at multiple frequencies, covering 17 decades of the electromagnetic spectrum, and variability timescales from weeks up to a decade. The data were collected from publicly available archives of observatories at radio from OVRO, optical and infrared (B, V, R, I, J, H, and K--bands), X-rays from the {\it Swift} and the {\it Rossi} X-ray Timing Explorer, high and very high energy $\gamma-$rays from the {\it Fermi} and Very Energetic Radiation Imaging Telescope Array System as well as the High Energy Stereoscopic System. Our results are: (1) the power-law form of the variability power spectra at radio, infra-red and optical frequencies have slopes $\sim$1.8, indicative of random-walk type noise processes; (2) the power-law form of the variability power spectra at higher frequencies, from X--rays to very high energy \,$\gamma$-rays, however, have slopes $\sim$1.2, suggesting a flicker noise type process; (3) there is significantly more variability power at X-rays, high and very high energy $\gamma$-rays on timescales $\lesssim$ 100 days, as compared to lower energies. Our results do not easily fit into a simple model, in which a single compact emission zone is dominating the radiative output of the blazars across all the timescales probed in our analysis. Instead, we argue that the frequency-dependent shape of the variability power spectra points out a more complex picture, with highly inhomogeneous outflow producing non-thermal emission over an extended, stratified volume.
\end{abstract}




\section{Introduction} \label{sec:intro}

Blazars form an extreme class of active galactic nuclei (AGN), for which the total radiative output is dominated by a highly magnetized and non-stationary nuclear relativistic jet launched from the center of a massive elliptical galaxy \citep[e.g.,][]{Urry95}. The blazar family includes flat-spectrum radio-loud quasars (FSRQs) and BL Lacertae (BL Lac) objects, depending on the prominence of emission lines in the optical spectrum. Their spectral energy distribution (SED) usually exhibits two broad peaks, resulting from non-thermal processes. Within the framework of the `leptonic' scenario, electron-positron ($e\pm$) pairs accelerated to GeV/TeV energies are believed to emit radiation from radio-to-optical/UV frequencies (sometimes up to X--rays in the case of BL Lac objects) via the synchrotron process, while the X--ray-to-TeV$\gamma-$ray emission arises from inverse-Comptonization by the  particle population producing the synchrotron radiation of the seed photons which are either produced locally (Synchrotron-Self Compton; SSC), or externally (External Compton; EC) to the jet \citep[e.g.,][and references therein]{Madejski16}. Alternatively, in the `hadronic' scenario for the blazar emission, the higher-frequency emission peak is believed to originate due to protons accelerated to $\geq$\, PeV-EeV energies and producing $\gamma$-rays via either direct synchrotron process, or meson decay and synchrotron emission by the secondaries produced in proton-photon interactions \citep[e.g.,][and references therein]{Bottcher13}.  The location of the synchrotron emission peak in the SED of BL Lac objects divides them further into high-frequency-peaked BL Lacs ($\nu_{\rm peak} > 10^{15}$ Hz; HBL), intermediate-frequency-peaked BL Lacs ($\nu_{\rm peak} \simeq 10^{14-15}$ Hz; IBL), and low-frequency-peaked BL Lacs ($\nu_{\rm peak} < 10^{14}$ Hz; LBL), with FSRQs sharing the space with the LBLs \citep[e.g,][]{Fossati98, Ghisellini17}.  

As a class, blazars are known to display strong variability ranging from radio to VHE\,$\gamma-$ray energies and on multiple timescales \citep[e.g.,][]{Aller99, Wagner95, Falomo14, Abdo10a} with a factor of few intensity changes on timescales as short as minutes which are found be especially prominent at high energies \citep{Cui04, Aharonian07, Albert07, Ackermann16, Zhu18}. Highly efficient particle acceleration, either due to the formation of shocks and turbulence in the jet flow \citep{Agudo18, Marscher14, Hughes11}, or annihilation of the magnetic field lines at the magnetic reconnection sites \citep[e.g.,][]{Giannios13, Sironi15, Petropoulou16} are considered to be prime candidates for the energy dissipation processes. The origin of rapid flux changes and particularly their relation to those observed on longer timescales of days-to-years is still widely debated. In this regard, the large amplitude variability on the shortest timescales poses additional challenge because jet bulk Lorentz factors needed to overcome the photon opacity barrier \citep[$\Gamma > 50$; e.g.,][]{Begelman08} are rather too extreme to be reconciled with the currently favored models of AGN jet acceleration (\citealt{McKinney09}). 

The power spectral density (PSD) of a blazar light curve typically has a power-law shape, $P({\nu_k}) \propto \nu_{k}^{-\beta}$, where $\nu_k$ ($\equiv$timescale$^{-1}$) is the temporal frequency and $\beta$ is the slope. This indicates that the blazar variability is a {\it stochastic} processes. Moreover, the typical slopes, $\beta \sim$ 1--3, on timescales from years (sometimes even decades) down to days (and in some cases, minutes), indicate that this stochastic process is of \emph{correlated} colored-noise type ($\beta=$ 1; flicker/pink and $\beta=$ 2; random-walk/red), unlike the \emph{uncorrelated} white-noise type process which  give rise to flatter ($\beta \sim $0) PSD slopes \citep{Press78}. Whether the {\it stochastic} process remains {\it stationery} on the timescales probed is crucial for understanding the physical processes operating on widely different spatial scales \citep[e.g.,][]{Kushwaha16, Giebels09}. Studies of different blazars performed at different wavelengths have indeed revealed breaks in the slopes of PSDs, which could be attributed to the size of the emission zone or, alternatively, to cooling timescales of the particles \citep[e.g.,][]{Hess10, Sobolewska14, Isobe15, Kastendieck11, Hess17, Ryan19, Zywucka20}. The emerging picture of the broad-band blazar variability is that at higher energies (i.e., X$-$ray, and GeV and TeV $\gamma-$rays) is characterized by {\it flicker/pink-}noise processes \citep[i.e.,][]{Hess17, Abdo10a, Isobe15}, whereas at lower energies (GHz band radio and optical frequencies) {\it damped/red-}noise type processes dominate \citep[i.e.,][]{Ciprini07, Kastendieck11, Park14, Max-Moerbeck14a, Nilsson18}. 

In \citet{Goyal17, Goyal18}, we presented the first attempts to uniformly characterize the statistical properties of the variability at radio--to--$\gamma-$rays, on timescales ranging from decades to weeks, by analyzing the GHz band radio, optical, X-ray and $\gamma-$ray light curves for the well-known blazars PKS\,0735$+$178 and OJ\,287. Our results conform to the findings mentioned above: the PSD shape at higher energies ($\beta \sim$1) differs markedly from that at lower energies ($\beta \sim$ 2). We interpreted our findings in terms of a single stochastic process, or a linear superposition of such, giving rise to the observed broad-band variability. We emphasized, that different statistical characters of the low and high energy variability, is difficult to explain within the framework of the widely discussed `one-zone' models, which invoke the same particle (electron) population within a single well-defined and homogeneous emission volume to account for both synchrotron and IC emission components, and in the framework of which one should expect the same slopes of the variability PSDs across all the wavelengths \citep{Finke14, Finke15}. Even though our results find support from the lack of clear correlations between the variations at different frequencies as inferred from cross-correlation studies of blazar light curves \citep[e.g.,][]{Max-Moerbeck14b, Lindfors16}, we note that many studies do find correlated variability among different frequencies, along the expectations of the widely used SSC scenario invoked for blazar sources \citep[e.g.,][]{Hovatta15, Fuhrmann16, Ramakrishnan16} 

Due to extreme photon deficiencies at TeV energies, augmented by the absorption process on the extragalactic background light for cosmologically distant sources, and sensitivity constraints of the currently operating Cherenkov telescopes with limited duty cycles, so far 73 blazars have been detected in the TeV range\footnote{http://tevcat.uchicago.edu/}; note that such problems are much less severe in the case of the all-sky monitoring by the {\it Fermi}-LAT at GeV energies \citep{Ackermann15}. Blazars are often targeted at TeV energies during short flares from lower energies; therefore, the observations are usually biased towards the flaring states \citep[see,][and references therein]{Ahnen17, Senturk13}. The blazars Mrk\,421 and PKS\,2155$-$304 are exceptions to this rule: thanks to their relatively high flux at TeV energies, these blazars could be successfully monitored on a nightly basis by the Very Energetic Radiation Imaging Telescope Array System (VERITAS) \citep[1995-2009;][]{Acciari14} and the High Energy Stereoscopic System (H.E.S.S.) observatories \citep[2002--2011;][]{Hess17}. For this reason, the TeV measurements in these cases can be taken as representative for energy dissipation processes operating in blazar jets at the highest energies in general, and {\it not only} during the highest flux states. Therefore, we focus in the present study on determination of the statistical properties of the stochastic processes giving rise to broadband variability up to TeV\,$\gamma-$ray energies, using good-quality, roughly decade-long, multiwavelength light curves. 

For this, we have assembled from the publicly available archives of both ground-based and space-borne observatories, light curves at very high energy (VHE)\,$\gamma-$rays from the VERITAS and the H.E.S.S.\citep[$>$200\,GeV;][]{Acciari14, Hinton04}, high energy (HE) $\gamma-$ray from the {\it Fermi-}Large Area Telescope \citep[LAT (0.1--300 GeV);][]{Thompson04}, X$-$rays from the {\it Rossi} X-ray Timing Explorer (RXTE) Proportional Counter Array \citep[PCA (3--20 keV);][]{Bradt93} and from the {\it Swift-}X$-$Ray Telescope \citep[XRT (0.3-10 keV);][]{Gehrels04}, infra-red (IR; J, H, and K--band), optical (B, V, R, and I--band) and 15\,GHz\, radio frequency from the Owens Valley Radio Observatory (OVRO). This has enabled us to present, for the first time, the PSD analysis of the multiwavelength light curves which defines/constrains the nature of stochastic variability of a blazar all the way up to VHE\,$\gamma-$ray energies and on timescales ranging from weeks to a decade.

Mrk\,421 is an archetypal blazar which due to its proximity \citep[J2000.0 R.A.\,$\rm=11^{h}04^{m}27\fs314$, Dec.\,$\rm=$+$38\degr12\arcmin31\farcs80$; redshift = 0.03002;][]{deVaucouleurs91} and the high flux, was the first blazar to be detected at TeV energies by the 10-m Whipple telescope \citep[][]{Punch92}. 2155$-$304, \citep[J2000.0 R.A.\,$\rm=21^{h}58^{m}52\fs065$, Dec.\,$\rm=$-$30\degr13\arcmin32\farcs11$; $z = 0.116$;][]{Bechtold02}, is one of the first few blazars to be detected at VHE\,$\gamma-$rays \citep{Chadwick99}. Mrk\,421 has been a subject of intense monitoring ever since its discovery in the VHE band, in particular, the relation between the variations at X-ray and VHE\,$\gamma-$rays have been probed widely \citep[][]{Blazejowski05, Fossati08, Hovatta15, Ahnen16, Gonzalez19}. For the blazar Mrk\,421, no quasi-periodic oscillations (QPOs) in the decade-long optical and HE\,$\gamma-$ray light curves were reported in the study of \citet{Sandrinelli17} while \citet{Fraija17} estimated a period of $\sim$16 years in the $\sim$100 yr long optical light curve. For the blazar, PKS\,2155$-$304, the variability analysis based on the full VHE\,$\gamma-$ray (H.E.S.S.; 2004--2012), partial HE\,$\gamma-$ray ({\it Fermi-}LAT; 2008--2012) and X$-$ray ({\it Swift-}XRT; 2006--2012) light curves have been presented in \citet{Hess17} and \citet{Kapanadze14}. PKS\,2155$-$304 is also known for a log-normal distribution of TeV flux densities, and also for a change in the PSD slope, such that $\beta \sim$ 1 and $\sim$ 2 on timescales longer and shorter than 1 day, respectively). This indicates a dominance of multiplicative over additive energy dissipation processes manifesting in the VHE regime \citep[][]{Aharonian07, Hess17, Chevalier19}. Interestingly, correlated variability between VHE\,$\gamma-$ray and X$-$ray \citep{Costamante08}, as well as VHE\,$\gamma-$rays and optical frequencies have also been claimed for this blazar \citep[][]{Aharonian09}. Recently, hints of QPOs, with a characteristic timescale of $\sim$1-2, yrs\, have been noted in the HE\,$\gamma$-ray and optical light curves \citep[see,][]{Sandrinelli16, Zhang17, Chevalier19}. Possible quasi-periodicities on similar timescales have also been reported in the decade-long optical/IR light curves \citep[e.g.,][]{Sandrinelli14b, Zhang14}. In addition, QPOs on comparatively shorter timescales have been claimed for this blazar, in its X$-$ray light curve ($\sim $4.6\,hrs\,; \citealt{Lachowicz09}) and in the polarized optical emission ($\sim$13 and 30 min\,; \citealt{Pekeur16}).

In Section~\ref{sec:obs} we describe in more detail all the data assembled here, and the data reduction procedures. The data analysis and the results obtained are outlined in Sections~\ref{sec:PSD}, and \ref{sec:result}, respectively, 
while a discussion and our main conclusions are presented in Section~\ref{sec:conclusion}.

\section{Data acquisition and analysis: multiwavelength light curves}
\label{sec:obs}

\subsection{VHE\,$\gamma-$rays: VERITAS and H.E.S.S.}
Fig.~\ref{fig:1}(a) presents the publicly available\footnote{https://veritas.sao.arizona.edu/veritas-science/mrk-421-long-term-lightcurve} nightly-binned TeV light curve of Mrk\,421 observed with the Whipple 10-m imaging Cherenkov telescope and the VERITAS array at energies $>$400\,GeV for the period 1995--2009 \citep{Acciari14}. PKS\,2155$-$304 has been a target of regular monitoring at the H.E.S.S. observatory due to its favorable location in the southern hemisphere and relatively high flux which, together with the telescope's large collecting area enables relatively high accuracy of measurements on a nightly basis. Fig.~\ref{fig:2}(a) presents the H.E.S.S. light curve for the period 2002--2009 \citep[published in][]{Hess17}. We note that the presented light curve does not include the portion of the data when the blazar underwent an extreme flare when its flux level increased by a factor of $\sim$100 \citep{Aharonian07}, compared to average flux levels of this blazar reported here \citep{Hess17}. The details regarding the observations and data analysis can be found in \citet{Acciari14} and \citet{Hess17}.

\subsection{HE\,$\gamma-$rays: {\it Fermi}-LAT}
{\it Fermi}-LAT data for the field containing the targets were analysed for the 0.1 -- 300\,GeV band, with an integration time of seven days from August 2008 until March 2018 \citep{Atwood09}. The details of the LAT data analysis can be found in \citet{Goyal17, Goyal18} and below we briefly recall the main features. We have generated the source light curve by performing the unbinned likelihood analysis using Fermi ScienceTools {\sc 10r0p5} with latest instrument response function {\sc p8r2\_source\_v6} source event selection and zenith angle $< 90^\circ$\footnote{\texttt{http://fermi.gsfc.nasa.gov/ssc/data/analysis/scitools/}}. All photons were counted within $20^\circ$ region centered at the blazar position to account for the broad point spread function (PSF) of the {\it Fermi-}LAT at the desired energies (0.1--300\,GeV). The analysis method starts with the selection of good data and time intervals using the tasks {\sc `gtselect'} and {\sc `gtmktime'} with selection cuts {\sc evclass=128 evtype=3}, followed by the creation of an exposure map in the region of interest (ROI) with $30^\circ$ radius for each time bin using tasks {\sc `gtltcube'}  and {\sc `gtexpmap'}. 

Next, we computed the diffuse source response (task {\sc `gtdifrsp'}) and finally modeled the data with the maximum-likelihood method (task {\sc `gtlike'}). In this last step, we included all point sources togather with the targets, Mrk\,421 and PKS\,2155$-$304 (189 and 167 other point sources, respectively) inside the ROI from the 3FGL catalog, as well as the standard templates for diffuse emission from our Galaxy ({\sc gll\_iem\_v06.fits}) and the isotropic $\gamma$-ray background ({\sc iso\_p8r2\_source\_v6\_v06.txt}) \citep{Acero16}. To account for the contamination in the source flux due to variable point-sources within the broad PSF of the LAT ($\sim$5$^\circ$ at 100 MeV), we checked for variable point sources within 5$^\circ$ of the target from the 3FGL source-list in the Fermi All-sky Variability Analysis \citep[FAVA;][]{Abdollahi17}. In the modeling, we fixed the photon indices and fluxes of all the point sources within the ROI other than the target and the FAVA sources at their 3FGL values. Lastly, the $\gamma$-ray spectrum was modeled with a power-law and log-parabolic functions for Mrk\,421 and PKS\,2155$-$304, respectively, as recommended by the 3FGL catalog by keeping the normalization and the spectral parameters free. Finally, a criterion of test statistic TS\,$\geq$\,10 was set to consider a measurement to be a successful detection $\geq 3\sigma$ \citep{Abdo09}. The generated light curves consisting of successful detections are shown in Fig.~\ref{fig:1}(b) and ~\ref{fig:2}(b) for Mrk\,421 and PKS\,2155$-$304, respectively.

\subsection{X$-$rays: {\it RXTE}-PCA and {\it Swift}-XRT} 
\label{sec:swift}
We downloaded all of 1,207 individual spectra for Mrk\,421 from the {\it RXTE} quick look analysis from the {\it heasarc} website\footnote{https://heasarc.gsfc.nasa.gov/db-perl/W3Browse/w3browse.pl}. The spectra were binned for 20 points in energy bins and fitted with a simple power-law in {\it xspec} with a $\chi^2$ statistics. In the spectral analysis, the Galactic absorption corresponding to neutral hydrogen column density ($N_{\rm H,\,Gal} =1.91 \times 10^{20}$\,cm$^{-2}$) were fixed in the direction of the source \citep{Kalberla05}. The unabsorbed energy fluxes were obtained within 3-20 keV energy range. The measured fluxes were then averaged using 1 d bins. Fig.~\ref{fig:1}(c) shows the resulting light curve. For the blazar PKS\,2155$-$304, the 457 individual spectra are sporadically observed between 1995--2012, therefore, we do not include them in the present analysis.

For the targets, we have analysed the archival data from the {\it Swift}-XRT observatory which consists of 1,070 (Mrk\,421) and 214 (PKS\,2155$-$304) pointed observations made between September 2005 and January 2019. The details of the XRT data analysis can be found in \citet{Goyal18} and below we briefly recall the main features. For the data reduction, we used the latest version of the calibration database ({\sc CALDB}) and version 6.19 of the {\sc heasoft} package\footnote{\texttt{http://heasarc.gsfc.nasa.gov/docs/software/lheasoft/}}. For each data set, we used the level 2 cleaned event files of the `photon counting' (PC) and `window timing' (WT) data acquisition mode generated using the standard {\sc xrtpipeline} tool. For the PC mode data, the source and background spectra were generated using a circular aperture with appropriate region sizes and grade filtering using the {\sc xselect} tool. The source spectra were extracted using an aperture radius of $47^{\prime\prime}$, centered at the source position, while four source-free regions of $118^{\prime\prime}$ radius each were used to estimate the background spectrum. For the WT mode data, the source region of $47^{\prime\prime}$ circular aperture and an annular region with inner and outer radii of $187^{\prime\prime}$ and  $281^{\prime\prime}$, respectively, for background estimation were selected to estimate the source and the background spectra. Appropriate corrections were made in the `BACKSCAL' keyword in the spectra to account for the different sizes of source and background regions, following the XRT tutorial\footnote{http://www.swift.ac.uk/analysis/xrt/backscal.php}. The ancillary response matrix was generated using the task {\sc xrtmkarf} for the exposure map generated by {\sc xrtexpomap}. All the source spectra were then binned for 20 points and corrected for the background using the task {\sc grppha}. All the observations were checked for the recommended pile-up limit for the PC ($\sim$0.5 counts\,s$^{-1}$) and WT ($\sim$100 counts\,s$^{-1}$) modes. In almost all the PC mode data, the count rate exceeded the pile-up limit for both blazars and we made appropriate corrections to the source count by modeling the XRT PSF with a Kings function following the standard procedure\footnote{http://www.swift.ac.uk/analysis/xrt/pileup.php}. On a few occasions, the source count exceeded the pile-up limit for the WT mode data for Mrk\,421 and appropriate corrections, i.e., rejecting the central 2-pixels while extracting the source spectrum, were applied following \citet{Romano06}. In none of the observations, however, did the source count rate exceed the recommended pile-up limit for PKS\,2155$-$304. 

\begin{figure*}
\centering
\includegraphics[width=\textwidth]{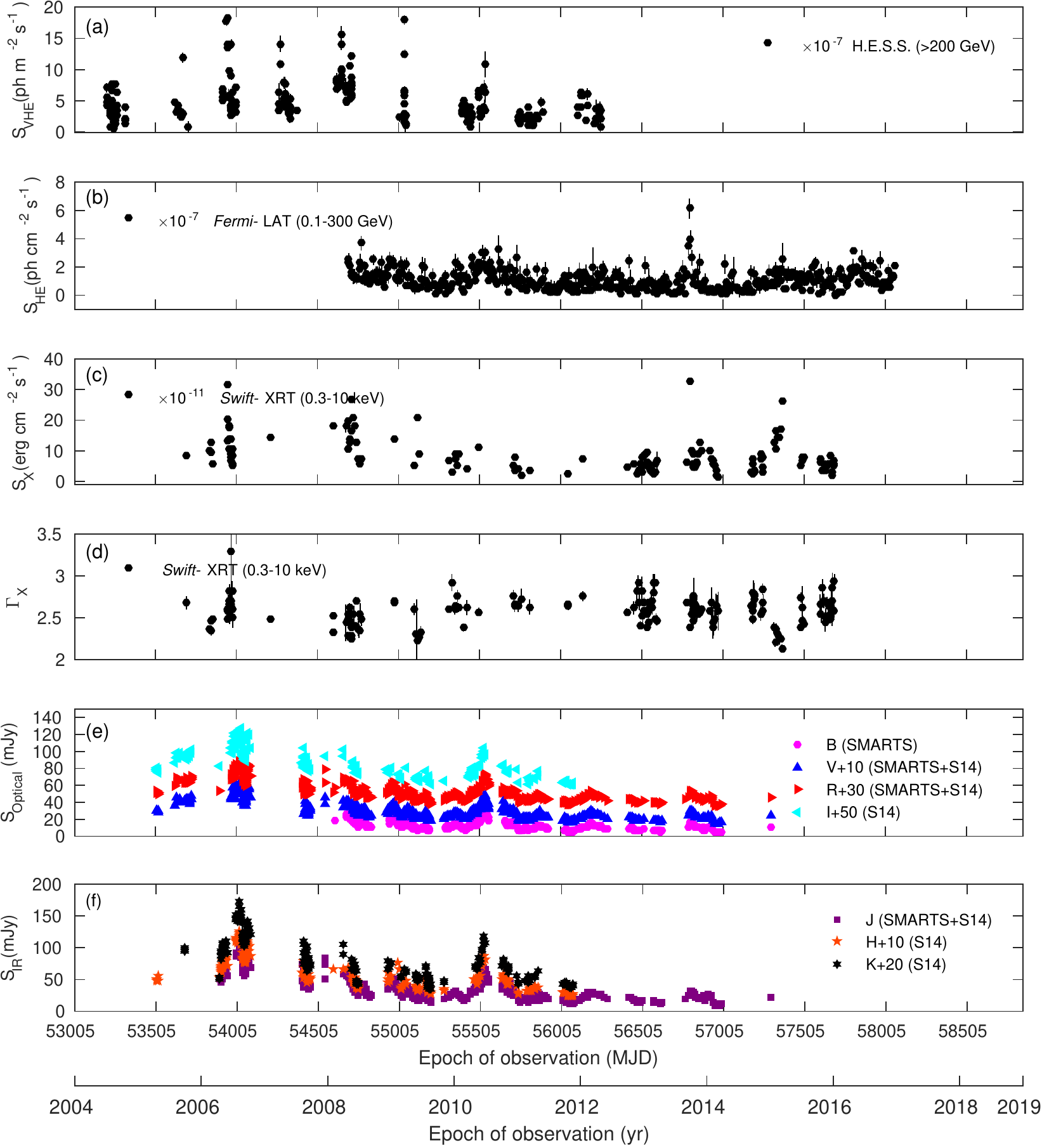}
\begin{minipage}{\textwidth}
\caption{The multiwavelength light curves of PKS\,2155$-304$. Panel (a) shows the VHE\,$\gamma-$ray data from the H.E.S.S at energies $>$200 GeV \citep{Hess17}. Panel (b) shows the HE\,$\gamma-$ray data from {\it Fermi-}LAT at energy range 0.1-300 GeV. Panel (c) shows the {\it Swift}-XRT light curve for energy range 0.3-10 keV while the panel (d) presents the run of $\Gamma$ for the X--ray data. Panel (e) shows the optical B, V, R, and I--band total flux light curves and panel (f) shows the IR J, H, and K--band total flux light curves.}
\label{fig:2}%
\end{minipage}
\end{figure*}

For each exposure, we used routines from the X$-$ray data analysis software {\sc ftools} and {\sc xspec} to calculate and to subtract the X$-$ray background model from the data. Spectral analysis was performed between 0.3 and 10 keV energy range by fitting a simple power-law moderated by the Galactic absorption with the corresponding $N_{\rm H, \, Gal}$ fixed to 1.91 $\times 10^{20}$ and 1.48 $\times$ 10$^{20}$\,cm$^{-2}$ in the directions of the blazars Mrk\,421 and PKS\,2155$-$304, respectively \citep{Kalberla05}. We used the 1 d binning interval to average the unabsorbed 0.3--10\,keV fluxes, to construct the source light curve for Mrk\,421 and PKS\,2155$-$304, respectively (panel d of Fig.~\ref{fig:1} and panel c of Fig.~\ref{fig:2}). For completeness, we also present the run of computed photon indices, $\Gamma$, for the {\it RXTE-}PCA and {\it Swift-}XRT data of the studied blazars in panels e of Fig.~\ref{fig:1} (Mrk\,421) and d of Fig.~\ref{fig:2} (PKS\,2155$-$304).

\subsection{Optical and IR: SMARTS and REM} 
\label{sec:opticallc}
Since 2005, PKS\,2155$-$304 has been a target of regular monitoring on daily timescales at multiple optical (B, V, R, and I) and IR (J, H, and K) bands. In the present study, we use publicly available light curves from the Small and Moderate Aperture Research Telescope System (SMARTS; \citealt{Bonning12}) programme coordinated by Yale University\footnote{http://www.astro.yale.edu/smarts/glast/home.php} and by Rapid Eye Mounting telescope (REM; \citealt{Zerbi01}), published in \citet[][S14 hereafter]{Sandrinelli14a}. The B, V, R, J, and K band SMARTS light curves for the source covers the period from $\sim$2008 until 2015. The V, R, I, J, H and K band light curves provided by S14 span 2005 through 2012. We refer the reader to \citet{Bonning12} and \citet{Sandrinelli14a} for details regarding the monitoring programmes and the data analyses. The data at V, R, J, and K bands have a great deal of overlap between the two programmes and they complement each other well at  V, R, J frequencies, with the pairs of magnitudes within 0.1 mag on the same nights, consistent with the calibration uncertainties provided by the two programmes. Therefore, we have combined the two datasets at V, R, and J frequencies to produce the daily binned light curves. The K--band light curve from the SMARTS programme has a magnitude offset $\geq$0.1 mag from that of provided by S14 on the same nights. On visual inspection, the K--band SMARTS light curve appears more variable than the S14 light curve on similar timescales. Therefore, in this study, we present the K--band PSD provided by S14 data alone. Next, all the B, V, R, I, J, H, and K magnitudes were converted to fluxes using $m_{zero}$\,$\times 10^{-0.4\times m}$, where $m_{zero}$ (=4063\,Jy, 3636\,Jy, 3064\,Jy, 2635\,Jy, 1590\,Jy, 1020\,Jy, and 600\,Jy for the B, V, R, I, J, H, and K-bands, respectively) refers to zero point magnitude flux of the photometric system \citep{Glass99}. The errors in fluxes were derived using standard error propagation \citep{Bevington03}. The resulting optical and IR-band light curves of PKS\,2155$-$304 are presented in Fig.~\ref{fig:2}(e) and (f).

\subsection{Radio: OVRO (15\,GHz)}

\begin{figure*}
\centering
\includegraphics[width=\textwidth]{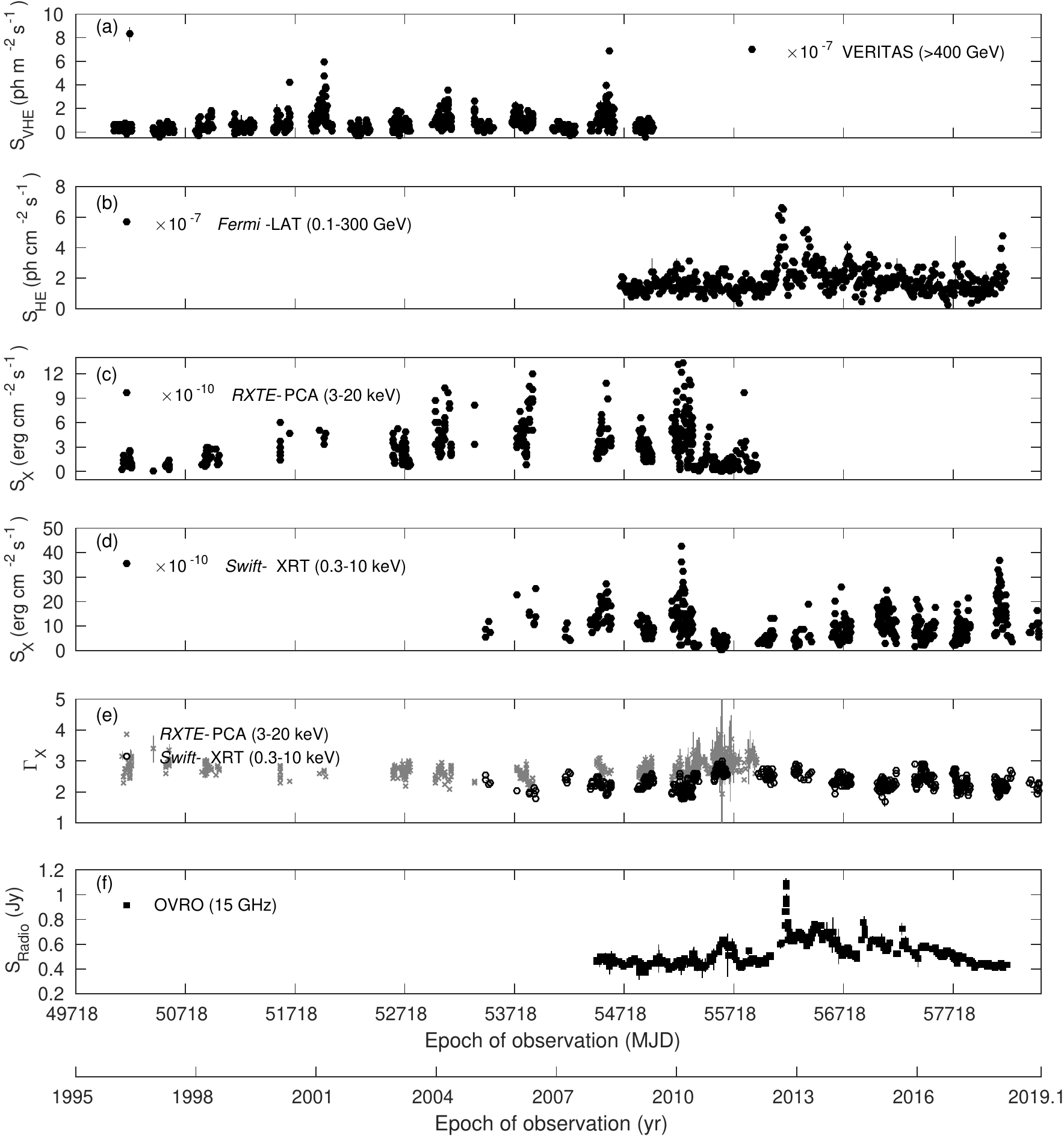}
\begin{minipage}{\textwidth}
\caption{The multiwavelength light curves of Mrk\,421. Panel (a) shows the VHE\,$\gamma-$ray data from the VERITAS at energies $>$400 GeV \citep{Acciari14}. Panel (b) shows the HE\,$\gamma-$ray data from {\it Fermi-}LAT at energy range 0.1-300 GeV.  Panel (c) shows the {\it RXTE}-PCA light curve for energy range 3-20 keV. Panel (d) shows the {\it Swift}-XRT light curve for energy range 0.3-10 keV. Panel (e) shows the run of $\Gamma$ for the {\it RXTE}-PCA (grey crosses) and the {\it Swift}-XRT (black open circles) light curves, respectively. Panel (f) shows the 15 GHz radio light curve from the OVRO monitoring programme.}
\label{fig:1}%
\end{minipage}
\end{figure*}

Fig.~\ref{fig:1}(f) shows the 15\,GHz radio light curve of Mrk\,421 for the period 2008--2018, gathered from the public archive of the OVRO monitoring programme \citep{Richards11}. 
 
\section{Power spectral analysis}
\label{sec:PSD}

The PSDs of the observed light curves of PKS\,2155$-$304, shown in Fig.~\ref{fig:1}, have been generated using the discrete Fourier transform (DFT) method outlined in \citet[][]{Goyal17, Max-Moerbeck14a, Vaughan03, Uttley02}. We describe the main features below. The periodogram of an evenly sampled light curve $f(t_i)$ observed at discrete times $t_i$, consisting of $N$ data points and the total monitoring duration $T$, is defined as the squared modulus of its DFT,
\begin{eqnarray}
| F(\nu_k) |^2 & = & \Bigg[ \sum_{i=1}^{N} f(t_i) \, \cos(2\pi\nu_k t_i)  \Bigg]^2  + \Bigg[ \sum_{i=1}^{N} f(t_i) \, \sin(2\pi\nu_k t_i)  \Bigg]^2 \, ,
\label{eq:dfteq}
\end{eqnarray}
where the mean $\mu$ is subtracted from the flux values, $f(t_i)$, in order to  remove the zero frequency power.  The DFT of a real function is a complex symmetric function with values at both negative and positive Fourier frequencies, ranging from $-N/2+1, -N/2+2,..,-1, 0, +1, +2,..., +N/2-1, +N/2$, in the case of an even number of data points \citep{Press92}. The DFT is computed for evenly spaced frequencies ranging from total duration of the light curve down to Nyquist sampling frequency ($\nu_{\rm Nyq}$). Specifically, the frequencies corresponding to $\nu_{k} = k/T$ with $k=1, ..., N/2$, $\nu_{\rm Nyq}= N/2T$, and $T = N (t_k-t_1)/(N-1)$ are considered. The power $P(\nu_k)$ at the given frequency $\nu_k$ is then obtained as the rms-squared-normalized periodogram

\begin{equation}
P(\nu_k) = \frac{2 \, T}{\mu^2 \, N^2} \, | F(\nu_k) |^2 \, .
\label{eq:psd}
\end{equation}

The normalized periodogram as defined in Eq.~\ref{eq:psd} corresponds to total excess variance when integrated over positive frequencies. Moreover, a constant amount of power (corresponding to white noise type process; $\beta\sim 0$) is expected to contribute in the estimated variability power due to measurement uncertainities alone. Such, `noise floor levels', are estimated as \citep[e.g.,][]{Isobe15, Vaughan03}
\begin{equation}
\rm P_{stat} = \frac{2 \, T}{\mu^2 \, N} \, \sigma_{\rm stat}^2 \, .
\label{eq:poi_psd}
\end{equation}
where, $\sigma_{stat}^2= \sum_{j=1}^{j=N} \Delta f(t_j)^2 / N$ is the mean variance of the measurement uncertainties on the flux values $\Delta f\!(t_j)$ in the observed light curve at times $t_j$, with  $N$ denoting the  number of data points in the original light curve. Because the observed light curves are not evenly sampled, we scale the noise floor level using typical (mean) sampling intervals \citep[see also,][]{Vaughan03}. The above described method requires, however, a regular sampling of the light curve at evenly spaced time intervals, otherwise the spectral window function corresponding to the sampling times gives non-zero response in the Fourier-domain\citep[e.g.,][]{Deeming75}. This will result in false powers in the DFT of the light curve (see Appendix~\ref{app:A} for discussion). Therefore, in order to perform the DFT of the observed light curves, we obtained the regular sampling only by linearly interpolating between the two consecutive observed data points on timescales typically 5--7 times smaller than the original observed sampling interval (see Table~\ref{tab:psd}). The distortions in the PSDs due to the discrete sampling and the finite duration of the light curve are described in detail in the Appendix of \citet{Goyal17} and the references therein. In our analysis, the resulting PSDs are generated down to the typical (mean) Nyquist sampling frequency of the observed light curves using the `Hanning' window function \citep[see also][]{Max-Moerbeck14a}. 

The periodogram obtained using equation~(\ref{eq:psd}), is known as the `raw' periodogram. By definition, it consists of independently distributed $\chi^2$ variables with two DOF as it is the sum of squares of Gaussian distributed variables \citep{TK95}. This means that each PSD estimate has a standard deviation around the true value that equals the value itself, providing a noisy estimate of the spectral power \citep{Papadakis93, Vaughan03}. Therefore, in order to obtain a reliable estimate of spectral power, a number of PSD estimates should be averaged where the mean spectral power estimate can resemble a Gaussian distribution provided enough number of data points are available for averaging. According to \citet{Papadakis93}, averaging about 20 data points in logarithmic space constrains the PSD estimate within 1$\sigma$ confidence interval. However, following this method would essentially remove a large number of data points from lower frequencies, instead, we bin the `raw' periodograms by averaging with a constant factor of 1.6 in frequency range (i.e., with a fixed step in logarithmic space) and evaluating the mean power at the representative frequency taken as the geometric mean of each bin, following \cite{Isobe15} and \cite{Goyal17}. This value is chosen to have at least two periodograms in each frequency bin except for the first bin.

The computed periodograms P($\nu_k$), for a noise-like process, are scattered around {\it true} value following a $\chi^2$ distribution with 2 degrees of freedom \citep{Papadakis93, TK95, Vaughan03}. Therefore,  the true power-spectrum is given as $P(\nu_k) = P_{true}({\nu_k}) \frac{\chi^2}{2}$. The transformation to log-log space, offsets the the observed periodograms as:  

\begin{equation}
\log_{10}[ P(\nu_k) ] = \log_{10}[ (P_{true}({\nu_k}) ] + \log_{10}\Bigl[ \frac{\chi^2}{2} \Bigr] .
\label{logpsd}
\end{equation}

This offset is constant and is equal $-$0.25068 which is the expectation value of $\chi^2$ distribution with 2 DOF in log-log space \citep{Papadakis93}. Finally, this value is added in computing the binned periodogram estimates.

Since the aim of the present study is to derive shapes of PSDs, the best-fit PSD parameters are determined using Monte Carlo (MC) simulations, following the `power spectral response' (PSRESP) method, introduced by \citet{Uttley02} and followed by  many others, including \citet{Chatterjee08, Max-Moerbeck14a, Isobe15, Meyer19}. We refer the readers to these studies for a detailed discussion about this approach while here we only briefly recall the main features. In this approach, a large number of light curves are simulated with a known underlying power-spectral shape. Each light curve is then rebinned to the observed sampling pattern and interpolated to have even sampling for the DFT application. The DFT of such light curve gives the distorted PSD due to various effects of rebinning, red--noise leak and aliasing. Averaging large number of such PSDs gives the mean of the distorted model (input) power spectrum. The standard deviation around the mean gives errors on the modeled power spectrum. The goodness of fit of the model is estimated by computing two functions, similar to $\chi^2$, defined as:

\begin{equation}
\chi^2_{\rm obs} = \sum_{\nu_{k}=\nu_{min}}^{\nu_{k}=\nu_{max}} \frac{[\overline{ \log_{10}P_{\rm sim}}(\nu_k)-\log_{10}P_{\rm obs}(\nu_k)]^2}{\Delta \overline{\log_{10}P_{\rm sim}}(\nu_k)^2}
\label{chiobs}
\end{equation}
and, 
\begin{equation}
\chi^2_{\rm dist, i} = \sum_{\nu_{k}=\nu_{min}}^{\nu_{k}=\nu_{max}} \frac{[\overline{ \log_{10}P_{\rm sim}}(\nu_k)-\log_{10}P_{\rm sim,i}(\nu_k)]^2}{\Delta \overline{\log_{10}P_{\rm sim}}(\nu_k)^2}
\label{chidist}
\end{equation}
where, $\log P_{\rm obs}$ and  $\log P_{\rm {sim, i}}$ are the observed and the simulated Log-binned periodograms, respectively. $\overline{ \log P_{\rm sim}}$ and $\Delta \overline{\log P_{\rm sim}}$ are the mean and the standard deviation obtained by averaging 1,000 PSDs; $k$ represents the number of frequencies in the log-binned power spectrum (ranging from $\nu_{min}$ to $\nu_{max}$), while $i$ runs over the number of simulated light curves for a given $\beta$.  

Where the $\chi^2_{\rm obs}$ determines the minimum $\chi^2$ for the model compared to the data and the $\chi^2_{\rm dist}$ values determine the goodness of the fit corresponding to the $\chi^2_{\rm obs}$. Note that $\chi^2_{\rm obs}$ and  $\chi^2_{\rm dist}$ are not the same as that of standard $\chi^2$ distribution because $\log_{10}P_{\rm obs}(\nu_k)$'s are not normally distributed variables since the number of power spectrum estimates averaged in each frequency bin is small ($<$20 for the first few frequency bins). Therefore, reliable goodness of fit is computed using the distribution of $\chi^2_{\rm dist}$ values. For this, the $\chi^2_{\rm dist}$ is sorted in ascending order. The probability, or p$_{\beta}$, that a given model can be rejected is then given by the percentile of $\chi^2_{\rm dist}$ distribution above which $\chi^2_{\rm dist}$ is found to be greater than  $\chi^2_{\rm obs}$ for a given $\beta$ \citep[also known as the success fraction;][]{Chatterjee08}. A large value of $p_{\beta}$ represents a good--fit in the sense that a large fraction of random realizations of the model (input) power spectrum are able to recover the shape and slope of the intrinsic (of which the observed PSD makes one realization) PSD. Therefore, this analysis essentially uses the MC approach toward a frequentist estimation of the quality of the model compared to the data. This is a well-known approach to describe the goodness of fit in the absence of well--understood fit statistics \citep[see, for details,][]{Press92}.   

In the present study, the light curves are simulated following the method of \citet{Emmanoulopoulos13}. This method has the advantage of preserving the probability density function of the flux distribution as well as the underlying power spectral shape, and not just the shape as given by \citet{TK95}. In the simulations, we have assumed single power-law PSDs with a given $\beta$ (to reproduce the PSD shape) and supplied mean and standard deviation of the logarithmically transformed flux values (to reproduce the flux distribution) which is, in general, a valid assumption for blazar light curves \citep{Hess17, Chevalier19, Liodakis17, Kushwaha17}. {In our simulations we have computed the mean and the standard deviation {$\sigma$} directly from the data as opposed to obtaining them from fitting a Gaussian function to the distribution. The computed mean and $\sigma$ from the data were found to be within 95\% confidence intervals of the mean and 
$\sigma$ obtained from the fitting. This is because the analysed light curves are relatively well--sampled and the fluxes are obtained with good precision (few percent measurement accuracies); therefore, any distortions due to limited sensitivity and finite sampling of the light curve are not significant. Each simulated light curve was then rebinned to have the same sampling pattern as that of the observed light curve. Finally, the measurement errors in the simulated flux values were incorporated by adding a Gaussian random variable with mean 0 and standard deviation equal to the mean error of the measurement uncertainties on the observed flux values \citep[][]{Meyer19}. In such a manner, 1,000 light curves are simulated in the $\beta$ range 0.2 to 3.0, with a step of 0.1 for each observed light curve. The periodograms are derived for each simulated light curve in the same manner as that of the observed light curve. The probability distribution curves as a function of $\beta$ for the analysed PSDs are given in the Appendix~\ref{app:B}. The best-fit PSD slope for the observed PSD is given by the one with the highest $p_{\beta}$ value. The errors on the best-fit PSD slope is obtained by fitting a Gaussian function to the $p_{\beta}$ curve and gives the full-width at half maximum \citep[FWHM;][]{Chatterjee08, Bhatta16b}. This gives a roughly 98 percent confidence interval on the best-fit PSD slopes.                  

All observed PSDs, along with the best-fit PSDs and the maximum $p_{\beta}$, are summarized in Table~\ref{tab:psd}. The PSDs are displayed in Figs.~\ref{fig:3} and ~\ref{fig:4} for the actual duration of the corresponding light curves, down to the observed (mean) sampling intervals. In our analysis, we have not subtracted the constant noise level (shown by the dashed horizontal lines in the figures), as some of the data points are below this level. In all the cases, the probability is higher than 10 percent (except for the OVRO light curve of the blazar Mrk\,421 where it is 8.4 percent), meaning that the rejection confidence (1--$p_{\beta}$ value) is lower than 90 percent for the modeled best-fit PSDs. This means that the single power-law PSD shape provides a good fit to the PSDs studied here. Furthermore, we also compute the square fractional variability by multiplying $\nu_k$ with the corresponding $P(\nu_k$) for the best-fit spectral shapes; such estimates are equivalent to fractional variability, F$_{var}$=$\frac{rms}{mean}$, where rms is the standard deviation of the light curve \citep[][]{Vaughan03}. Fig.~\ref{fig:5} presents the composite square fractional variability vs. variability frequency estimates for the blazars Mrk\,421 (panel a) and PKS\,2155$-$304 (panels b and c), respectively. In such a representation, one can explicitly compare variability amplitudes at different wavelengths on different variability timescales. 

\begin{table*}
\caption{Parameters of the observed light curves and PSD analysis\label{tab:psd}}
\scriptsize
\begin{tabular}{cccccccccccc}\hline
\hline
Light curve & Monitoring epoch  &  $T_{\rm obs}$  &  $N_{obs}$  & $\Delta T_{\rm min}$  &  $\Delta T_{\rm max}$  & $T_{\rm mean}$  &  $T_{\rm int}$ & $\rm \log_{10}(P_{stat})$ & $\log_{10}(\nu_k) $ range &  {$\beta \pm err$}  & $p_\beta$ \\    
             &                     & (yr)            &             &   (d)                 &    (d)                 &    (d)          &   (d)       & ($\frac{\mathrm rms}{\mathrm mean})^2$d & (d$^{-1}$)  &    &  \\
(1)                            &     (2)               &   (3)       &  (4) &  (5)   &  (6)  &  (7)  &  (8) &  (9)       &  (10)        & (11)    & (12)   \\\hline
\multicolumn{12}{c}{Mrk\,421} \\
\hline
VERITAS ($>$400 GeV)           &  1995 Dec 19--2009 May 29   & 13.4      & 783   &  0.5     &  210  &  6.2  &  0.5 &  $+$0.07  &     $-$3.7 to $-$1.2   & 1.1$\pm$0.5  &  0.381   \\
{\it Fermi}--LAT (0.1--300 GeV)&  2008 Aug 8--2018 Mar 16    & 9.6       & 497   &  1     &  21   &  7.1  &  1.0 &  $-$0.96  &     $-$3.5 to $-$1.3   & 1.1$\pm$0.4  &  0.156   \\
{\it RXTE}--PCA (3--20 keV)    &  1996 Mar 1--2011 Dec 13    & 15.8      & 549   &  0.5   &  620  &  10.5 &  0.5 &  $-$1.83  &     $-$3.7 to $-$1.5   & 1.1$\pm$1.6  &  0.995   \\
{\it Swift}--XRT (0.3--10 keV) &  2005 Mar 31--2019 Jan 17   & 13.8      & 576   &  0.5    &  288  &  8.7  &  0.5 &  $-$2.16  &     $-$3.7 to $-$1.5   & 1.3$\pm$0.7  &  0.996   \\
OVRO  (15 GHz)                 &  2008 Jan 8--2018 Apr 5     & 10.2      & 516   &  1     &  86   &  7.2  &  1.0 &  $-$1.48  &     $-$3.6 to $-$1.3   &1.6$\pm$0.3  &  0.084   \\ \hline
\multicolumn{12}{c}{PKS\,2155$-$304} \\                                                                                                                             
\hline                                                                                                                                                              
H.E.S.S. ($>$200 GeV)          &  2004 Jul 14--2012 Nov 15    &  8.4      & 232   &   0.5    & 347   &  13   &  0.5 &  $-$0.37  &     $-$3.5 to $-$1.6   &0.6$\pm$1.4  &  0.959   \\
{\it Fermi}-LAT (0.1--300 GeV) &  2008 Aug 8--2017 Nov 3      &  9.2      & 471   &   1    & 21    &  7.1  &  1.0 &  $+$0.24  &     $-$3.5 to $-$1.3   & 1.3$\pm$0.6  &  0.653   \\
{\it Swift}-XRT (0.3--10 keV)  &  2005 Nov 17--2016 Oct 25    &  10.9     & 144   &   0.5    & 386   &  27   &  0.5 &  $-$0.72  &     $-$3.6 to $-$1.9   & 1.3$\pm$2.1  &  0.419   \\
Optical (B)                    &  2008 May 17--2015 Feb 10    &  7.4      & 462   &   0.5  & 303   &  5.8  &  0.5 &  $-$3.95  &     $-$3.4 to $-$1.2   &1.6$\pm$0.5  &  0.950   \\
Optical (V)                    &  2005 May 14--2015 Feb 10    &  10.9     & 657   &   0.5  & 325   &  5.7  &  0.5 &  $-$2.47  &     $-$3.6 to $-$1.3   & 1.7$\pm$0.6  &  0.542   \\
Optical (R)                    &  2005 May 14--2015 Feb 10    &  10.4     & 670   &   0.5  & 321   &  5.7  &  0.5 &  $-$2.96  &     $-$3.6 to $-$1.1   & 1.7$\pm$0.6  &  0.806   \\
Optical (I)                    &  2005 May 14--2012 May 28    &  7.0      & 281   &   0.5  & 321   &  9.1  &  0.5 &  $-$2.33  &     $-$3.4 to $-$1.4   &1.8$\pm$0.8  &  0.984   \\
IR      (J)                    &  2006 Jun 10--2015 Feb 10   &  9.3      & 601   &   0.5    & 321   &  5.6  &  0.5 &  $-$2.95  &     $-$3.5 to $-$1.1    &1.8$\pm$1.1  &  0.769   \\
IR      (H)                    &  2005 May 18--2012 May 29    &  7.0      & 258   &  0.5    & 380   &  9.9  &  0.5 &  $-$2.65  &     $-$3.4 to $-$1.6   &1.5$\pm$0.7  &  0.993   \\
IR      (K)                    &  2005 Apr 11--2012 May 30    &  6.5      & 242   &  0.5    & 324   &  9.8  &  0.5 &  $-$2.07  &     $-$3.4 to $-$1.5   &1.9$\pm$1.3  &  0.686   \\
\hline
\end{tabular}
\begin{minipage}{\textwidth}
Columns : (1) Light curve (observed photon energy/frequency);
(2) the epoch of monitoring for the light curve (start--end); 
(3) the total duration of the observed light curve;
(4) the number of data points in the observed light curve;
(5) the minimum sampling interval for the observed light curve; 
(5) the maximum sampling interval for the observed light curve; 
(7) the mean sampling interval for the observed light curve (light curve duration/number of data points); 
(8) the sampling interval for the interpolated light curve; 
(9) the noise level in PSD due to the measurement uncertainty;
(10) the temporal frequency range covered by the  binned logarithmic power spectra;
(11) the best-fit power-law slope of the PSD along with the corresponding errors representing 98 per cent confidence intervals (see Sec.~\ref{sec:PSD});
(12) corresponding $p_\beta$.    

\end{minipage}
\end{table*}

\begin{figure*}
\centering
\hbox{
\includegraphics[width=0.33\textwidth]{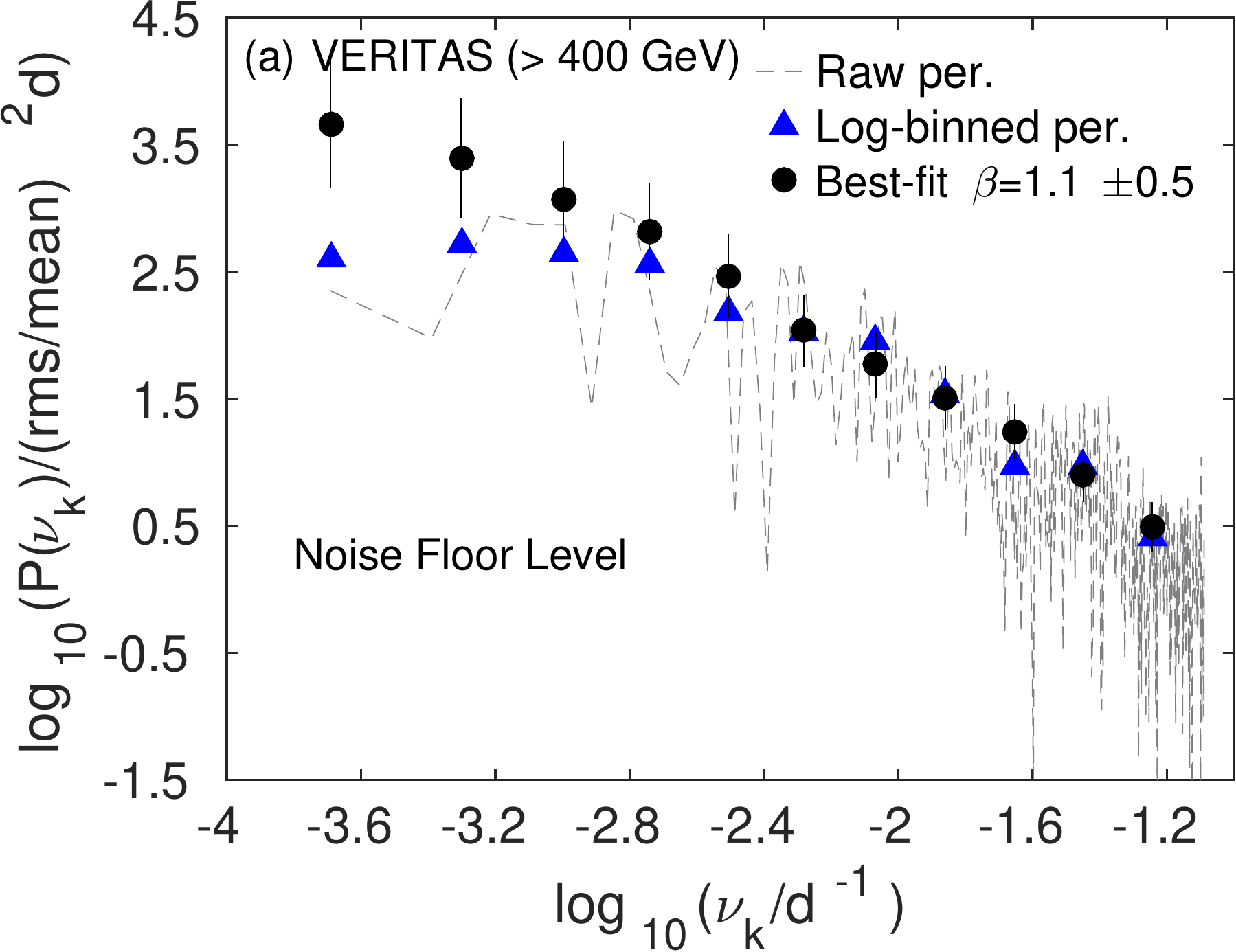}
\includegraphics[width=0.33\textwidth]{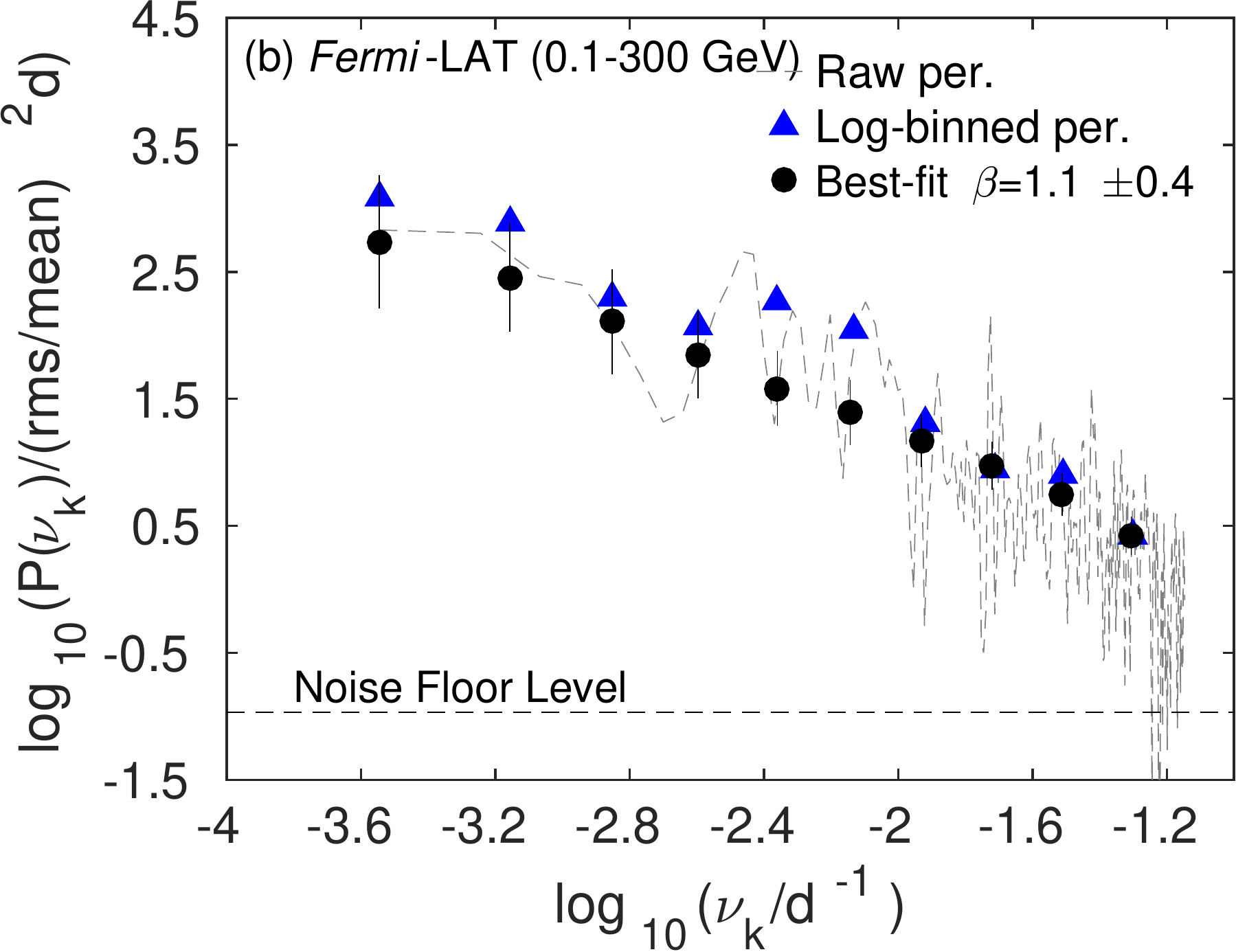}
\includegraphics[width=0.33\textwidth]{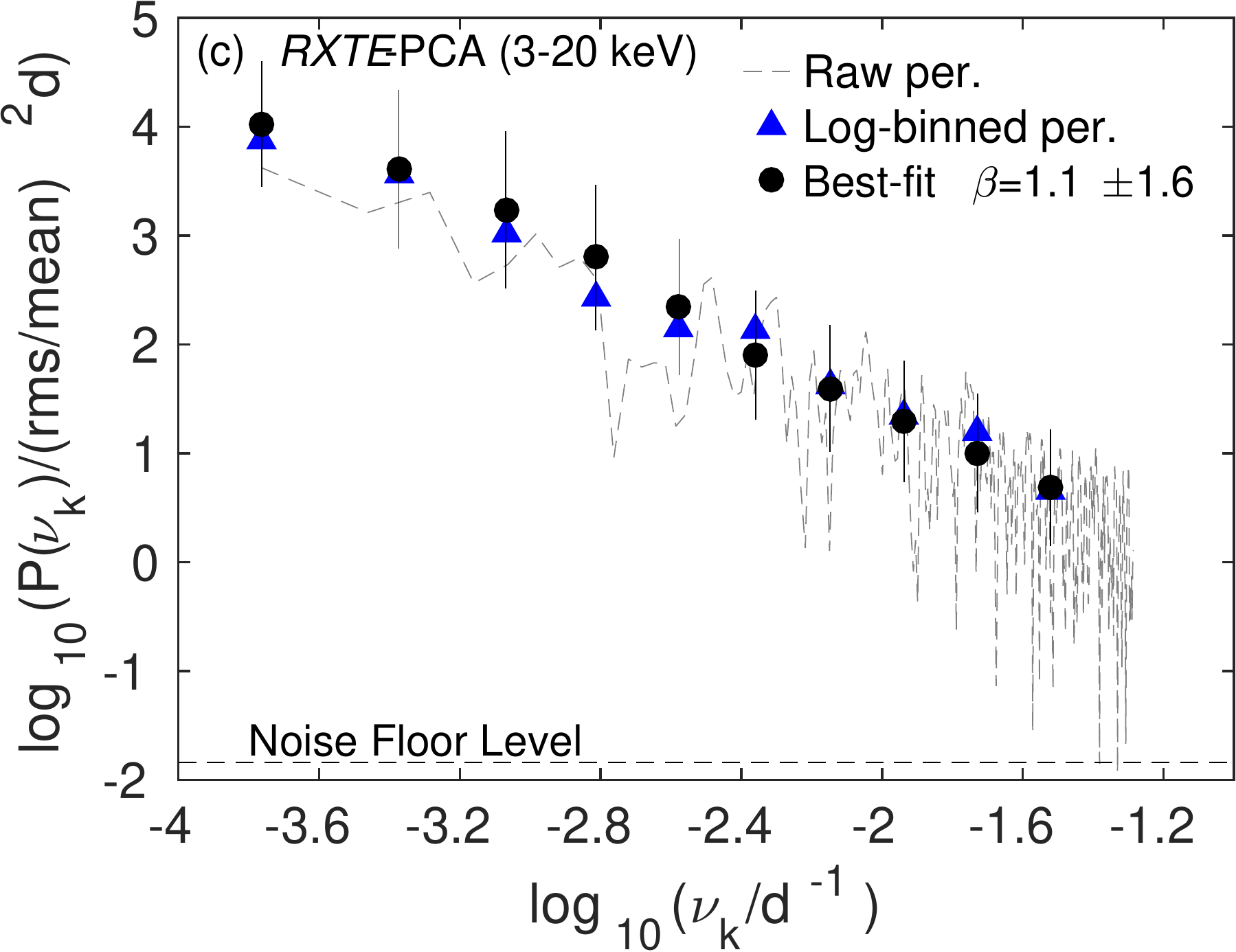}
}
\includegraphics[width=0.33\textwidth]{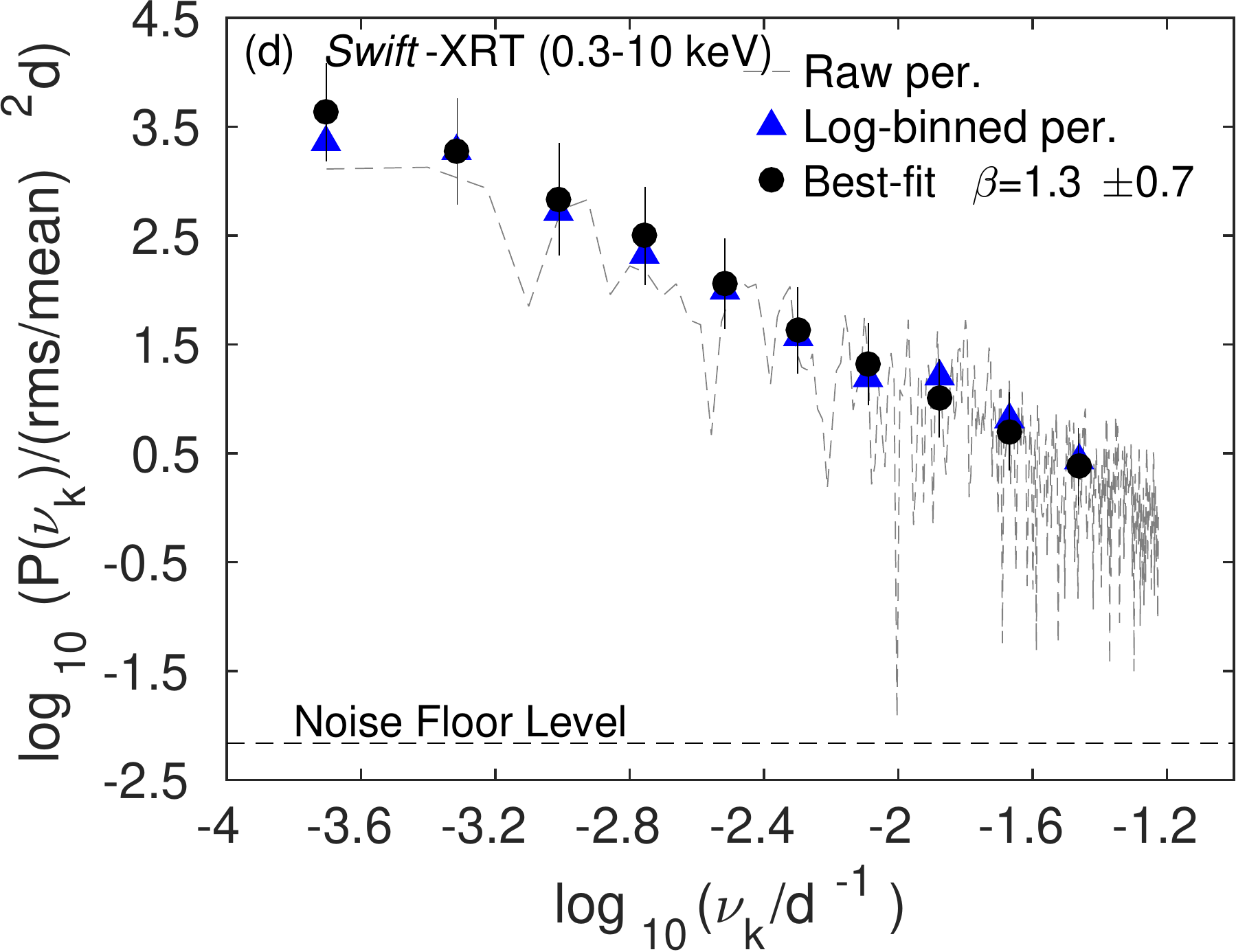}
\includegraphics[width=0.33\textwidth]{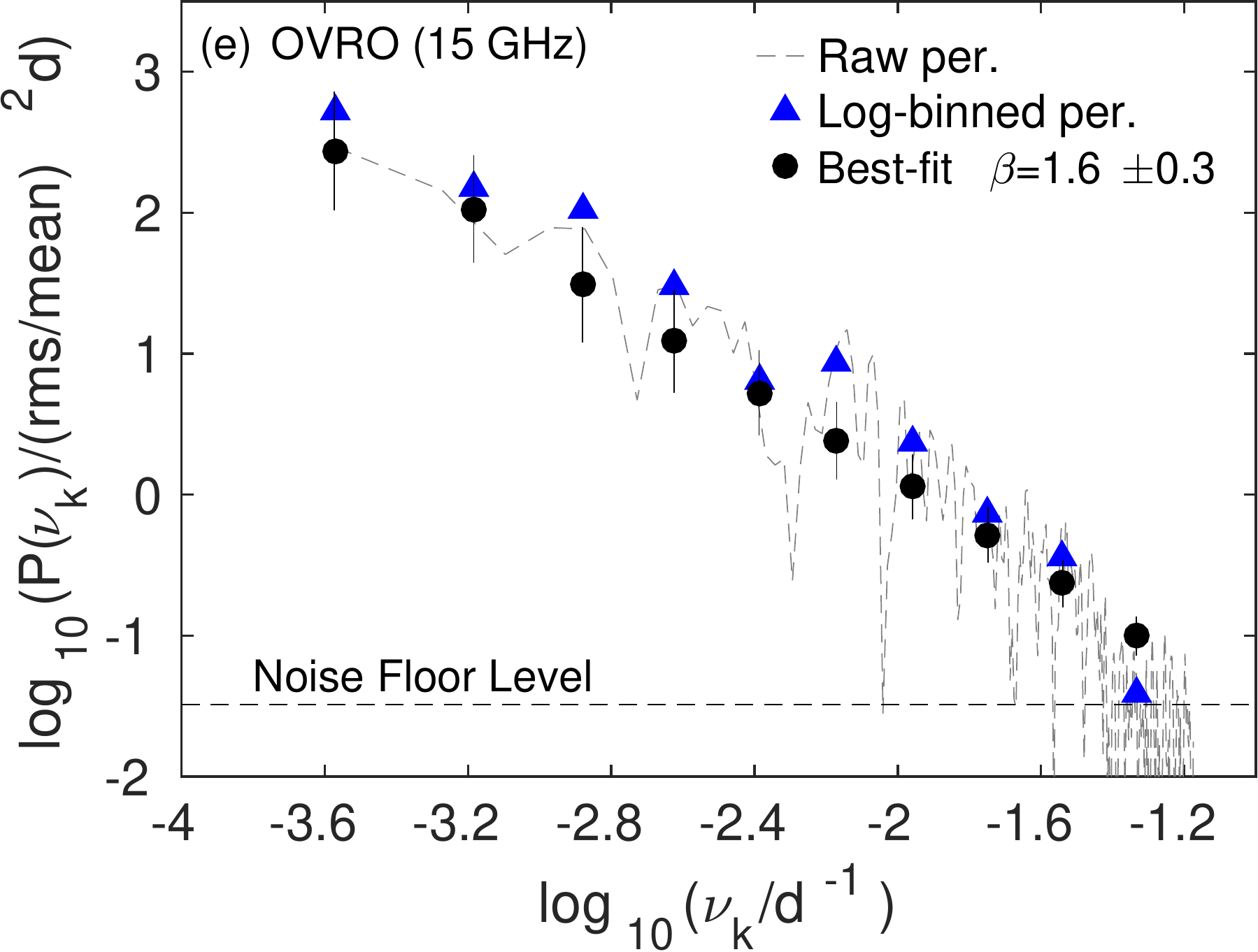}
\begin{minipage}{\textwidth}
\caption{PSDs derived using the multiwavelength light curves shown in Fig.~\ref{fig:1} for the blazar Mrk\,421. 
The dashed gray line and blue triangles and open circles denote the raw and log-binned periodogram estimates, respectively, while the solid black circles represent the mean values of 1,000 simulations for the best-fit PSD model, obtained using the PSRESP method (see Section~\ref{sec:PSD}). The error bars on the mean values are the standard deviation of the distribution of simulated PSDs. The dashed horizontal line indicates the noise floor level due to the measurement error achieved (Table~\ref{tab:psd}).}
\label{fig:3}%
\end{minipage}
\end{figure*}

\begin{figure*}
\centering
\hbox{
\includegraphics[width=0.33\textwidth]{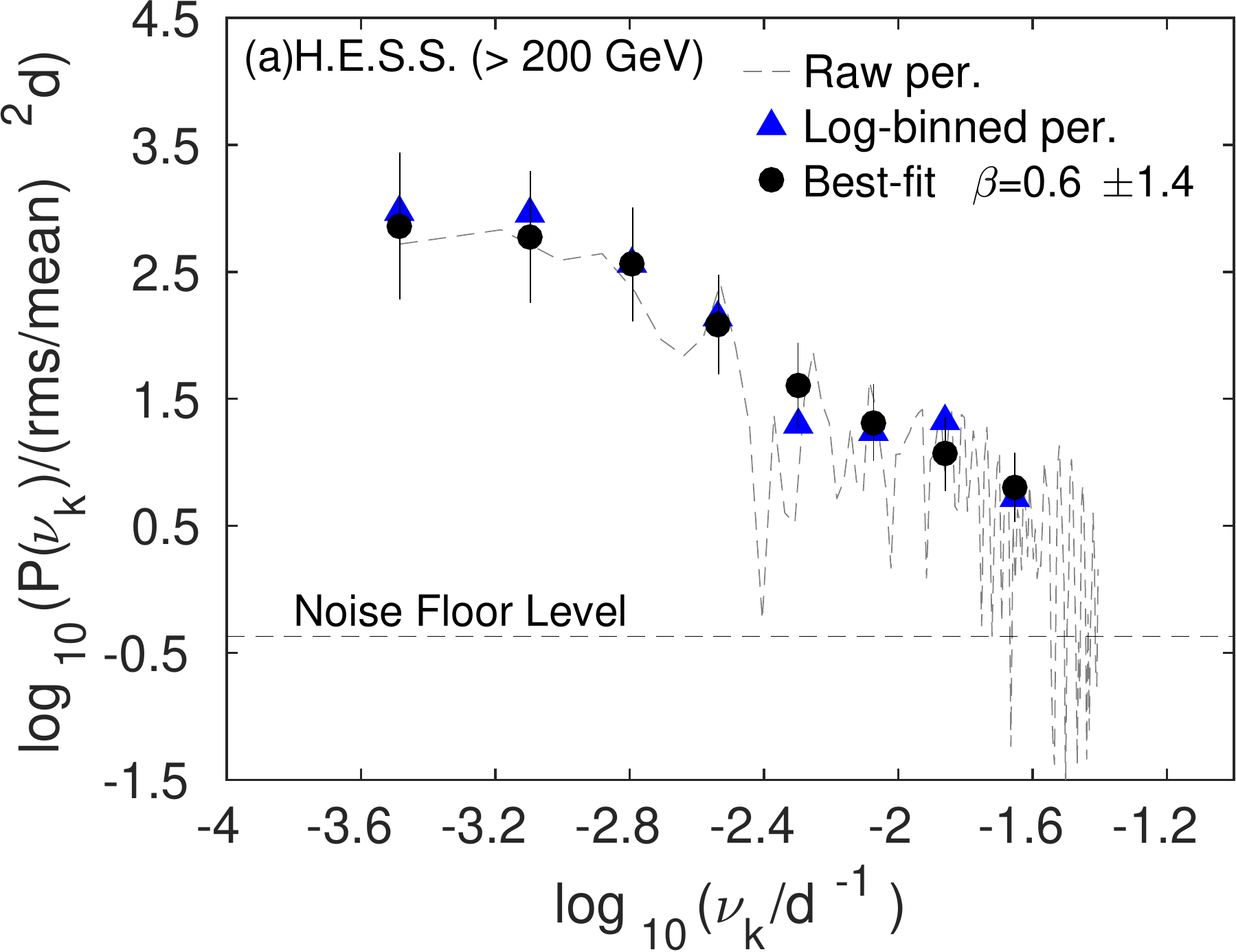}
\includegraphics[width=0.33\textwidth]{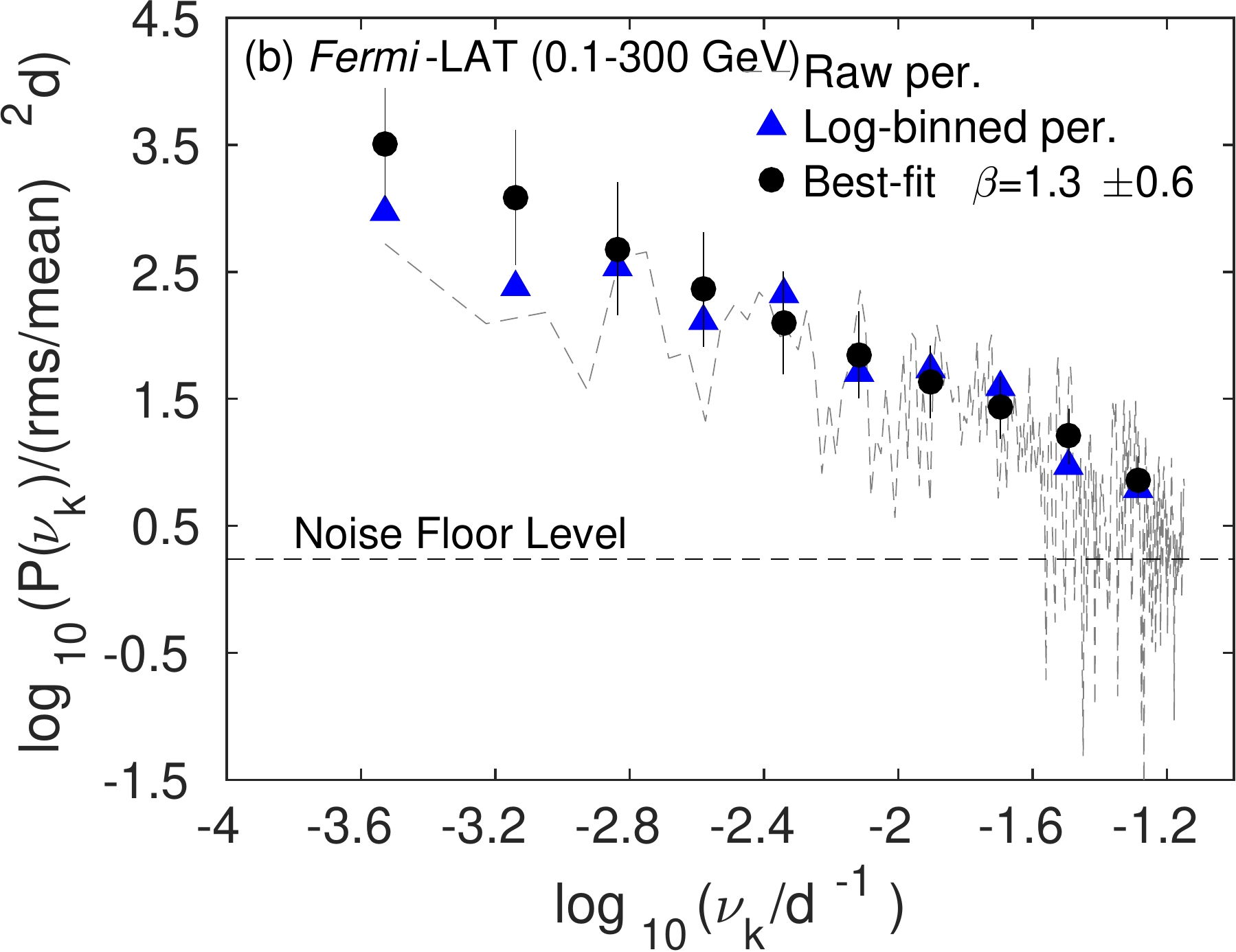}
\includegraphics[width=0.33\textwidth]{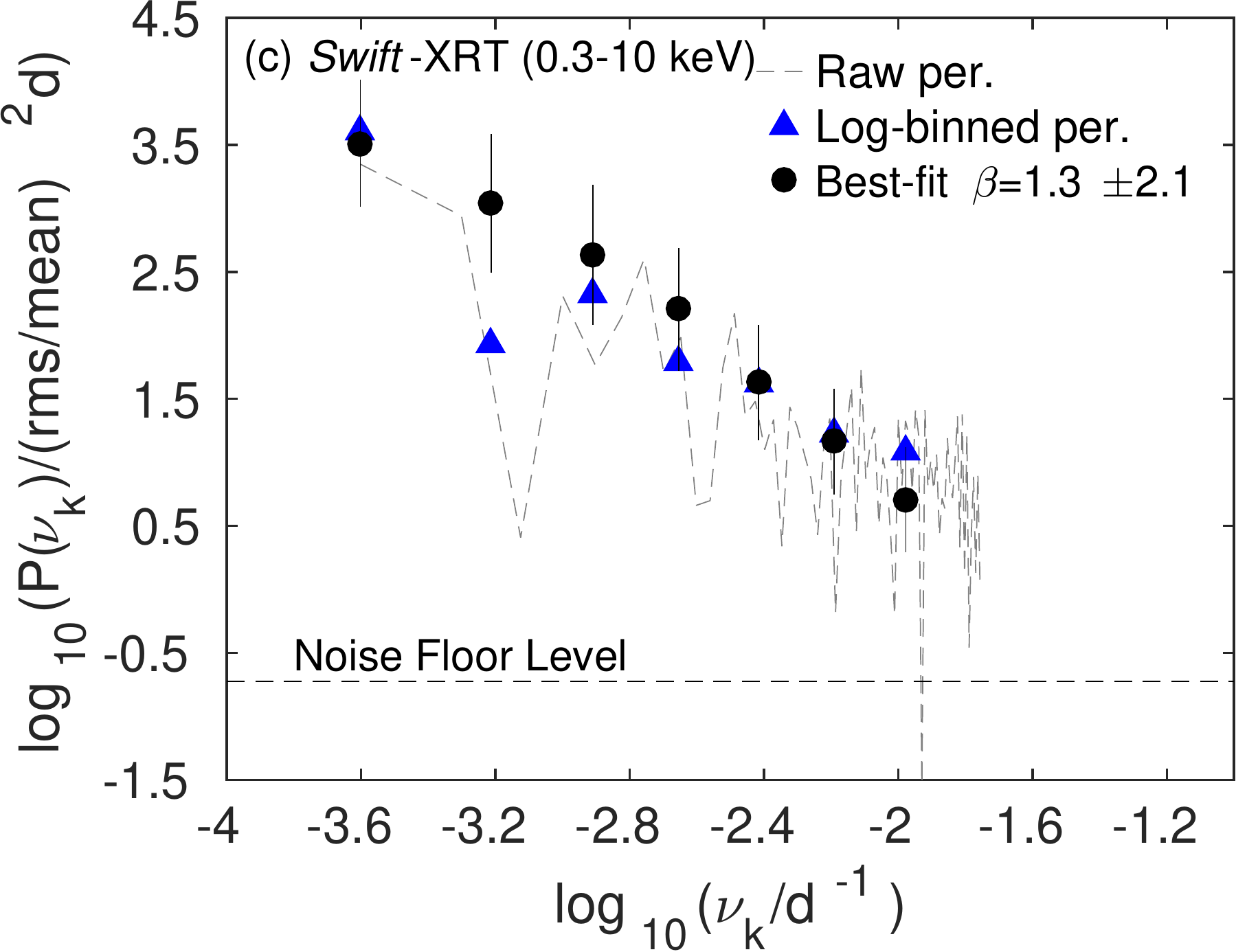}
}
\hbox{
\includegraphics[width=0.33\textwidth]{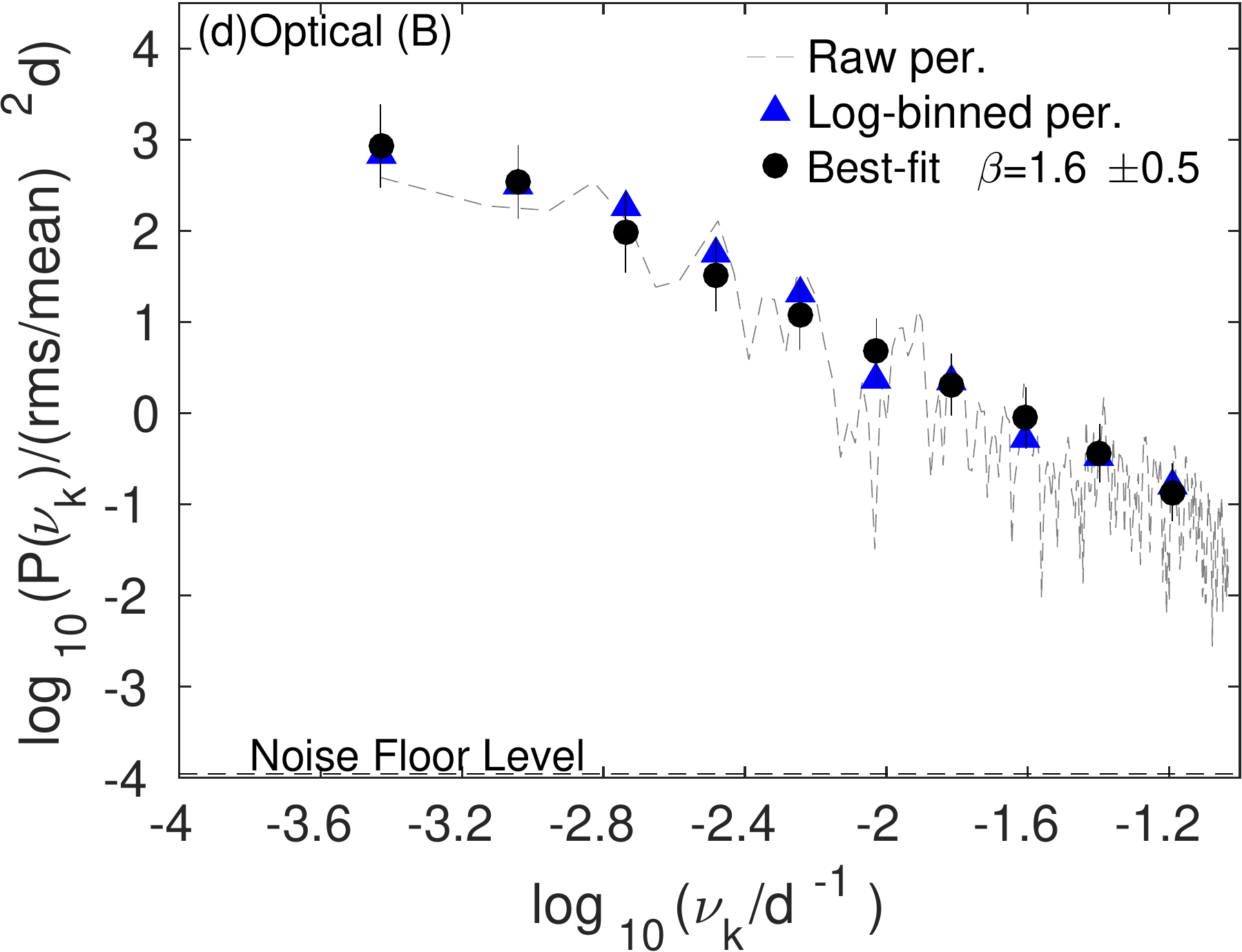}
\includegraphics[width=0.33\textwidth]{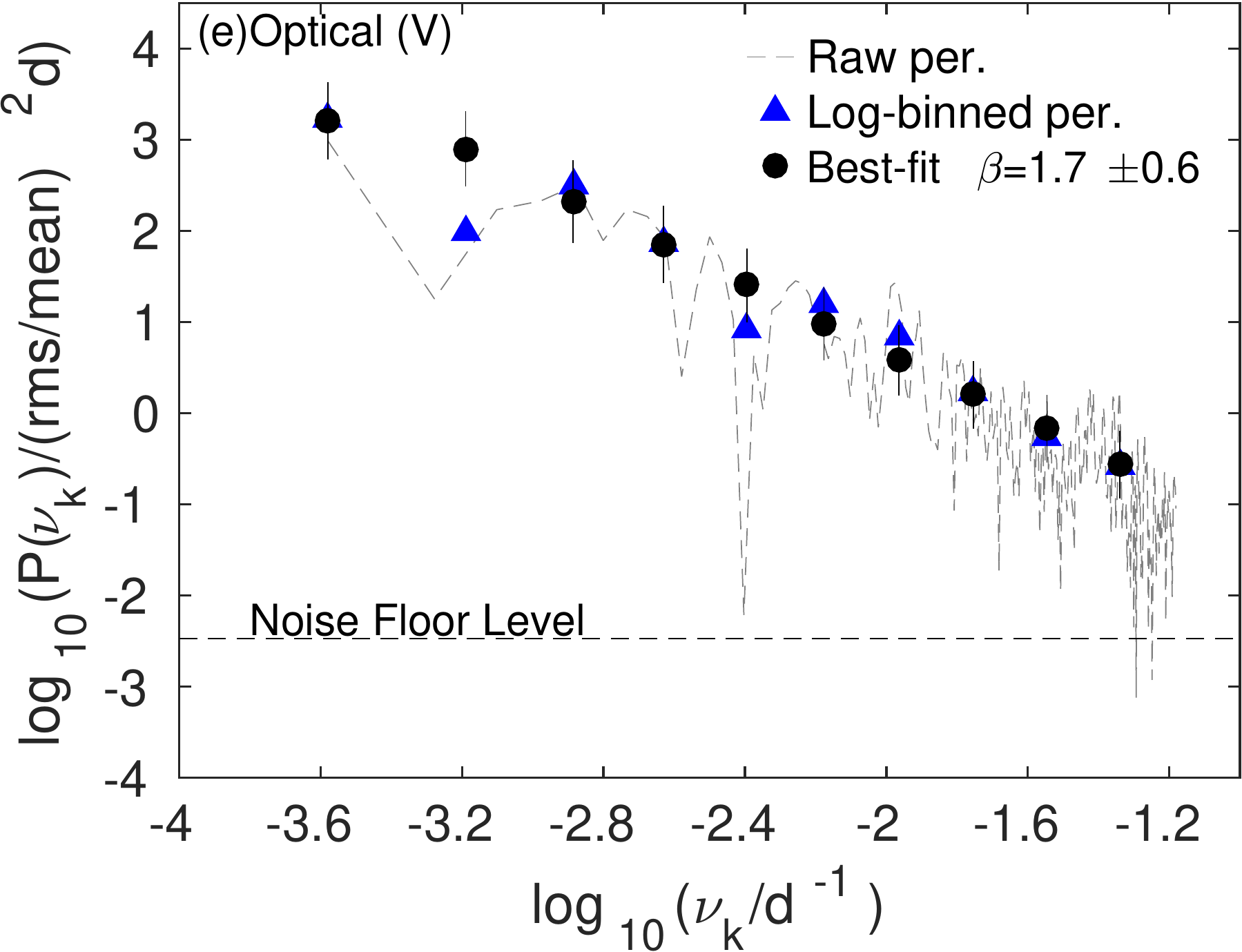}
\includegraphics[width=0.33\textwidth]{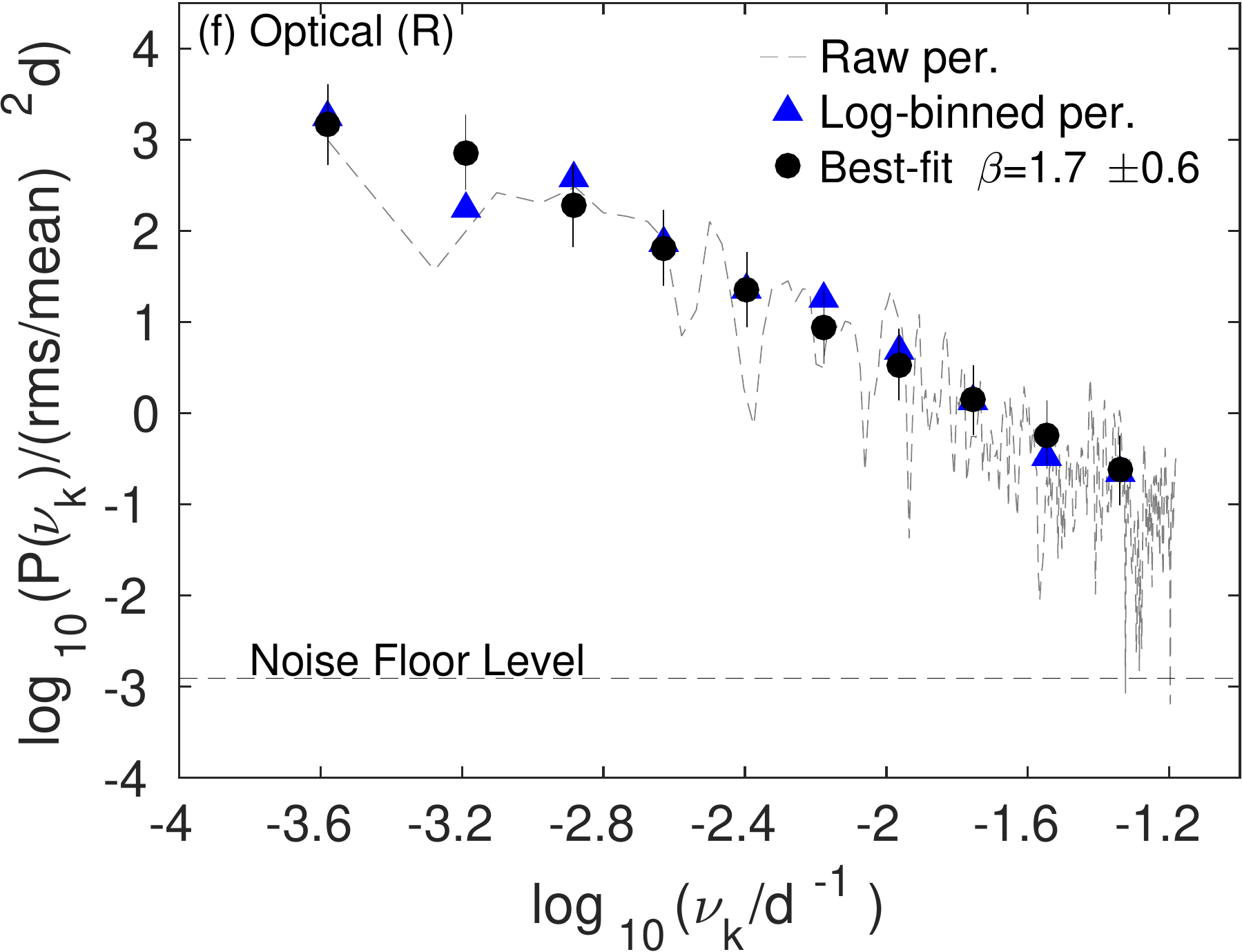}
}
\hbox{
\includegraphics[width=0.33\textwidth]{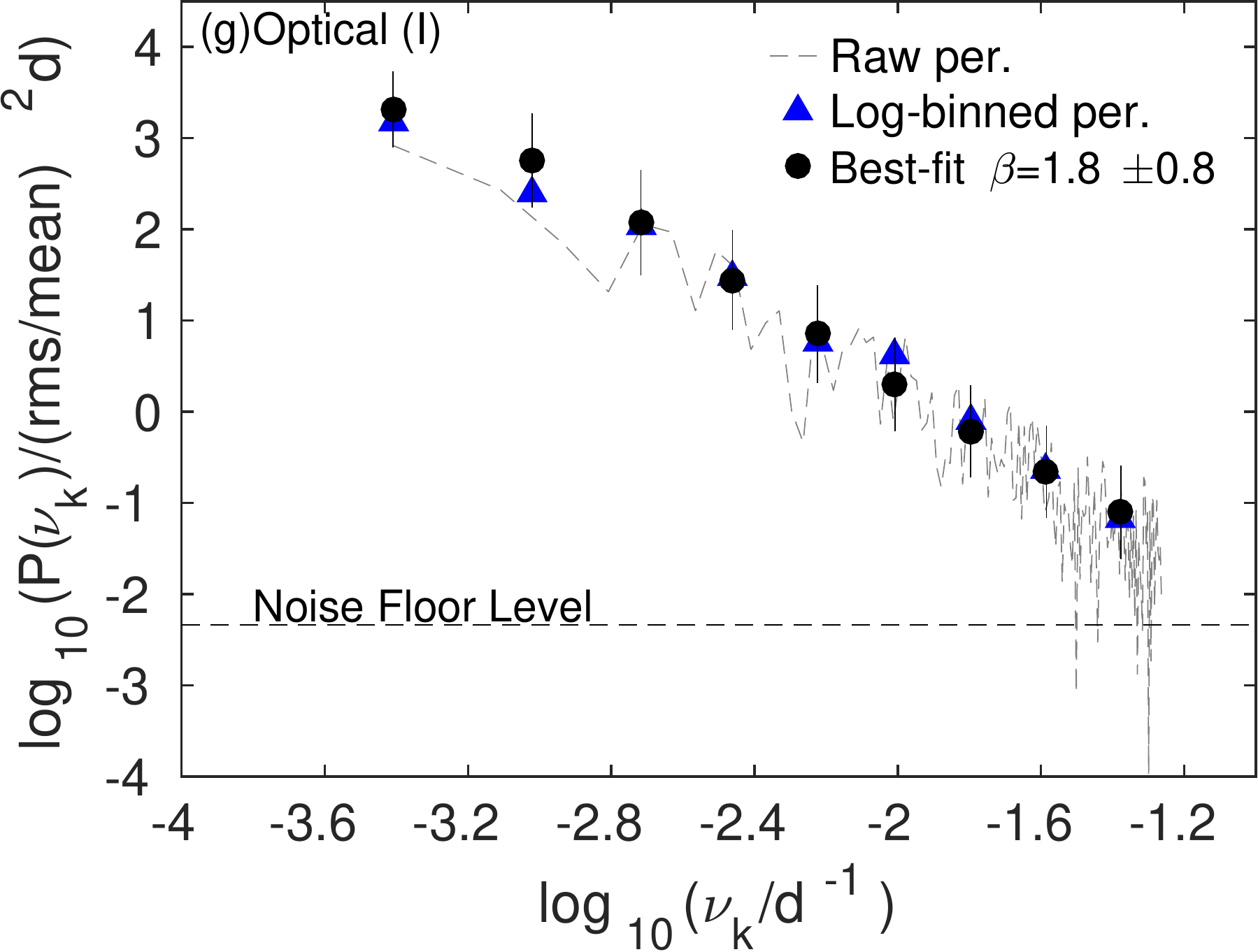}
\includegraphics[width=0.33\textwidth]{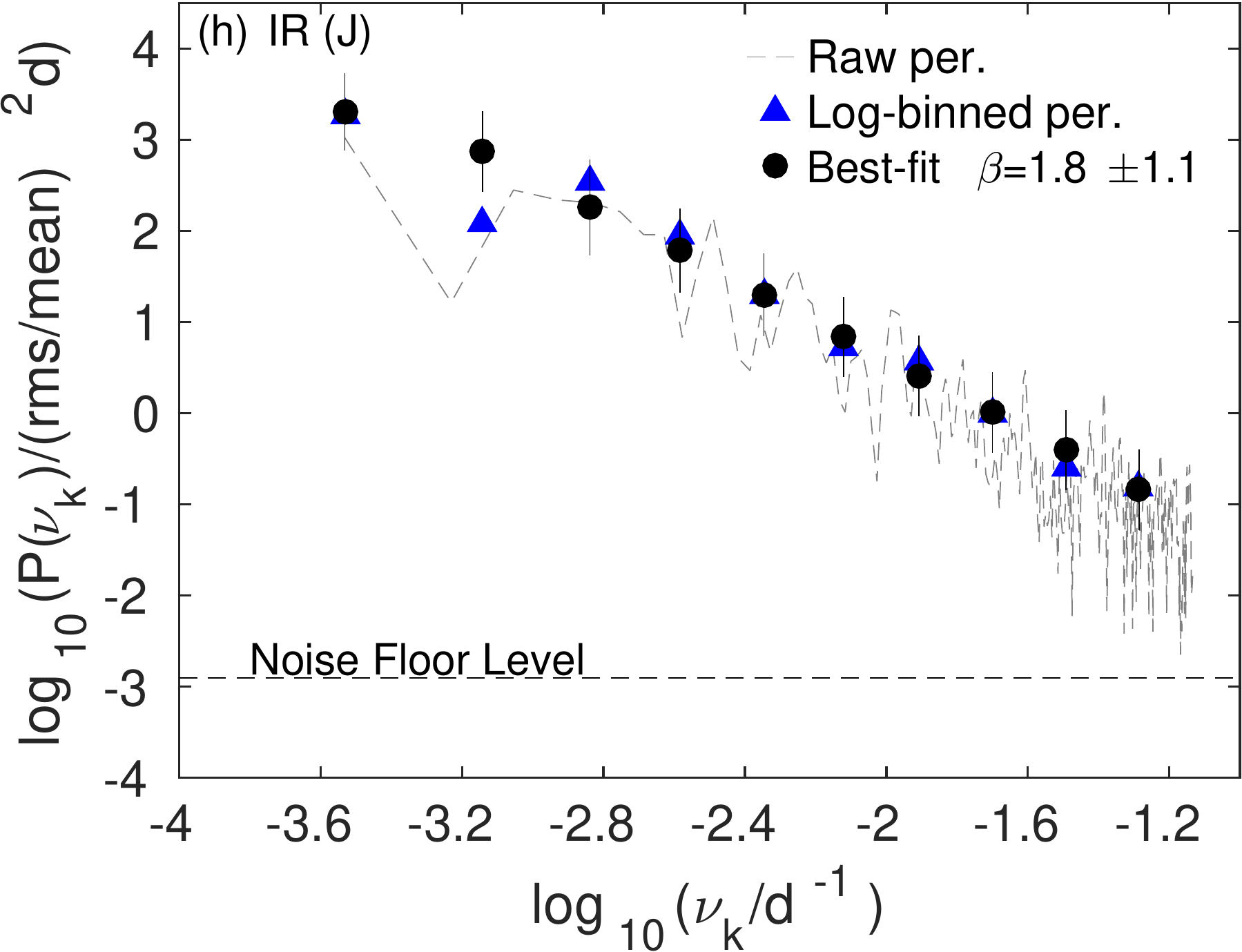}
\includegraphics[width=0.33\textwidth]{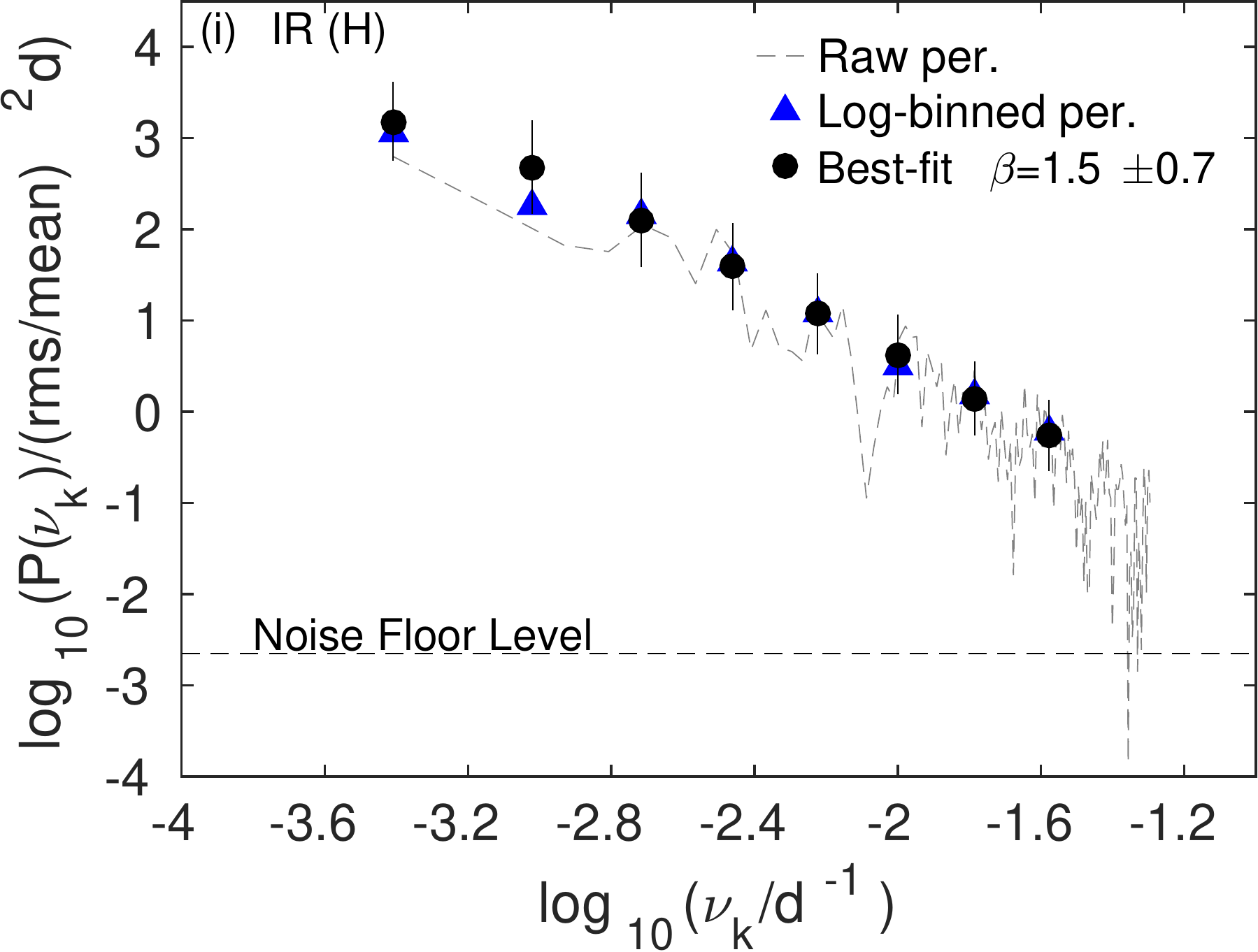}
}
\includegraphics[width=0.33\textwidth]{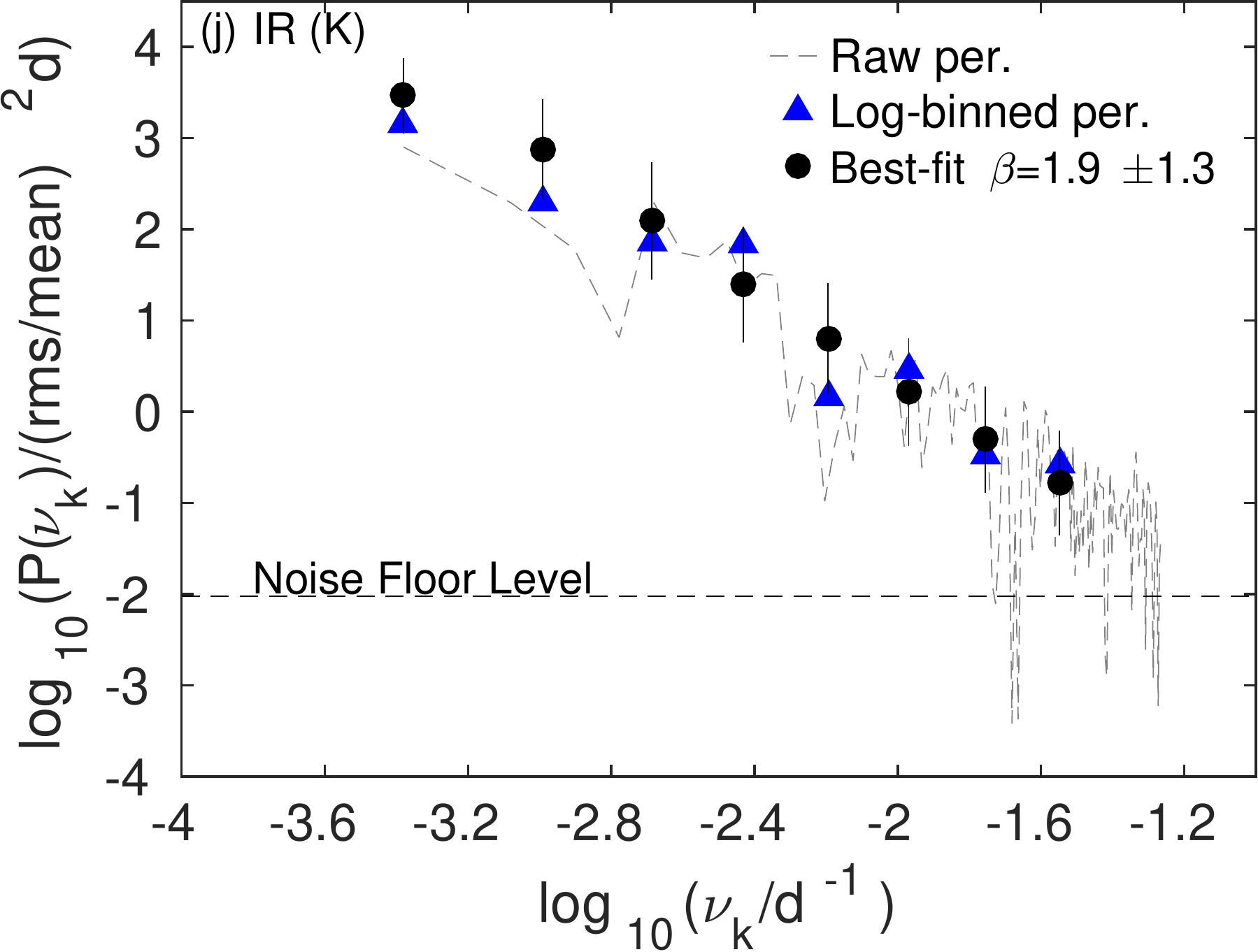}
\begin{minipage}{\textwidth}
\caption{PSDs derived using the multiwavelength light curves shown in Fig.~\ref{fig:2} for the blazar PKS\,2155$-$304. The dashed gray line and blue triangles or open circles denote the raw and log-binned periodogram estimates, respectively, while the solid circles represent the mean values of 1,000 simulations for the best-fit PSD model, obtained using the PSRESP method (see Section~\ref{sec:PSD}). The error bars on the mean values are the standard deviation of the distribution of simulated PSDs. The dashed horizontal line indicates the noise floor level due to the measurement error achieved (Table~\ref{tab:psd}).}
\label{fig:4}%
\end{minipage}
\end{figure*}

\section{Results}
\label{sec:result}

Here we present the results of our PSD analysis of multiwavelength light curves of the TeV blazars Mrk\,421 and PKS\,2155$-$304 on timescales ranging from about a decade down to weeks or days. For the analysis we have used good-quality, roughly decade-long light curves sampled around once a week at 12 different frequencies, covering $\sim$17 decades of the electromagnetic spectrum. Our main findings are:

\begin{enumerate}   

\item{For Mrk\,421, the VHE, HE\,$\gamma-$ray and X-ray PSDs on the timescales from years to months, can all be well represented by a single power-law function with the slopes  $\beta\simeq1.1\pm0.5$, $\simeq1.1\pm0.4$, $\simeq1.1\pm1.6$ and  $\simeq1.3\pm0.7$, respectively (panels a, b, c and d of Fig.~\ref{fig:3} and Table~\ref{tab:psd}). Similarly, for PKS\,2155$-$304 and on the corresponding timescales, the PSDs are represented by a single power-law function with the slopes $\beta\simeq0.6\pm1.4$, $\simeq1.3\pm0.6$, and $\simeq1.3\pm2.1$, respectively  (panels a, b and c of Fig.~\ref{fig:4} and Table~\ref{tab:psd}). The variability thus is indicative of essentially ``pink/flicker noise'' type at these higher frequencies of the electromagnetic spectrum. We note that the PSD slopes derived here for the X-ray light curves are in good agreement with $\beta \sim$1.2--1.5 reported in \citet{Isobe15, Chatterjee18} for the blazar Mrk\,421. Similarly, the PSD slopes derived at VHE and HE\,$\gamma-$ray light curves by means of the DFT method, are in good agreement with $\beta \sim$1.2 and $\sim$1.1 reported in \citet[][]{Hess17}, who used other methods for the PSD characterization. }

\item{The gathered optical-IR monitoring data for PKS\,2155$-$304 reveal large changes in intensities on vastly different timescales. The roughly daily sampled (barring the interval when the target was close to the Sun), light curves at B, V, R, I, J, H, and K bands closely resemble one another, with a factor of a few changes in intensity on weeks/months timescales, and a few percent on the timescales of days. The overall optical and IR PSDs, derived on the timescales from years to weeks, can be well represented within errors by a single power-law function of slope $\beta \simeq 1.8$, meaning a ``pure red noise'' type of the source variability close to peak of the synchrotron component of the electromagnetic spectrum (panels d, e, f, g, h, i, and j of Fig.~\ref{fig:4}, and Table~\ref{tab:psd}). } 

\item{The PSD slope derived using the 15\,GHz radio light curve for the blazar Mrk\,421 on timescales ranging from years to weeks is $\beta\sim$1.6$\pm$0.3, indicative of a statistical character of red noise type of source variability at these frequencies (panel e of Fig.~\ref{fig:3} and Table~\ref{tab:psd}).}

\item{An explicit comparison of squared fractional variability (i.e., $\nu_k P(\nu_k)$ versus $\nu_k$) between higher (VHE and HE $\gamma-$ray) energy bands and lower (radio--to--optical) of the electromagnetic spectrum, reveals more variability power at the higher frequency bands on timescales $\leq$100 days (panels a and b of Fig.~\ref{fig:5}). Simiar behaviour is shown by variability at X--rays energies for Mrk\,421 as compared to radio energies (panel a of Fig.~\ref{fig:5}). The X-ray PSD for PKS\,2155$-$304 could not be constrained on such short variability timescales due to the very sparse sampling of the X-ray light curve.}

\item{The PSDs generated using our long-duration optical, IR, and {\it Fermi}-LAT light curves do not reveal the QPOs reported in \citet{Sandrinelli16} and \citet{Zhang17} for PKS\,2155$-$304 as the observed power spectra are distributed within 1$\sigma$ confidence regions of best-fit PSDs. This could be either due to the marginal significance of the detected features ($\lesssim$3$\sigma$), or it could be indicative of a transitory nature of the QPOs \citep[see,][for a detailed discussion]{Vaughan16, Bhatta16a}. We note, however, that we have not followed Lomb-Scargle periodogram or Weighted Wavelet Z-transform methods which are used to detect periodicities in the original publications which could potentially cause the difference. }
 
\end{enumerate}

\begin{figure*}
\hbox{
\includegraphics[width=0.33\textwidth]{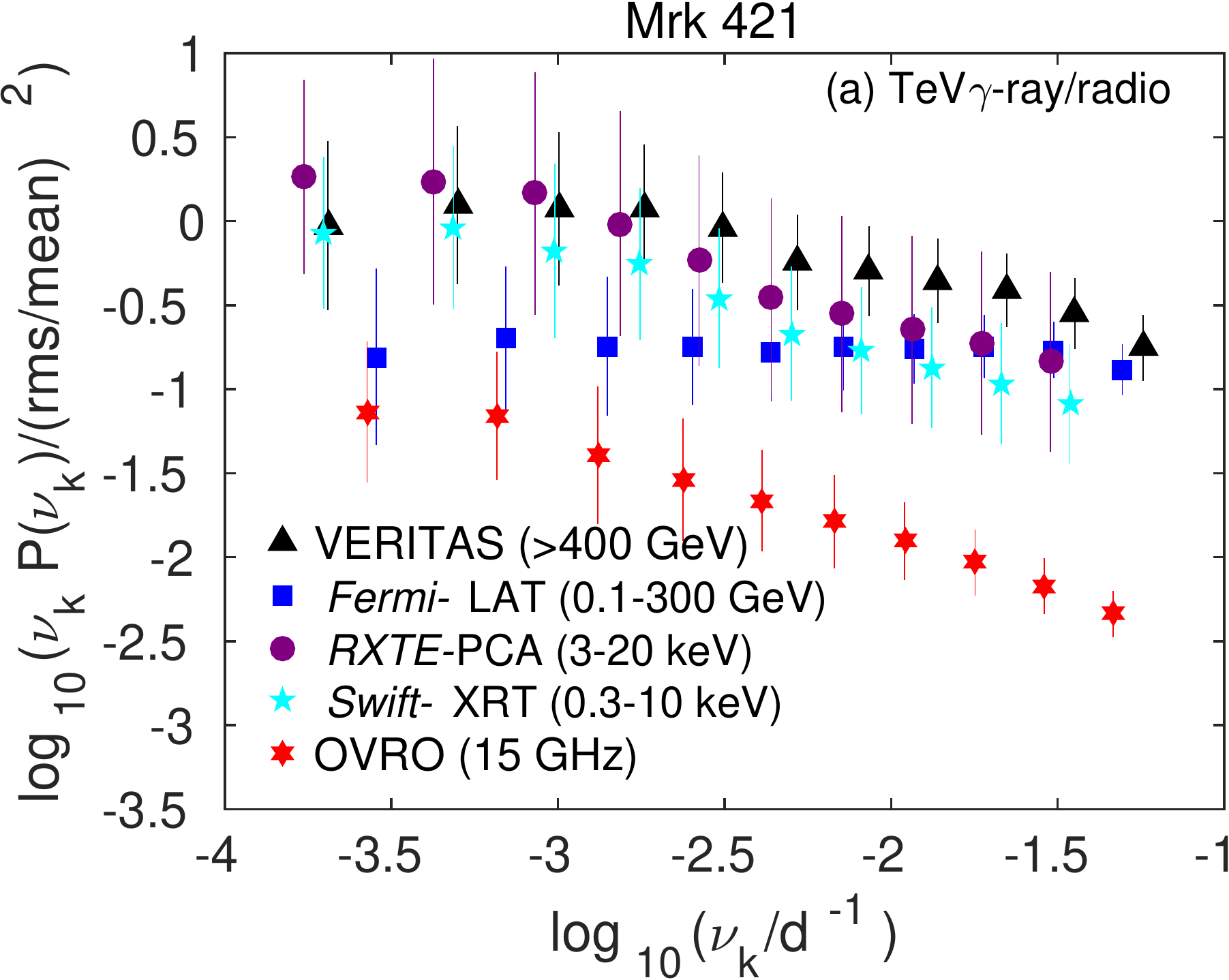}
\includegraphics[width=0.33\textwidth]{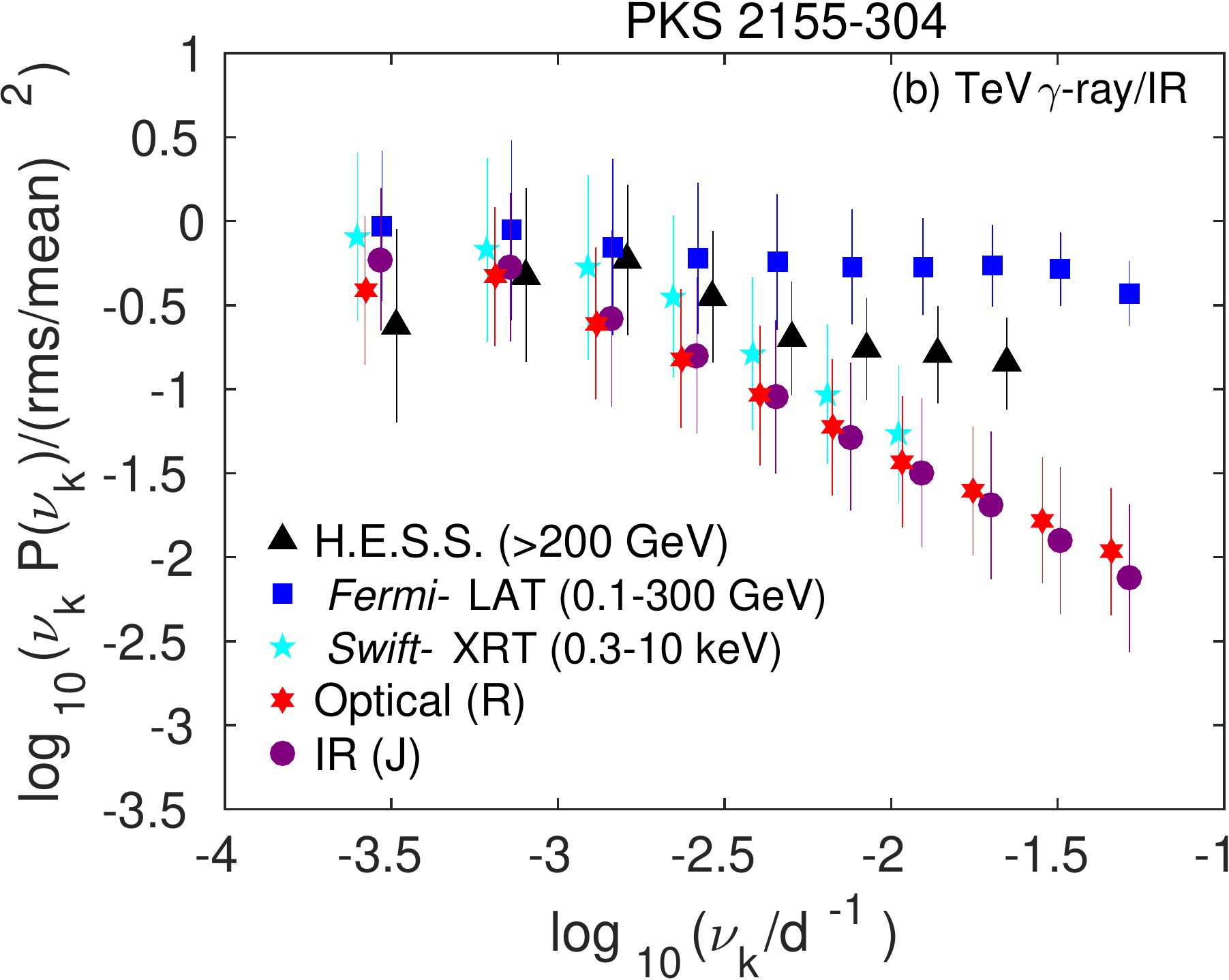}
\includegraphics[width=0.33\textwidth]{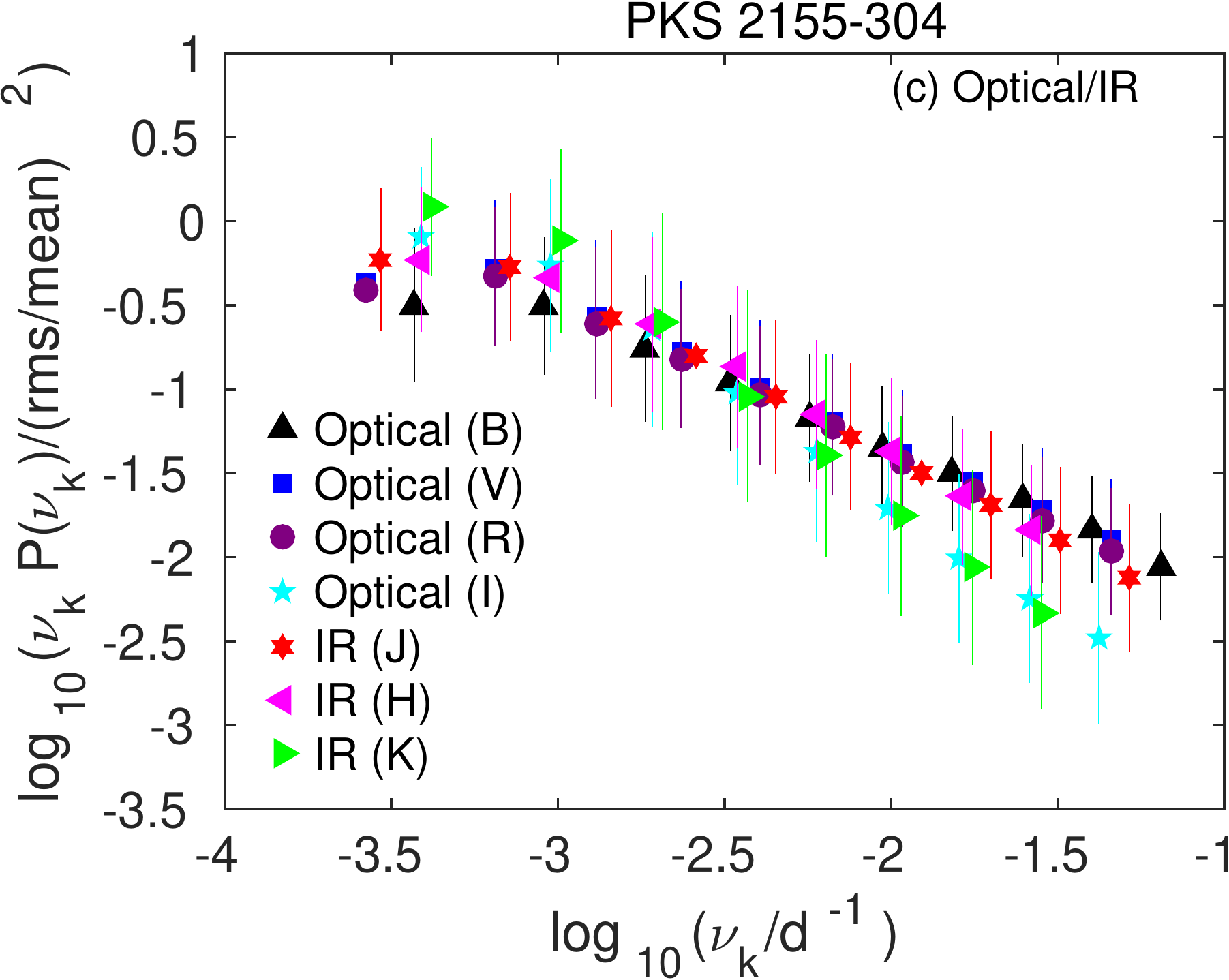}
}
\begin{minipage}{\textwidth}
\caption{The composite multiwavelength square fractional variability of Mrk\,421 (panel a) and PKS\,2155$-$304 (panels b and c).} 
\label{fig:5}
\end{minipage}
\end{figure*}

\section{Discussion and Conclusions}
\label{sec:conclusion}
  
Based on our PSD analyses of multi-band (GHz band radio, R-band, and $\it Fermi-$LAT) light curves of the two LBLs, namely PKS\,0735$+$178 and OJ\,287 \citep{Goyal17, Goyal18}, we have argued that the observed broad-band emission of the blazars is generated in the \emph{extended} volumes of structured, highly turbulent jets (as opposed to the scenarios postulating a well-defined single ``emission zone''). That is because the Fourier decomposition of the analysed light curves did not reveal any single well-defined variability timescale, but instead a wide range of such, manifesting as the colored-noise type of the derived power spectra.  We emphasize, that a single/broken power-law shape of the blazar PSDs is a general finding of the timing analysis performed by various authors \citep[e.g.,][]{Kataoka01, Hess10, Abdo10a, Kastendieck11, Sobolewska14, Park14, Isobe15, Kushwaha16, Hess17,Chatterjee18}, and that the presence of QPOs in blazar lightcurves is still an open issue due to typically low significance of the claimed features (see in this context \citealt{Covino19}, and references therein; also, see the discussion on the 12-year-long periodicity in the optical light curve of OJ\,287 in Appendix~C of \citealt{Goyal18}). Moreover, in order to account for the specific colours of the noise observed, in \citet{Goyal17, Goyal18} we have proposed that the variability in the synchrotron--dominated spectral region is driven by a single stochastic process with relatively long relaxation timescale $\tau_1 \gtrsim 100$\,days, whereas in the IC--dominated spectral domain, by a linear superposition of stochastic processes with vastly different relaxation timescales ranging from $\tau_1$ down to $\tau_2 \lesssim 1$\,day. In particular, following the discussion in \citet{Kelly09,Kelly11}, we speculated that stochastic (Gaussian) fluctuations in the local jet conditions (e.g., velocity fluctuations in the jet plasma, and/or magnetic field variations) lead to energy dissipation over a wide range of spatial scales. The resulting acceleration of jet particles, and their following radiative response, is however delayed with respect to the input perturbations, and this forms the {\it damped/red-noise} (i.e., $\beta \sim$ 2) segment of the PSD at the synchrotron--dominated frequencies on the timescales shorter than $\tau_1$, which may be identified with some global MHD timescale characterizing the extended jet volume. Over the IC--dominated spectral segment, on the other hand, due to density inhomogeneities in the local photon populations available for the upscattering, the PSD is modified further to form the flickering/pink noise ($\beta \sim 1$) down to the relaxation timescale $\tau_2$, which should be of the order of the maximum light travel timescale across the jet where the observed emission is being produced. 

Our new results for the  HBLs Mrk\,421 and PKS\,2155$-$304 are based on good-quality, densely sampled, decade-long multiwavelength light curves (except of X-rays in case of PKS\,2155$-$304) on timescales ranging from weeks to a decade.  We find that the PSDs above the noise floor level (which becomes dominant at high temporal frequencies), are  well represented by single power-laws with slopes $\beta \sim 2$ at synchrotron--dominated frequencies (radio and B, V, R, I, J, H, and K--bands), and $\beta \sim 1$ at the IC--dominated frequencies (i.e., VHE\, and HE\,$\gamma-$ rays). The power-law slope is also $\beta \sim 1$ at X-rays, even though in these sources, within the single-zone SSC scenario, the keV photons are expected to be generated in the synchrotron process by the high energy tail of the particle distribution  while the HE and VHE energy emission results from the up-scattering of seed photons in the EC part of the spectrum \citep[see Section~\ref{sec:intro};][]{Hess12, Petropoulou14}.
Note, on the other hand, that due to the particularly sparse sampling of the XRT monitoring data, the X-ray power spectrum of PKS\,2155$-$304 could be constrained only down to the variability timescales of $\sim 100$\,days, while in the multi-wavelength comparison Figure\,\ref{fig:5}(b), we see that up to this timescale the variability amplitudes at X-rays are comparable to those seen at optical, and also $\gamma$-ray, photon energies; that is to say, with the available X-ray monitoring data for the blazar, we do not probe the variability timescales at which the difference between $\gamma$-ray (inverse-Compton) and optical (synchrotron) power spectra are significant.

The other important result comes from a systematic comparison of the squared fractional variability at the selected synchrotron and IC light curves. On timescales of months ($\lesssim 100$ days, down to the sampling intervals of the $\gamma$-ray light curves, i.e. $\sim$weeks), the observed variability amplitudes in the $\gamma$-ray range are significantly larger --- up to one order of magnitude --- than those observed at radio, IR or optical frequencies. For the blazar PKS\,2155$-$304, quite surprisingly, within this exact time domain, the variability amplitudes seem larger at HE $\gamma$-rays than in the VHE $\gamma$-ray range (1$\sigma$ error bars), although it should be stressed out here once again that the sampling of the light curves within these three bands is very different (half-a-year observing window for the H.E.S.S. instruments versus a continuous monitoring with the {\it Fermi}-LAT). This behaviour is not shown by Mrk\,421 where the variability amplitudes at X-rays ({\it RXTE} and {\it Swift}), HE and  VHE $\gamma$-ray range seem to follow closely each other on timescales $\leq$100 days. Whether this tentative finding contrasts with the variability behavior of other blazars is to be investigated by means of a systematic analysis of multiwavelength light curves that covers many cycles of ``quiescence'' and ``flaring'' states for a larger sample of sources. At this point we only mention that in the framework of the leptonic scenarios for BL Lacs, the GeV emission results from IC scattering of seed photons that proceeds in  mainly in the Thompson regime, while in the case of the TeV emission the Klein-Nishina effects become typically more relevant, depending on the exact shape of the source broad-band spectrum, and this may be one of the factors shaping the amplitudes of the observed flux changes \citep[see in this context, e.g.,][]{Aleksic15a}.

 All in all, the analysis results for the HBLs Mrk\,421 and PKS\,2155$-$304 presented in this paper, which are in accord with the results for the LBLs PKS\,0735$+$178 and OJ\,287 presented in \citet{Goyal17, Goyal18}, strengthen our main conclusion, namely that commonly employed one-zone scenarios for the blazar emission cannot account for all the complexity of the blazar variability observed on various timescales. Indeed, in the framework of the simplest one-zone SSC model, different statistical patterns (in particular, red vs. pink noise-type PSDs) are not expected at different frequencies within the synchrotron--dominated spectral region (IR/optical vs. X-rays), neither should they appear between the synchrotron and IC--dominated spectral regions \citep[IR/optical vs. $\gamma-$rays; see][]{Finke14}. This, in our opinion, points out to a more complex picture of blazars, with a highly inhomogeneous outflow producing non-thermal emission over an extended, stratified volume, as in this way stochastic, uncorrelated fluctuations in the local jet conditions, leading to the locally enhanced energy dissipation, could result in a correlated (colored noise-type) variability of the overall emission continuum \citep[see in this context the discussion in][and references therein]{Kelly09,Kelly11}. In other words, while the one-zone emission scenarios could still be applied in a meaningful way when dealing with well-defined, and possibly shorter flaring events, they should be considered with caution when modelling the blazar emission continua averaged over longer monitoring epochs, or when comparing the blazar flux changes on vastly different timescales (say, years vs. hours and minutes).

Finally, we note that  Mrk\,421 and PKS\,2155$-$304 are the only blazars for which broad-band variability power spectra could be generated right up to TeV\,$\gamma-$ray energies, {\it without} the VHE measurements being  biased towards high flux states. The present work is also relevant to future studies of blazar variability up to VHE using the upcoming Cherenkov Telescope Array (CTA), which is expected to raise the number of currently known TeV-emitting blazars by an order of magnitude \citep{2017arXiv170907997C}.

\clearpage
\section*{Acknowledgements}

I thank the anonymous referee for a careful reading and constructive comments on the manuscript which improved the content and the presentation. I also thank David Sanchez (H.E.S.S.) for kindly providing the H.E.S.S. light curve of the blazar PKS\,2155$-$304 in electronic form. Useful discussions with {\L}ukasz Stawarz, Micha{\l} Ostrowski, Paul Wiita, Gopal-Krishna, and Marian Soida are gratefully acknowledged. I acknowledge support from the Polish National Science Centre (NCN) through the grant 2018/29/B/ST9/02298 and partial support from the grant UMO-2016/22/E/ST9/00061. Some of the light curve simulations were performed at the PL grid using the Zeus cluster under the computing grant `lcsims'. I also thank Staszek Zola, Krzysztof Chy{\.z}y, Marek We{\.z}gowiec and Dorota Kozie{\l}-Wierzbowska for allowing to perform a few simulations on their computers.  

This paper has made use of up-to-date SMARTS optical/near-infrared light curves that are available at www.astro.yale.edu/smarts/glast/home.php. 

The \textit{Fermi} LAT Collaboration acknowledges generous ongoing support from a number of agencies and institutes that have supported both the development and the operation of the LAT as well as scientific data analysis. These include the National Aeronautics and Space Administration and the Department of Energy in the United States, the Commissariat \`a l'Energie Atomique and the Centre National de la Recherche Scientifique / Institut National de Physique Nucl\'eaire et de Physique des Particules in France, the Agenzia Spaziale Italiana and the Istituto Nazionale di Fisica Nucleare in Italy, the Ministry of Education, Culture, Sports, Science and Technology (MEXT), High Energy Accelerator Research Organization (KEK) and Japan Aerospace Exploration Agency (JAXA) in Japan, and the K.~A.~Wallenberg Foundation, the Swedish Research Council and the Swedish National Space Board in Sweden.
 
Additional support for science analysis during the operations phase is gratefully acknowledged from the Istituto Nazionale di Astrofisica in Italy and the Centre National d'\'Etudes Spatiales in France. This work performed in part under DOE Contract DE-AC02-76SF00515.

The   support   of   the   Namibian   authorities   and   of   the University of Namibia in facilitating the construction and operation of H.E.S.S. is gratefully acknowledged, as is the support by the German Ministry for Education and Research (BMBF), the Max Planck Society, the  French Ministry for Research, the CNRS-IN2P3 and the Astroparticle Interdisciplinary Programme of the CNRS, the U.K. Particle Physics and Astronomy Research Council (PPARC), the IPNP of  the Charles University, the South African Department of Science and Technology and National Research Foundation, and by the University of Namibia. We appreciate the excellent work of the technical support staff in Berlin, Durham, Hamburg, Heidelberg, Palaiseau, Paris, Saclay, and in Namibia in the construction and operation of the equipment.
 
This research has made use of data from the OVRO 40-m monitoring program (Richards, J. L. et al. 2011, ApJS, 194, 29) which is supported in part by NASA grants NNX08AW31G, NNX11A043G, and NNX14AQ89G and NSF grants AST-0808050 and AST-1109911


\appendix

\section{Response of spectral window function for our unevenly sampled data series }
\label{app:A}

The spectral window function is a diagnostic of deleterious effects of Fourier transform when subjected to unevenly sampled time series. It is defined to be the Fourier transform of the sampling times, which in case of evenly sampled time series, is 1 at zeroth Fourier frequency and 0 otherwise and moreover is unknown for unevenly sampled data series \citep{Deeming75}. Figs~\ref{fig:6} and ~\ref{fig:7} show the distributions of sampling interval of the analysed light curves and the spectral window functions of the respective data series (panels a--e and f--j, respectively for Mrk\,421 and panels a--j and k--t, respectively, for PKS\,2155$-$304). As shown, the distribution of sampling intervals is neither sharply peaked and concentrated in one bin (as is the case for evenly sampled data series) nor uniform over the intervals (as for randomly sampled data series). For the {\it Fermi}-LAT data, where the fluxes were measured using one-week integration times, the distribution is at intervals corresponding to integer multiples of seven days, as expected. The spectral window function for the data series is computed down to typical (mean) Nyquist sampling frequencies (Table~\ref{tab:psd}). In each case, the spectral window function shows considerable power (other than 0) at all the Fourier frequencies precisely due to the uneven sampling of the light curves which could be mistaken for real power in the periodograms. Moreover, a mild peak at frequencies corresponding to $\sim$200-300 days is noted in the spectral window functions of the VERITAS, H.E.S.S., optical and IR light curves. This peak is due to systematic (seasonal) gaps attained in the light curves for the source from the ground--based observatories when the target is too close to the Sun. A more prominent peak at $\sim$28 days are noted in the spectral window functions of the VERITAS and the H.E.S.S. light curves because of systematics gaps due to full Moon periods during which the above facilities do not operate. This peak is more prominent than the $\sim$300 day peak because the data series contains more periods than the yearly gaps in the decade-long light curves. The spectral window function for the {\it RXTE-}PCA and the {\it Swift-}XRT light curve also show a mild peak around $\sim$300 days, perhaps because of scheduling of these particular observations. The spectral window function for the {\it Fermi-}LAT light curve shows a fairly uniform response, indicative of no systematic gaps. Therefore, a careful examination of spectral window function is essential as it can reveal periodicities that occur precisely due to regular gaps in the sampling of the time series which could be mistaken as a QSOs. We note that aliasing folds back roughly equal amount of power to the frequency range probed which is dominated at the Nyquist frequency \citep{Uttley02}, thus could not create peaks in the spectral window function.

\begin{figure*}
\hbox{
\hspace*{0.1cm}\includegraphics[width=0.33\textwidth]{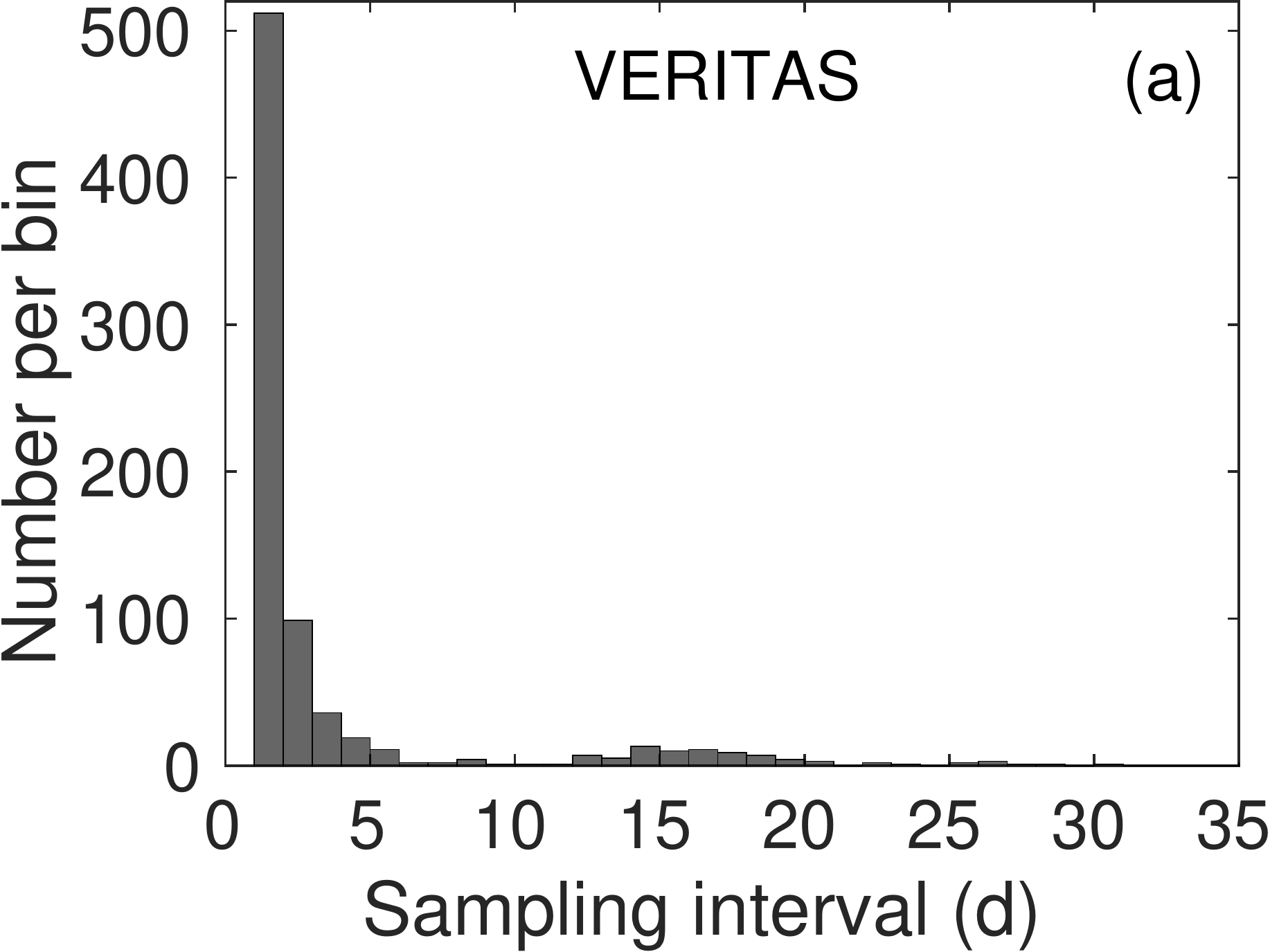}
\hspace*{0.1cm}\includegraphics[width=0.33\textwidth]{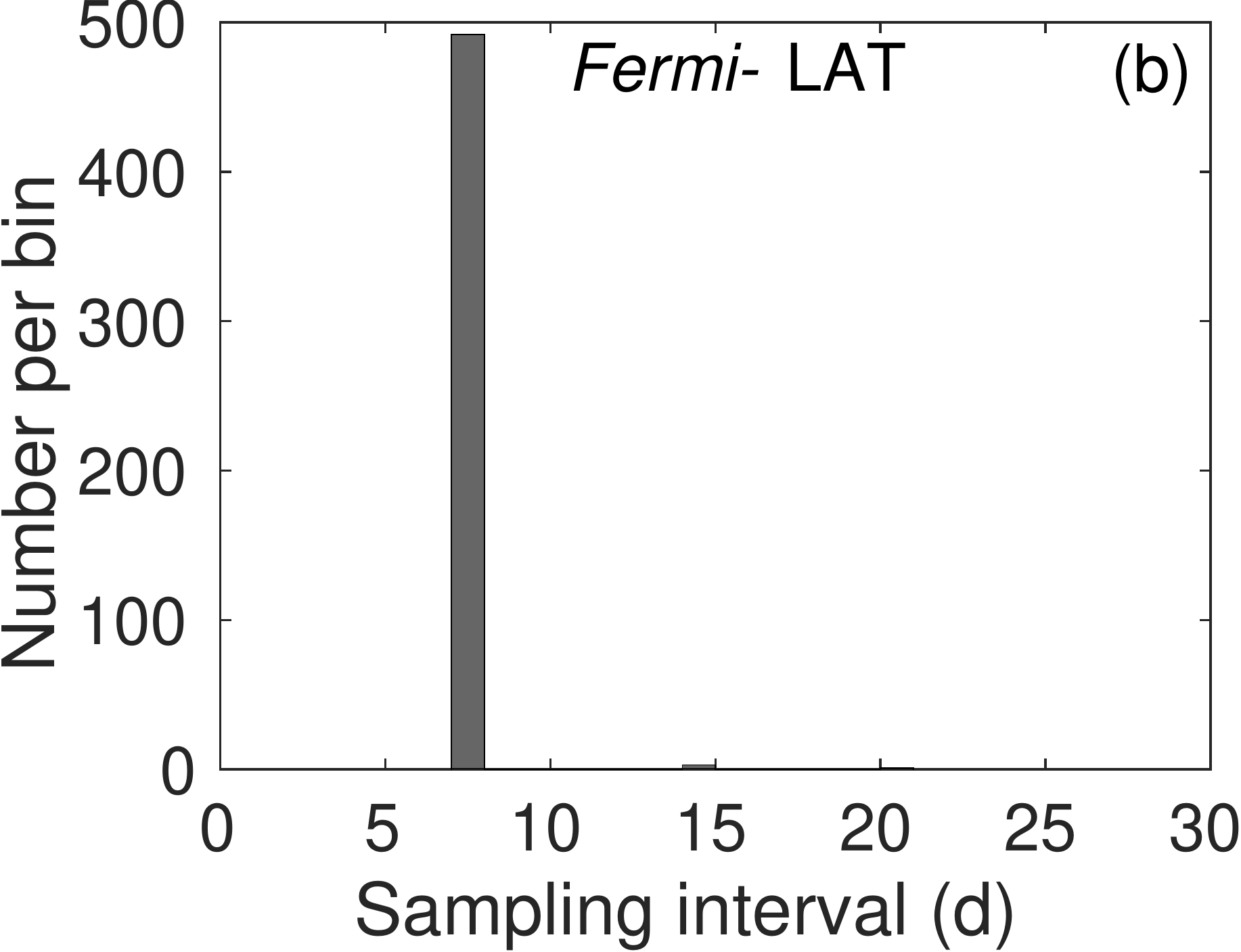}
\hspace*{0.1cm}\includegraphics[width=0.33\textwidth]{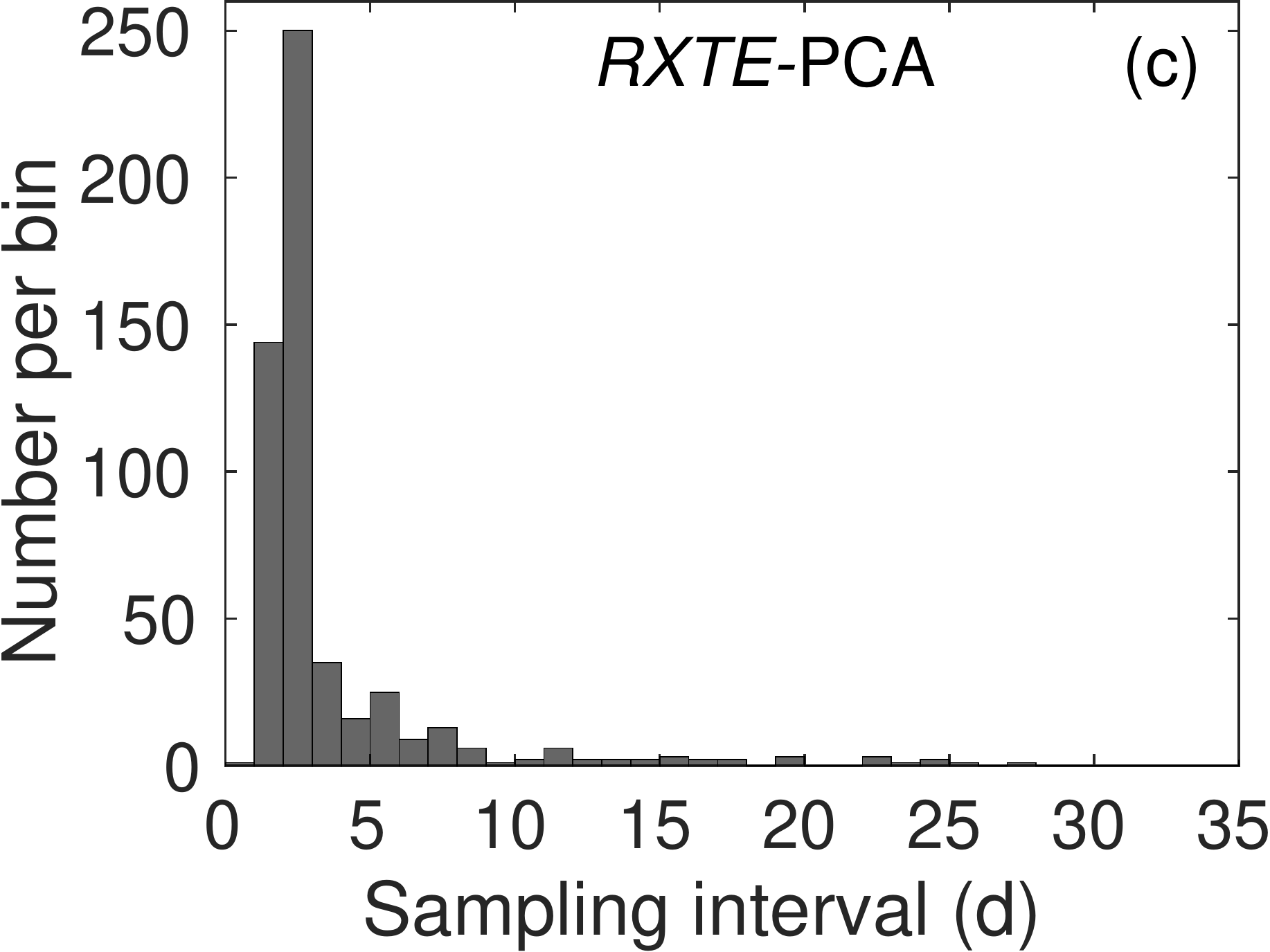}
}
\hbox{
\hspace*{0.1cm}\includegraphics[width=0.33\textwidth]{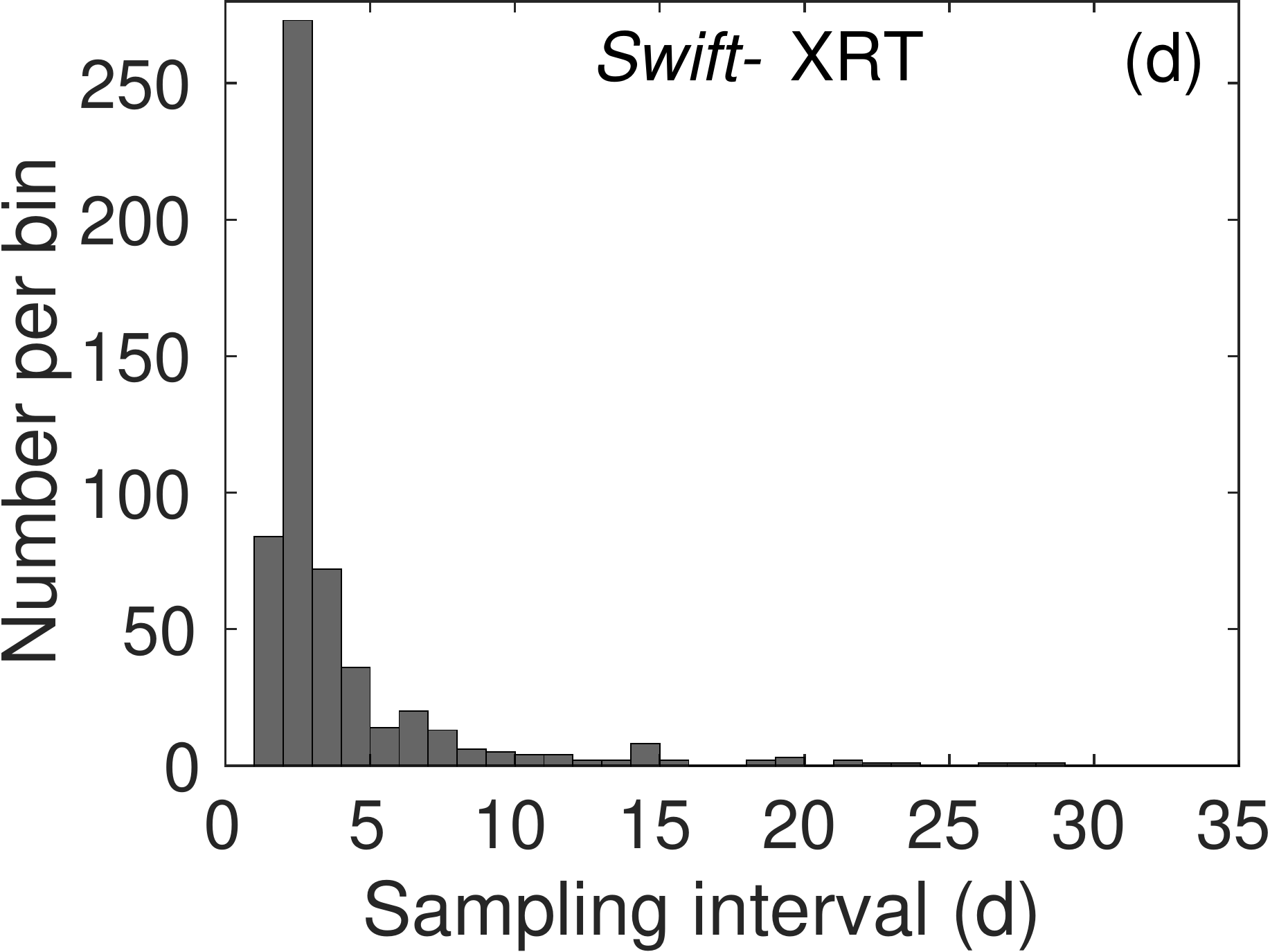}
\hspace*{0.1cm}\includegraphics[width=0.33\textwidth]{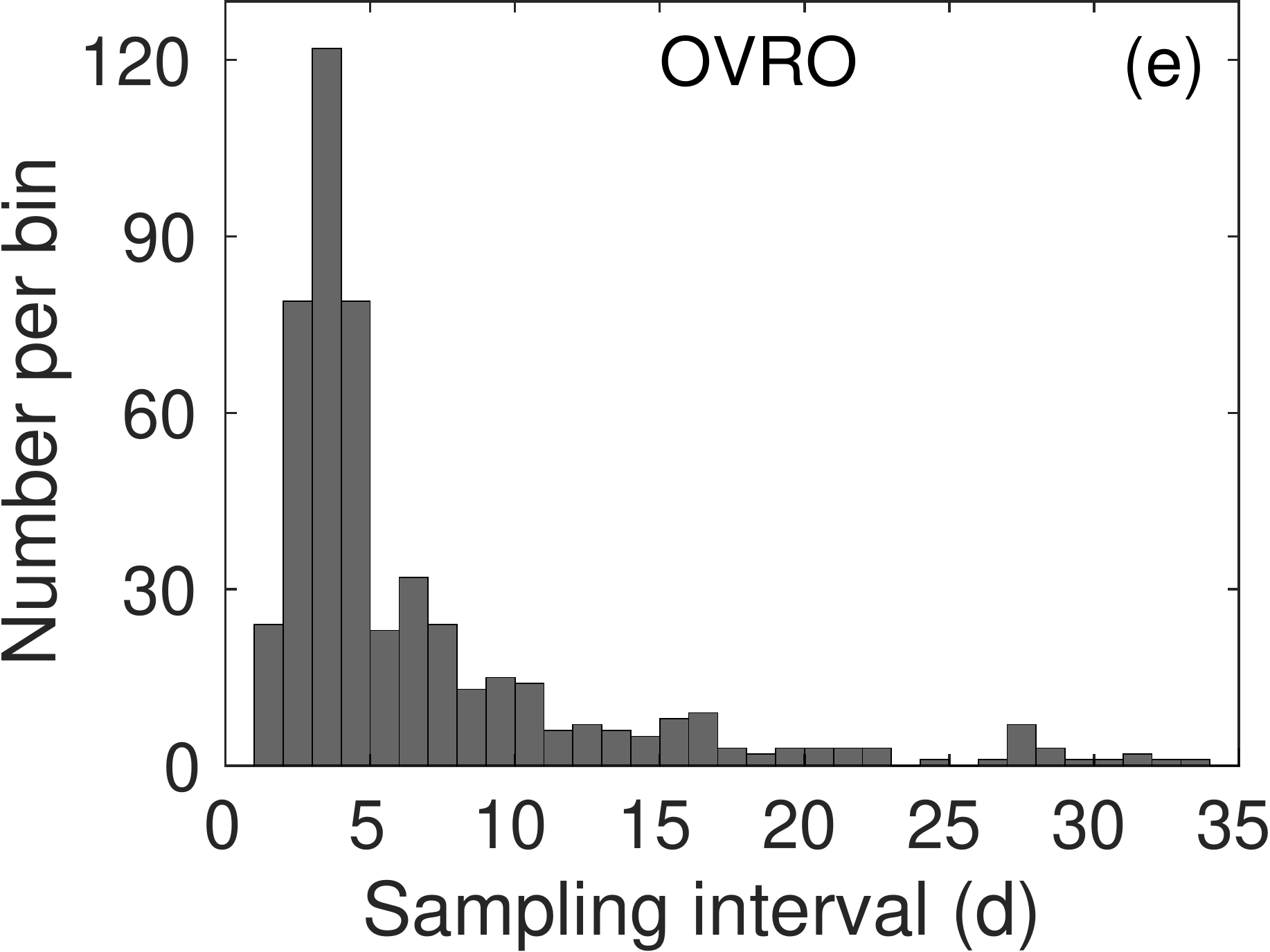}
\hspace*{0.1cm}\includegraphics[width=0.33\textwidth]{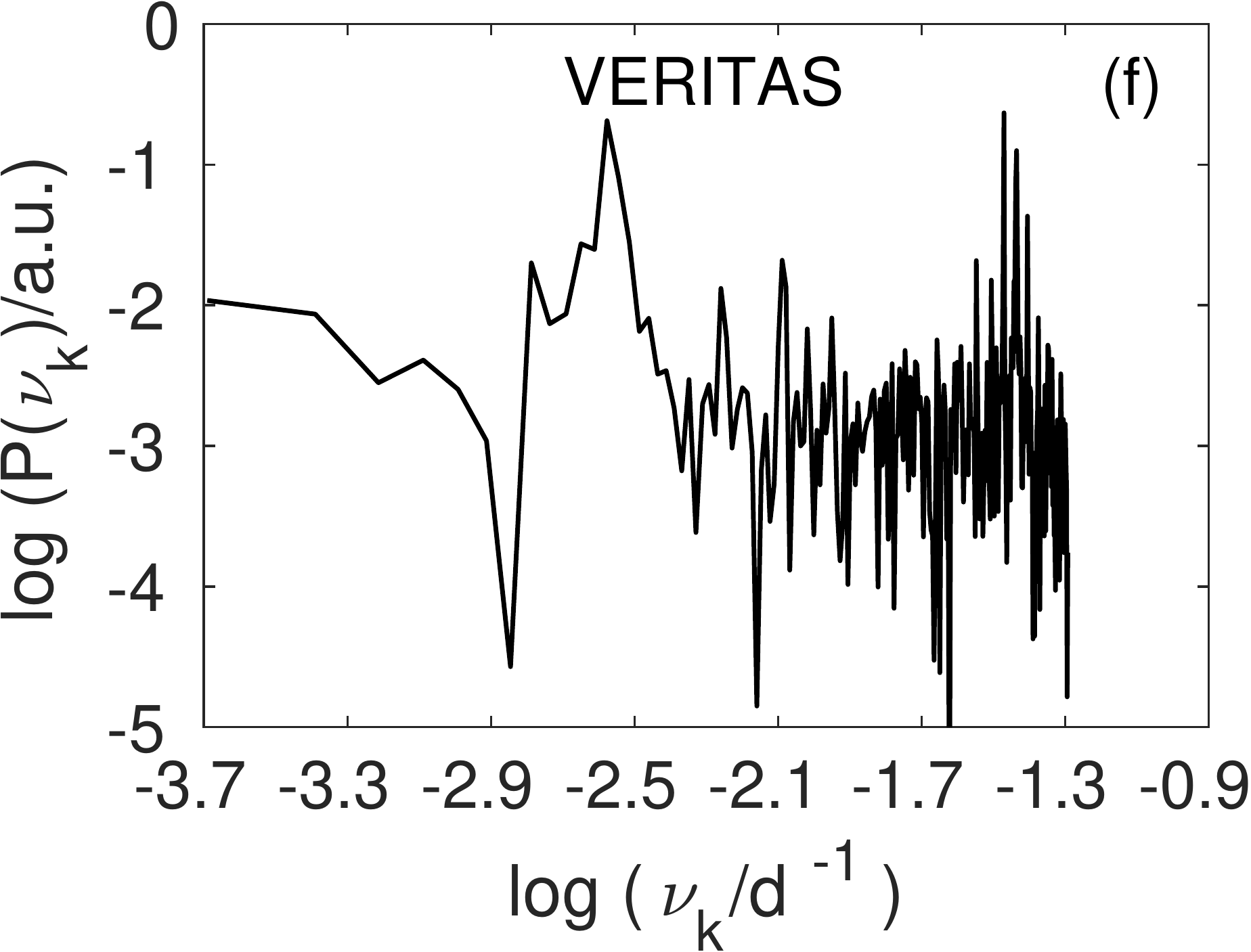}
}
\hbox{
\hspace*{0.1cm}\includegraphics[width=0.33\textwidth]{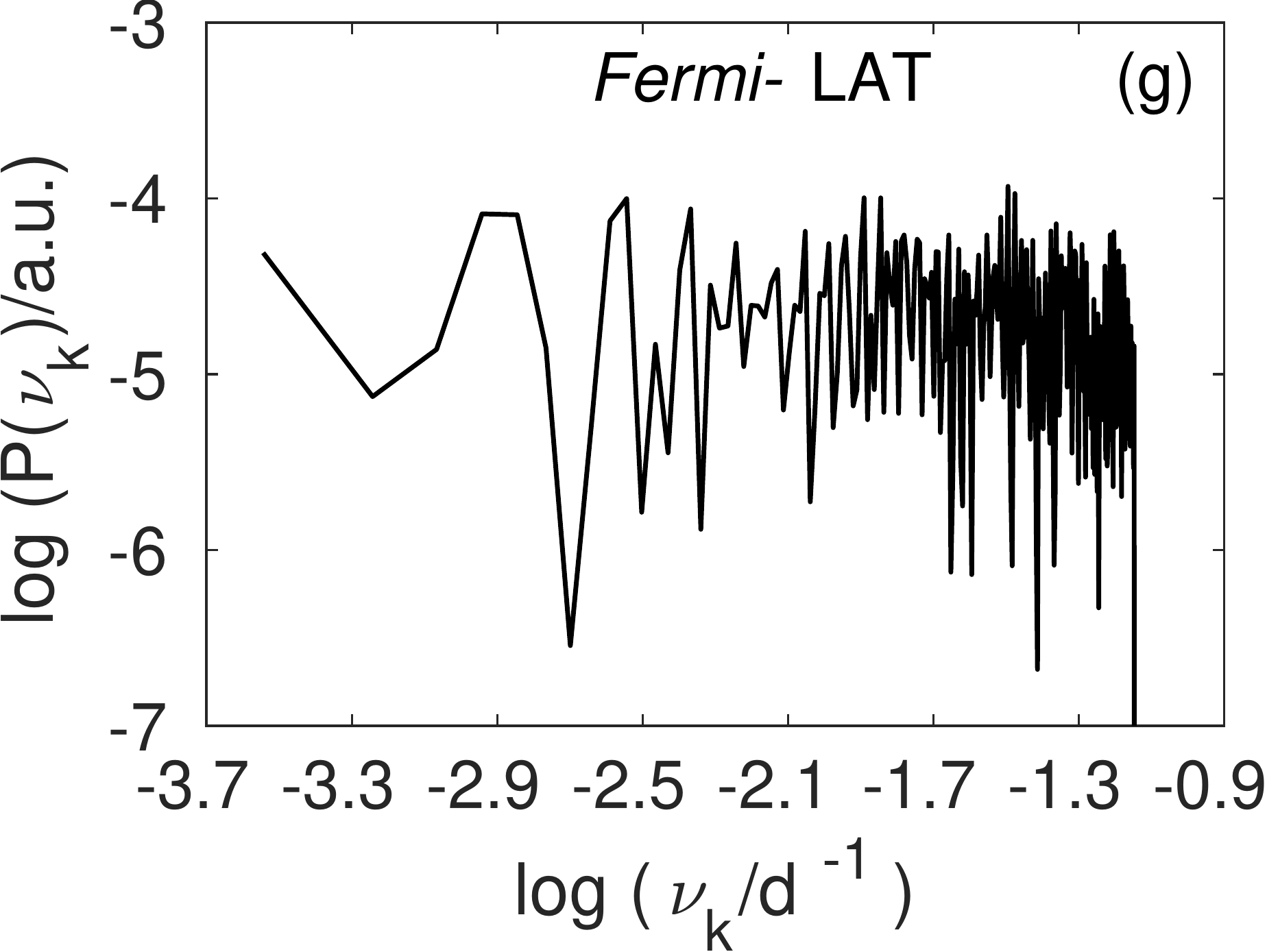}
\hspace*{0.1cm}\includegraphics[width=0.33\textwidth]{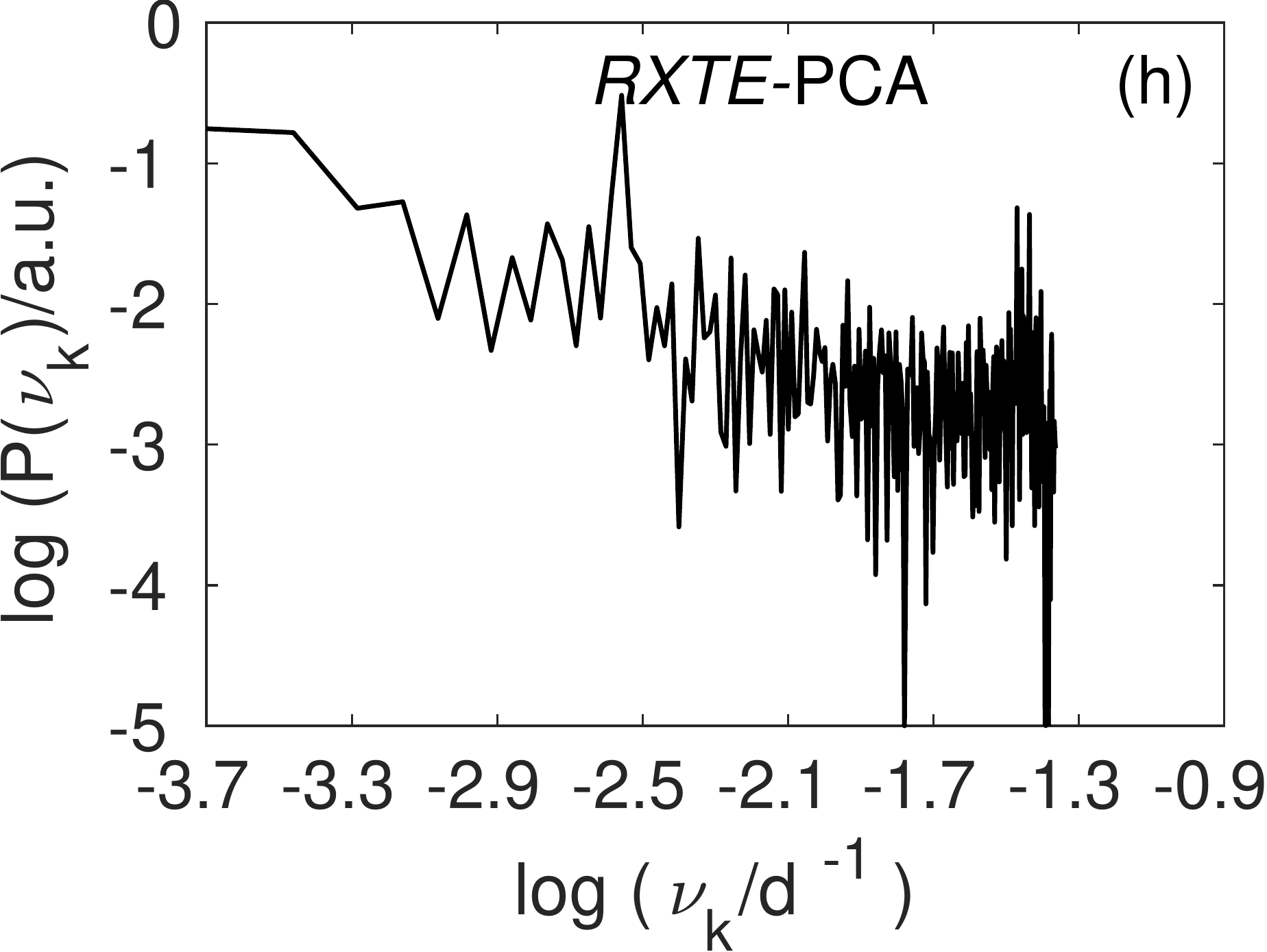}
\hspace*{0.1cm}\includegraphics[width=0.33\textwidth]{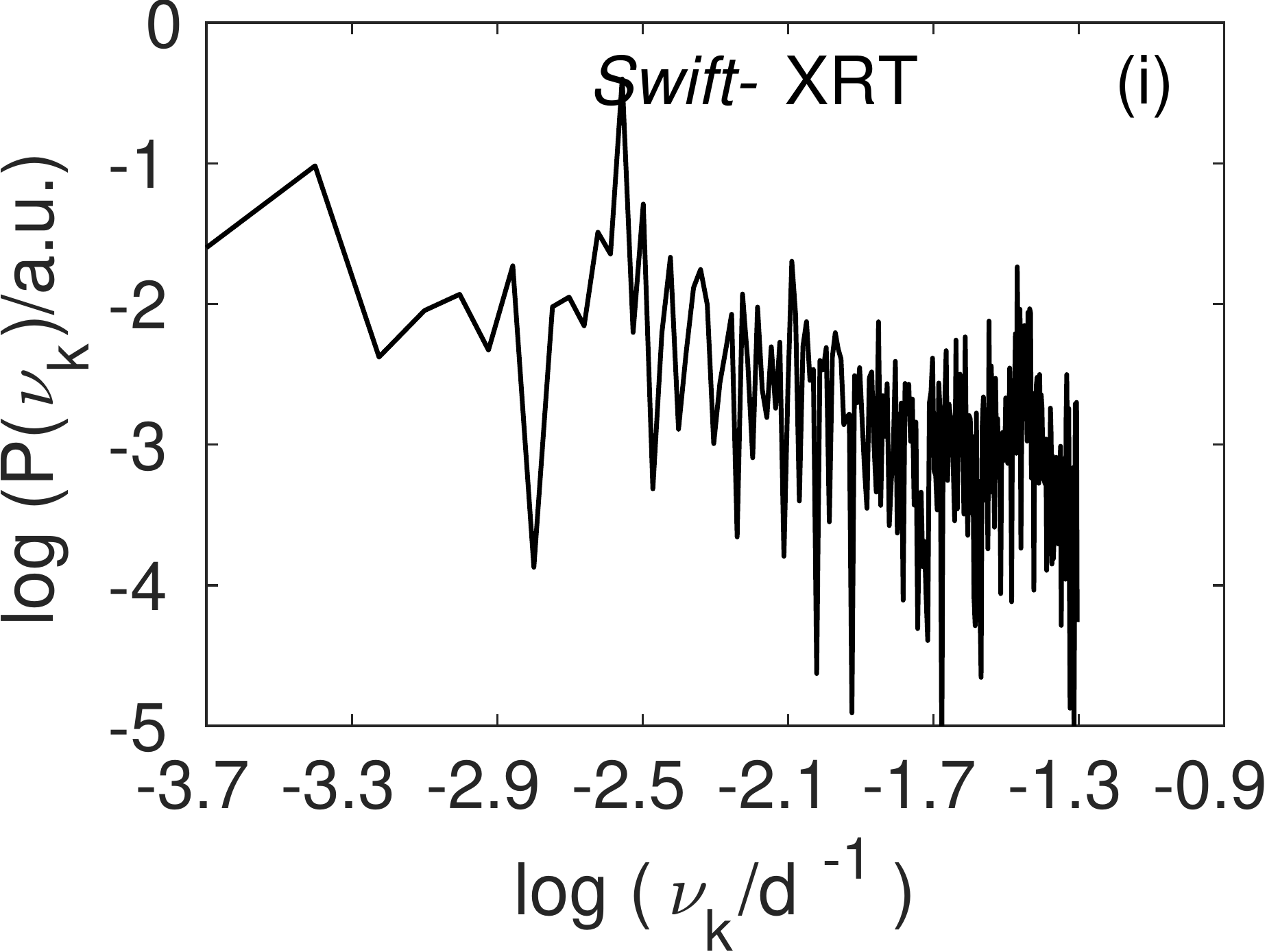}
}
\hspace*{0.1cm}\includegraphics[width=0.33\textwidth]{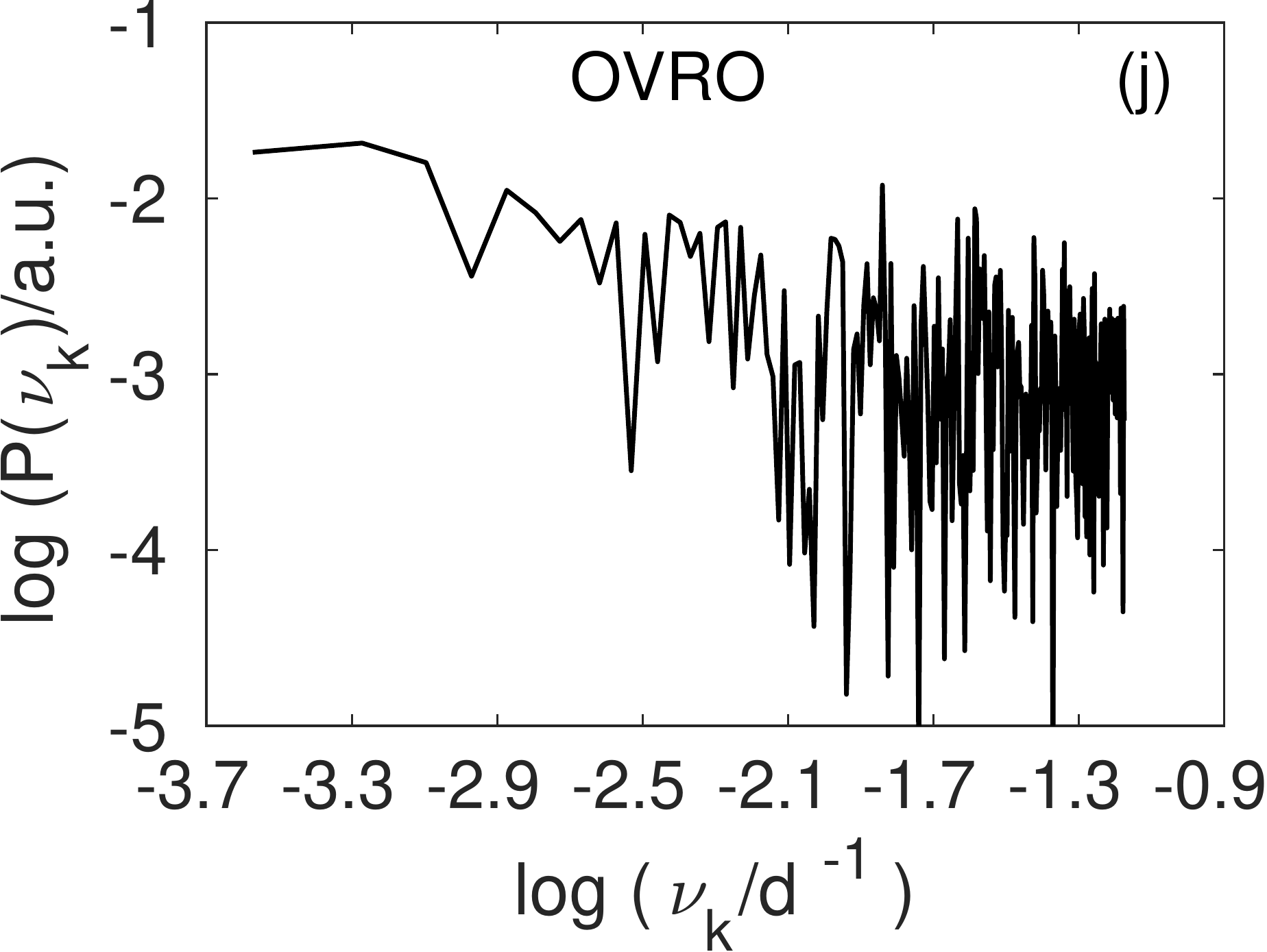}
\caption{({\it a-e)}: Distributions of the sampling intervals of the analysed light curves for the blazar Mrk\,421.  For clarity, we have truncated the histograms above the sampling intervals with only two or fewer data points. We note that the sampling interval extends up to 210 days, 21 days, 620 days, 288 days, and 324 days for the VERTIAS, {\it Fermi-}LAT, {\it RXTE-}PCA, {\it Swift-}XRT and the OVRO light curves, respectively (Table~\ref{tab:psd}). ({\it f-j}): Corresponding spectral window functions. }

\label{fig:6}
\end{figure*}

\begin{figure*}
\hbox{
\hspace*{0.1cm}\includegraphics[width=0.33\textwidth]{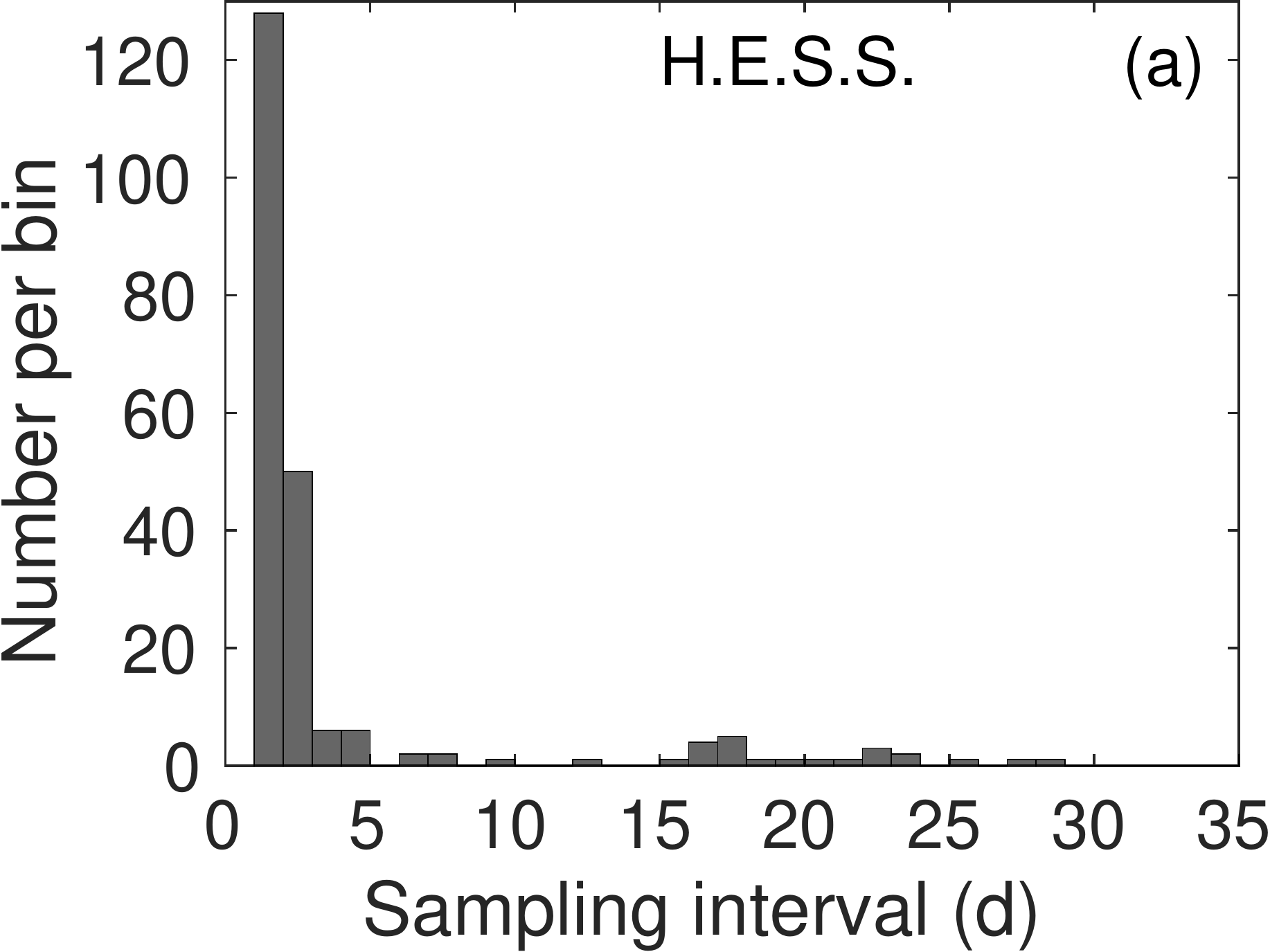}
\hspace*{0.1cm}\includegraphics[width=0.33\textwidth]{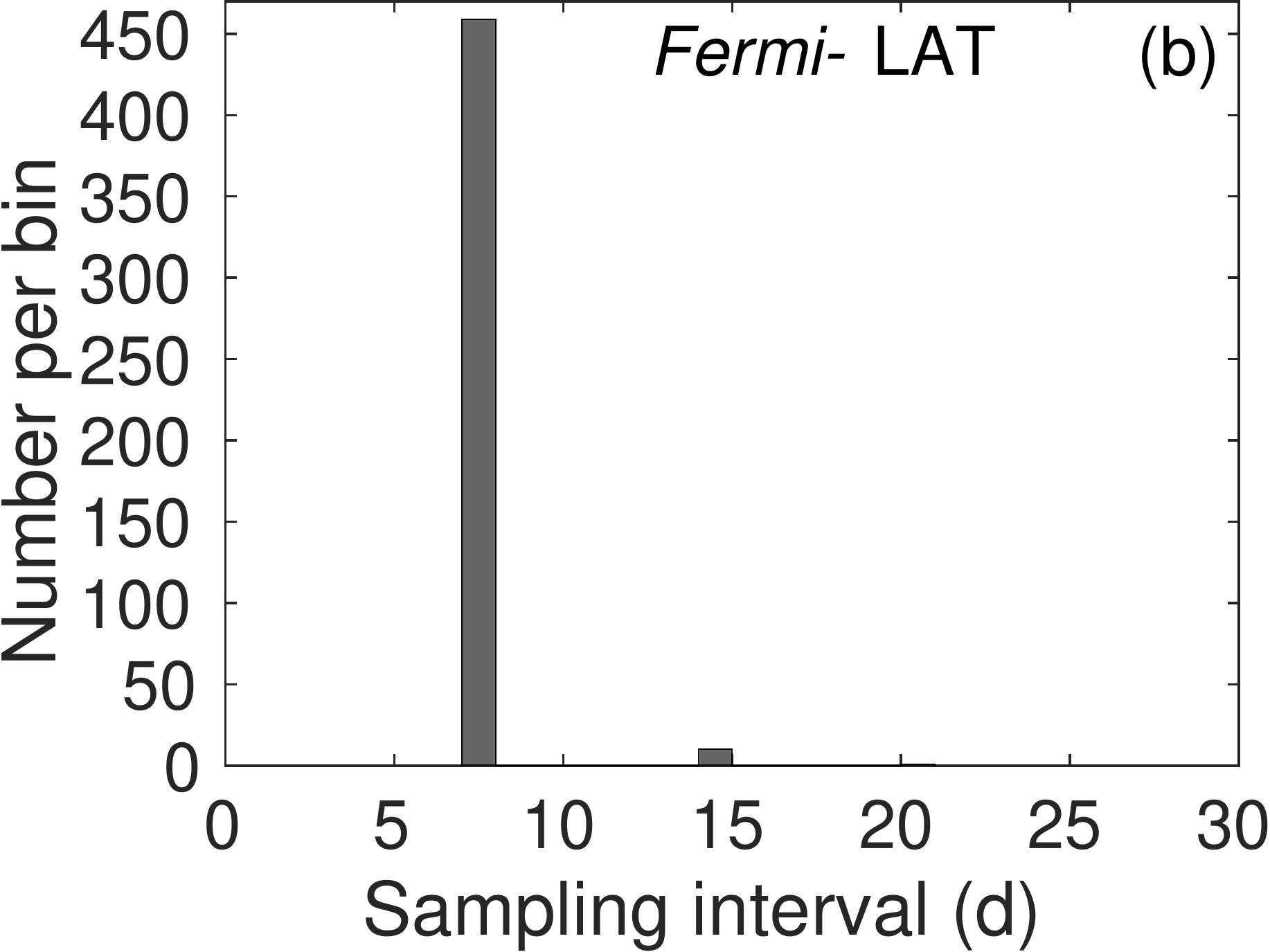}
\hspace*{0.1cm}\includegraphics[width=0.33\textwidth]{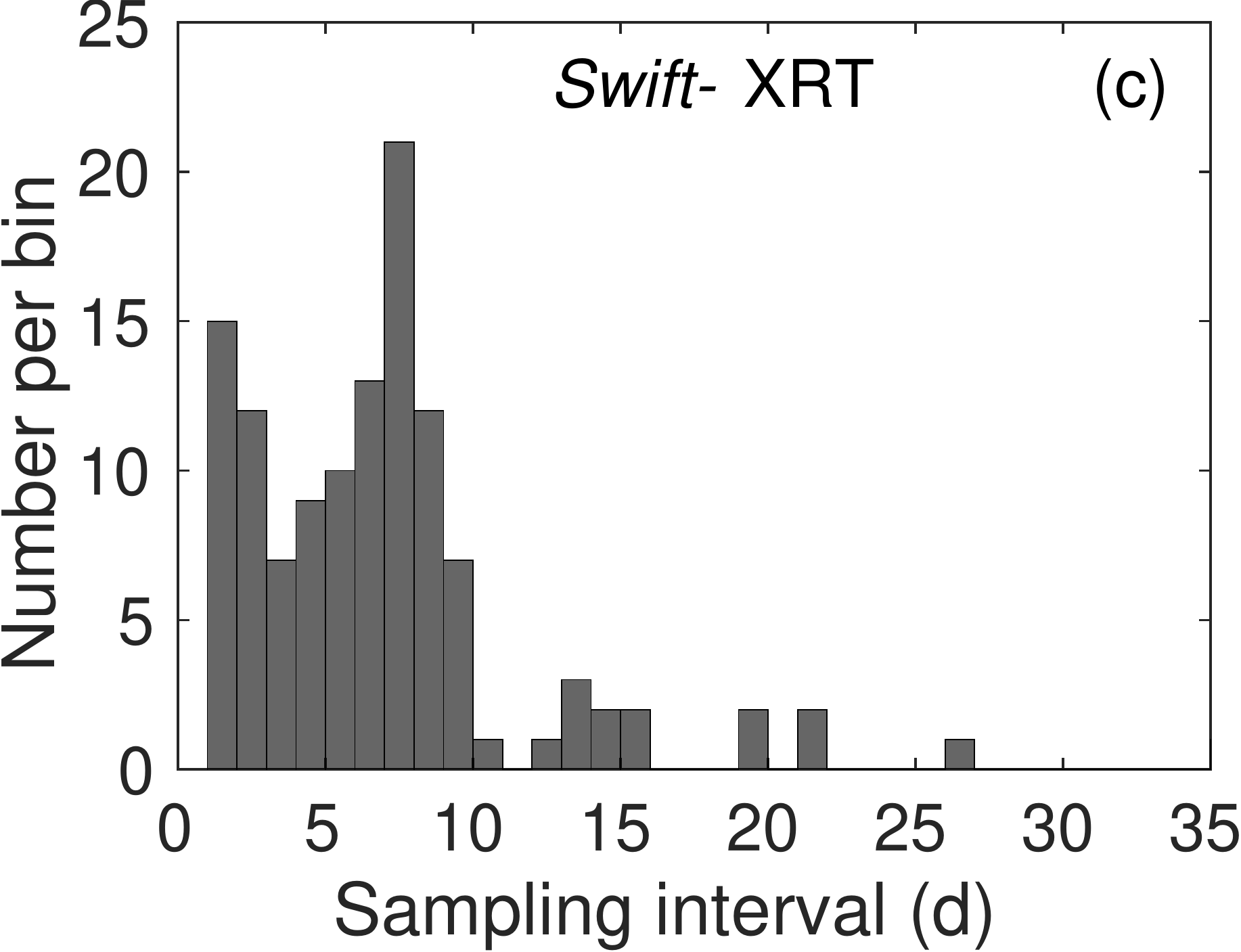}
}
\hbox{
\hspace*{0.1cm}\includegraphics[width=0.33\textwidth]{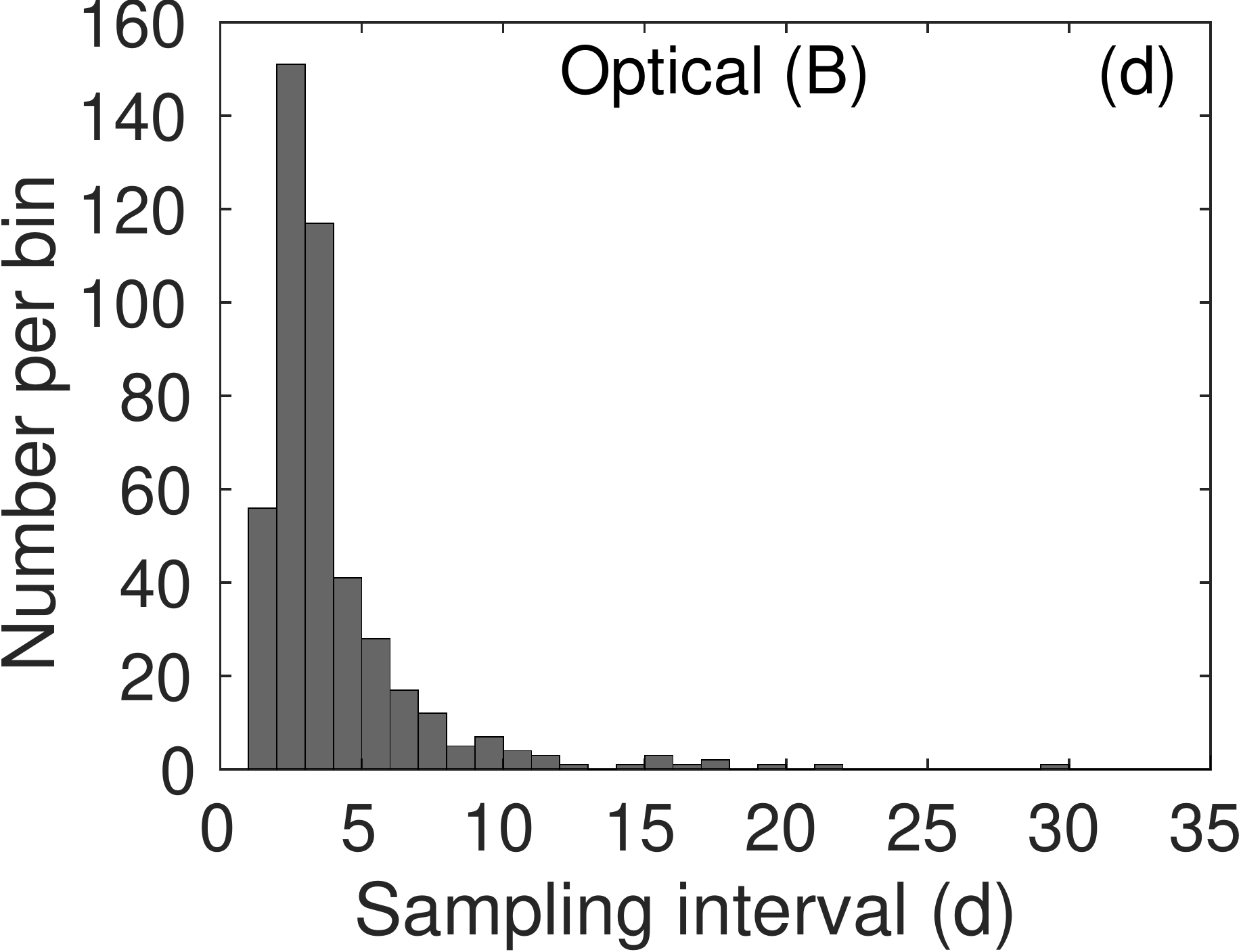}
\hspace*{0.1cm}\includegraphics[width=0.33\textwidth]{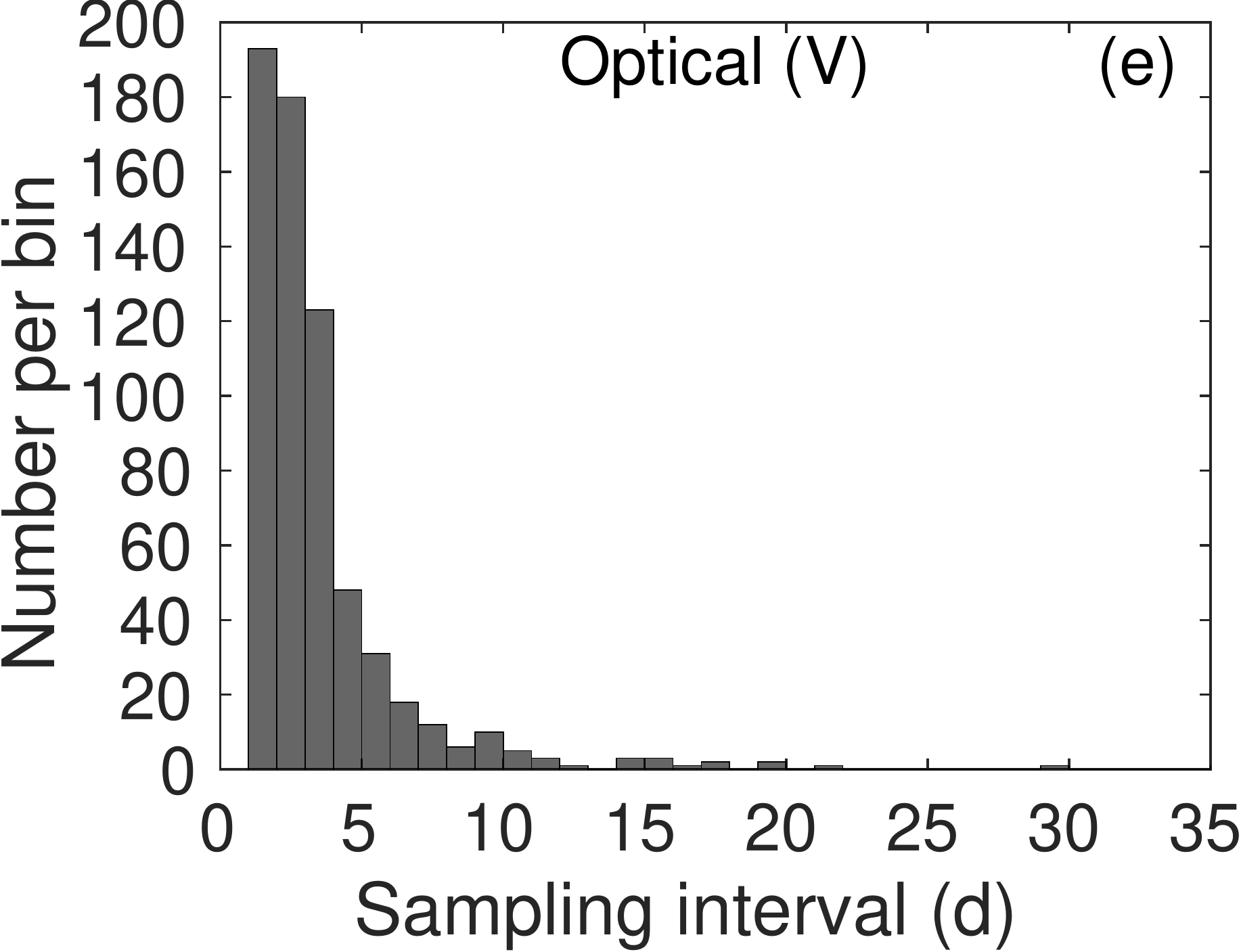}
\hspace*{0.1cm}\includegraphics[width=0.33\textwidth]{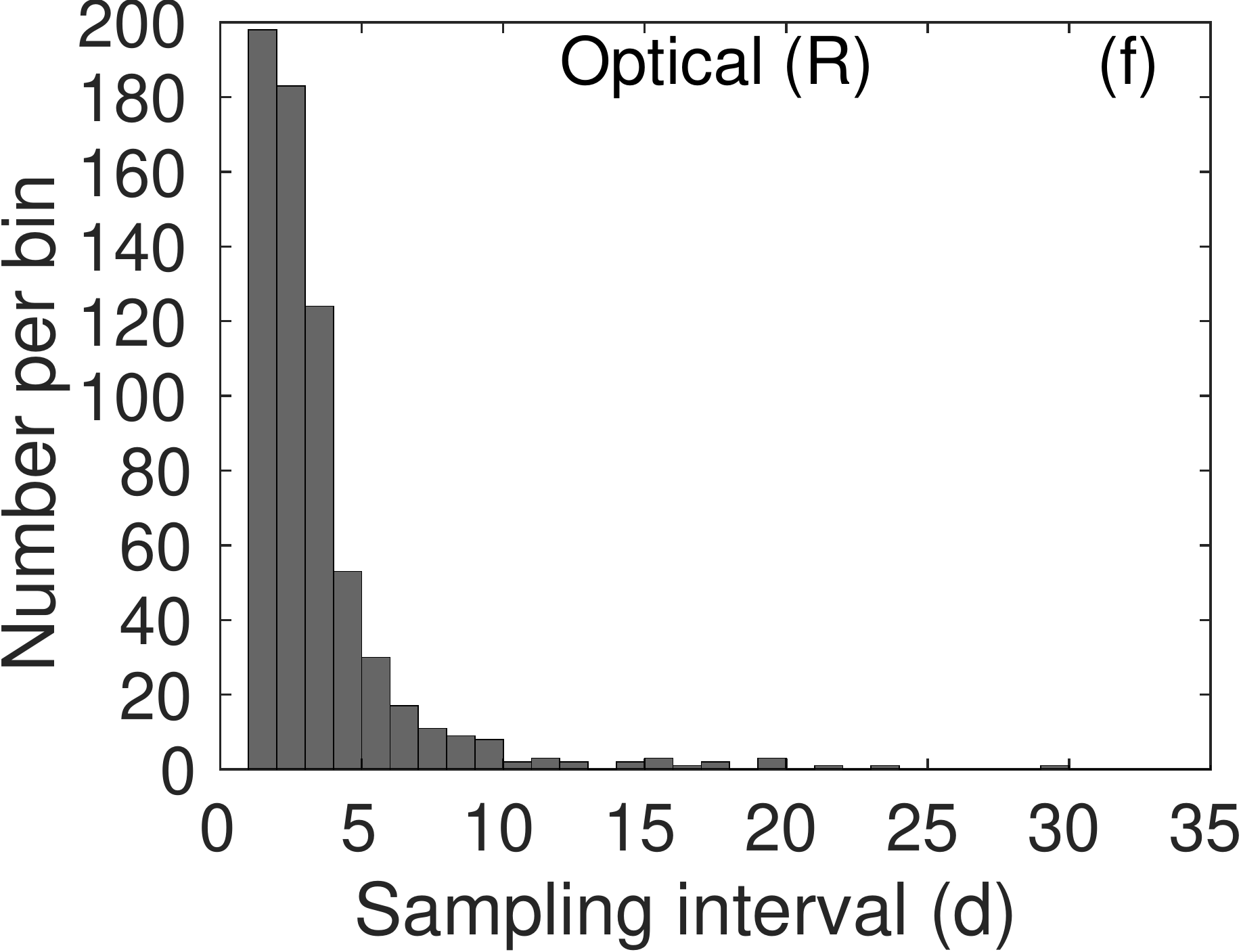}
}
\hbox{
\hspace*{0.1cm}\includegraphics[width=0.33\textwidth]{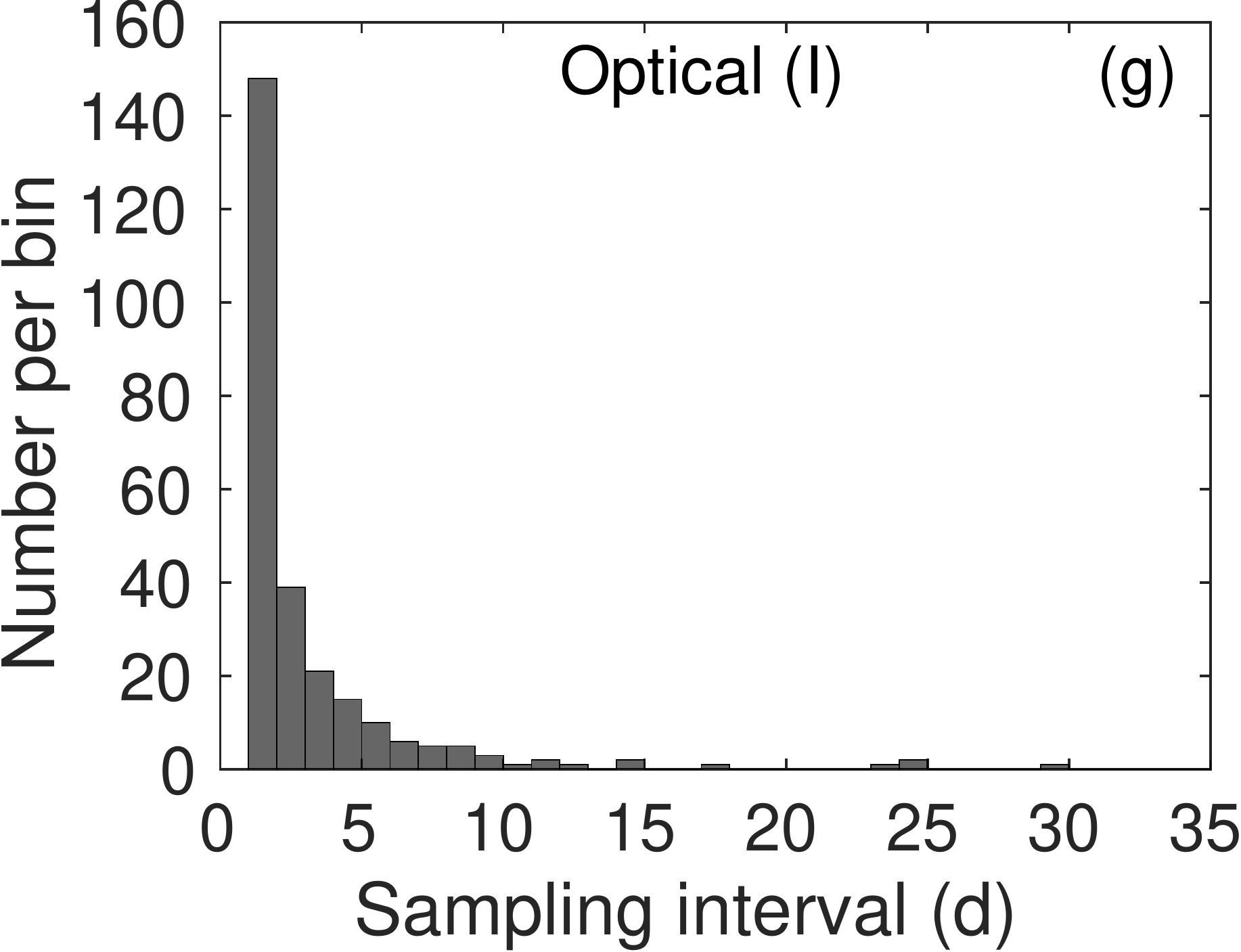}
\hspace*{0.1cm}\includegraphics[width=0.33\textwidth]{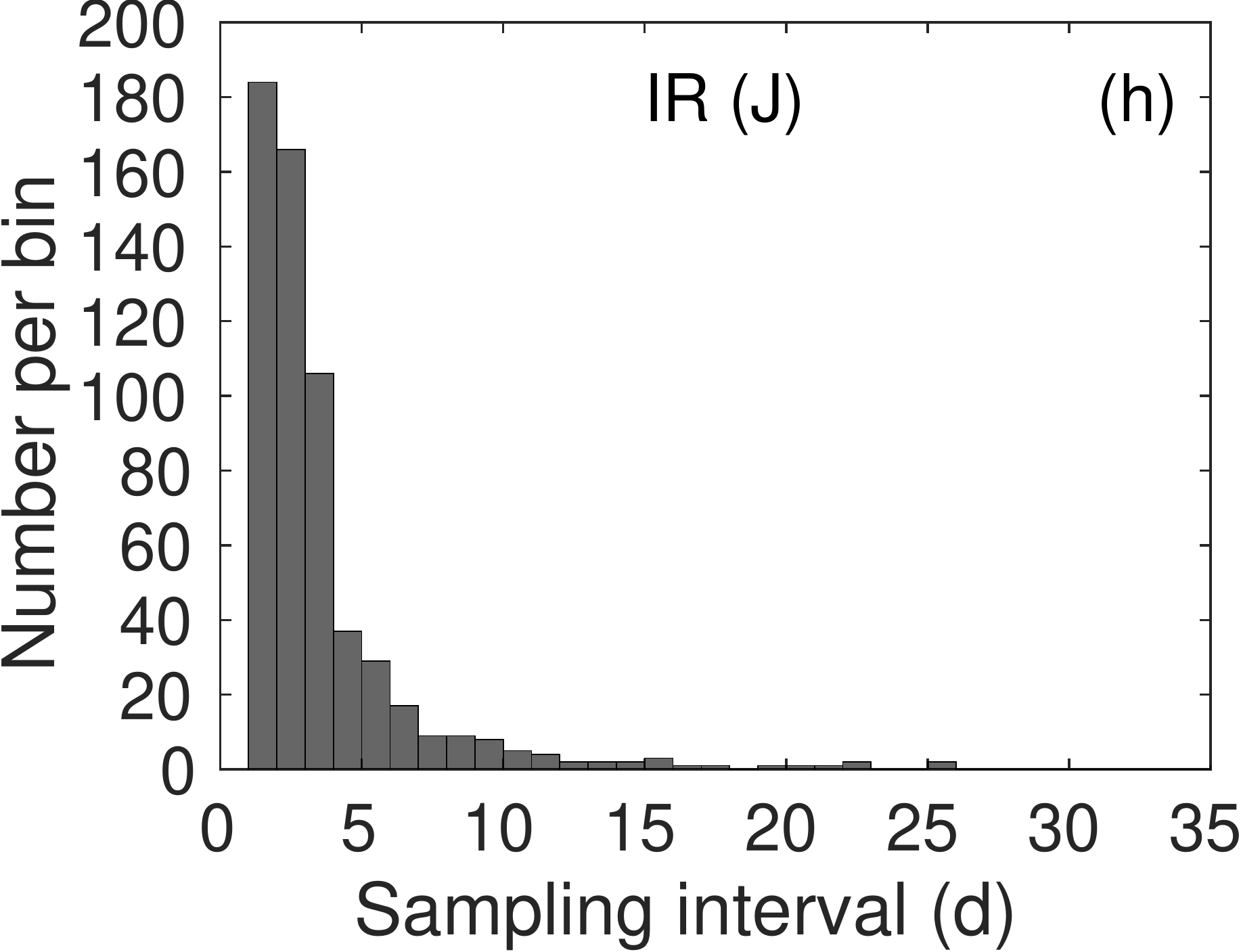}
\hspace*{0.1cm}\includegraphics[width=0.33\textwidth]{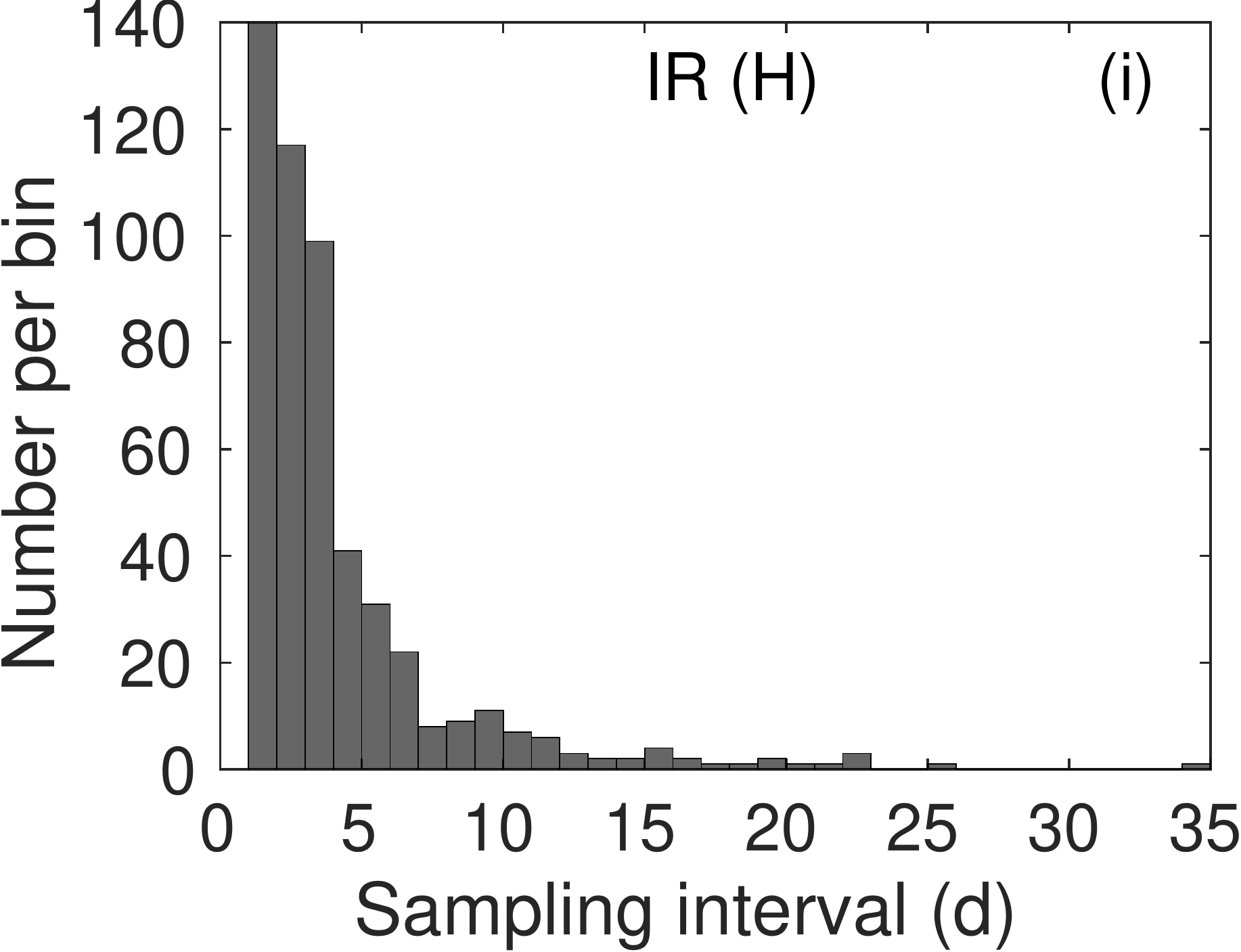}
}
\hbox{
\hspace*{0.1cm}\includegraphics[width=0.33\textwidth]{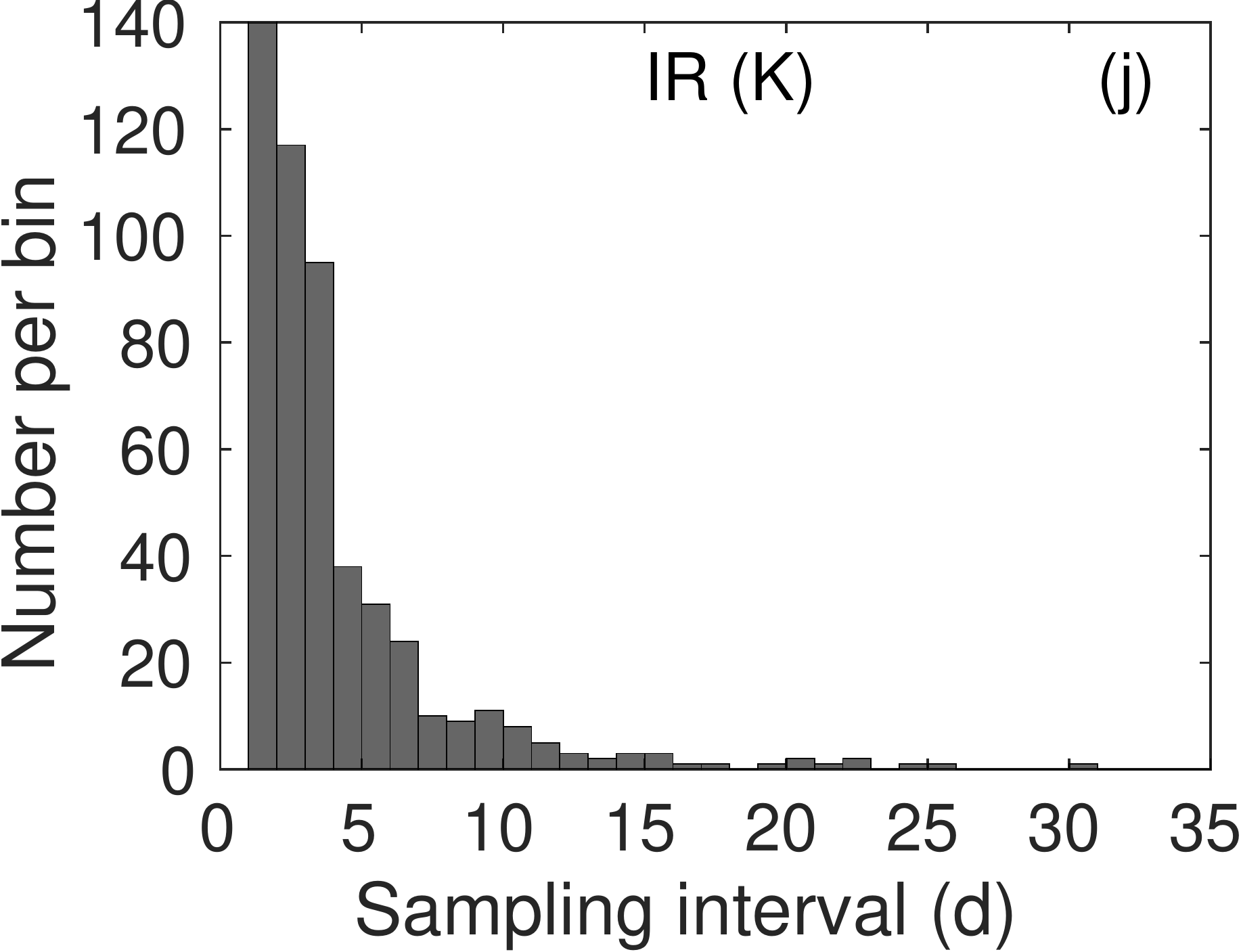}
\hspace*{0.1cm}\includegraphics[width=0.33\textwidth]{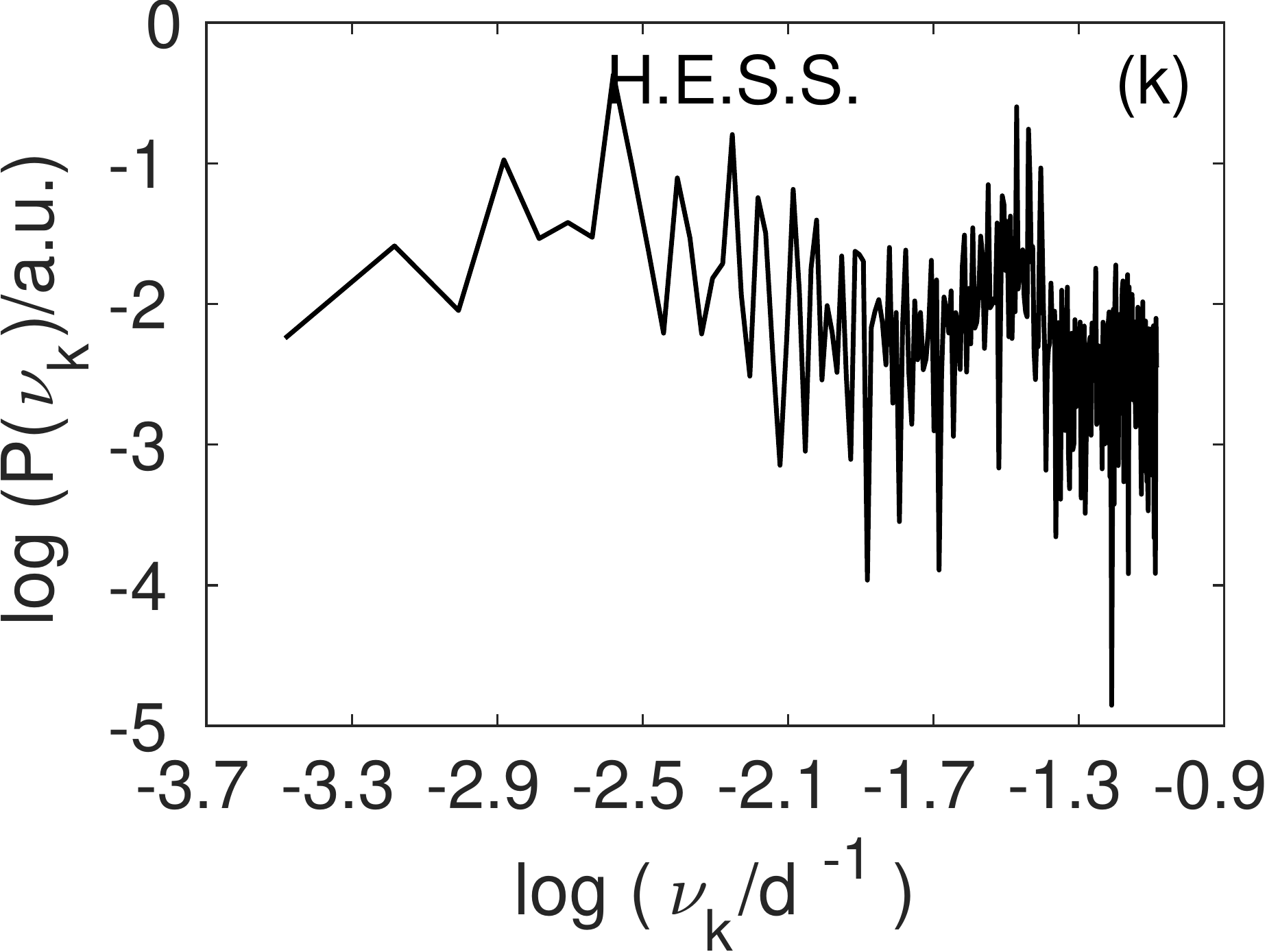}
\hspace*{0.1cm}\includegraphics[width=0.33\textwidth]{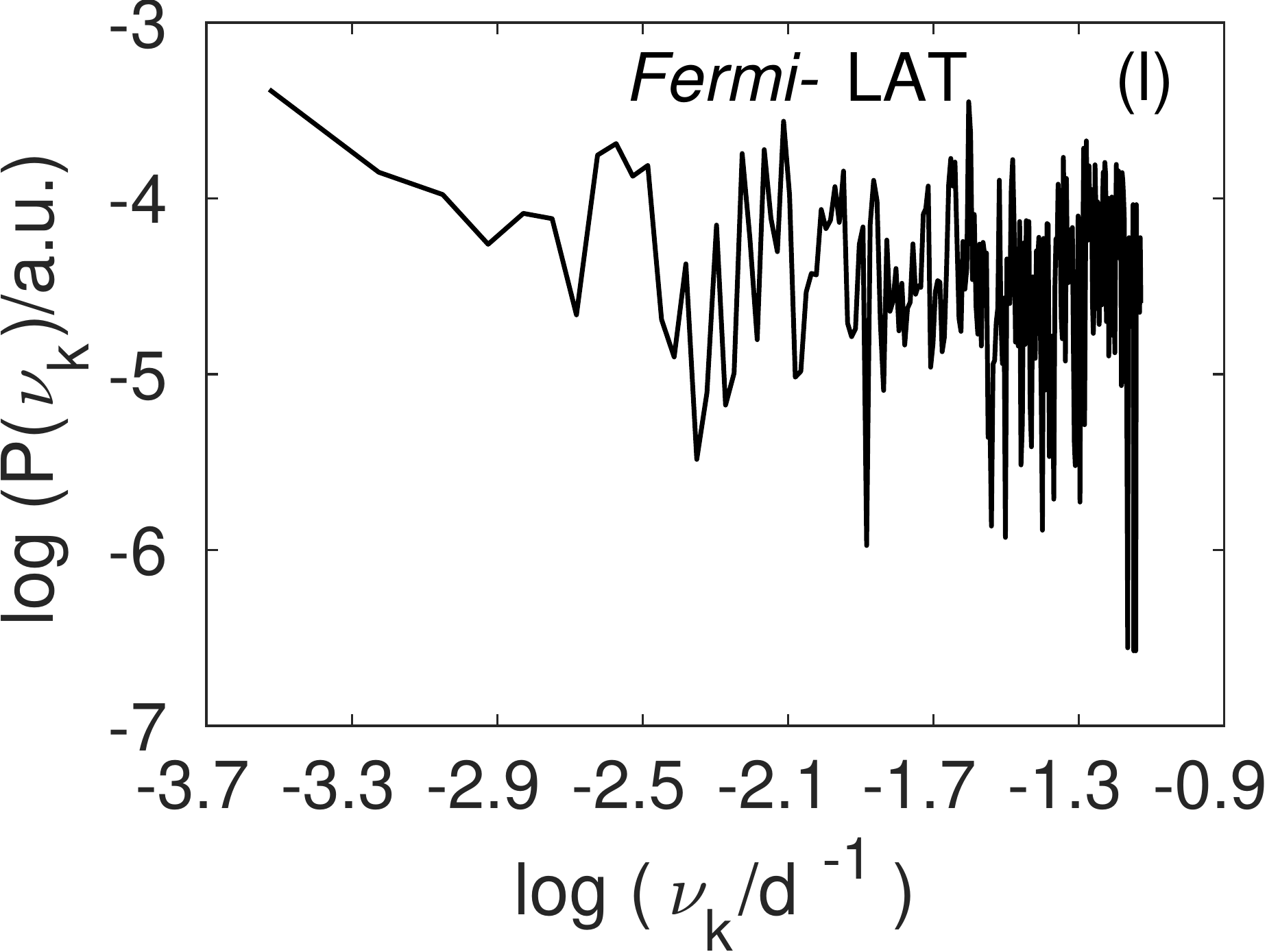}
}
\hbox{
\hspace*{0.1cm}\includegraphics[width=0.33\textwidth]{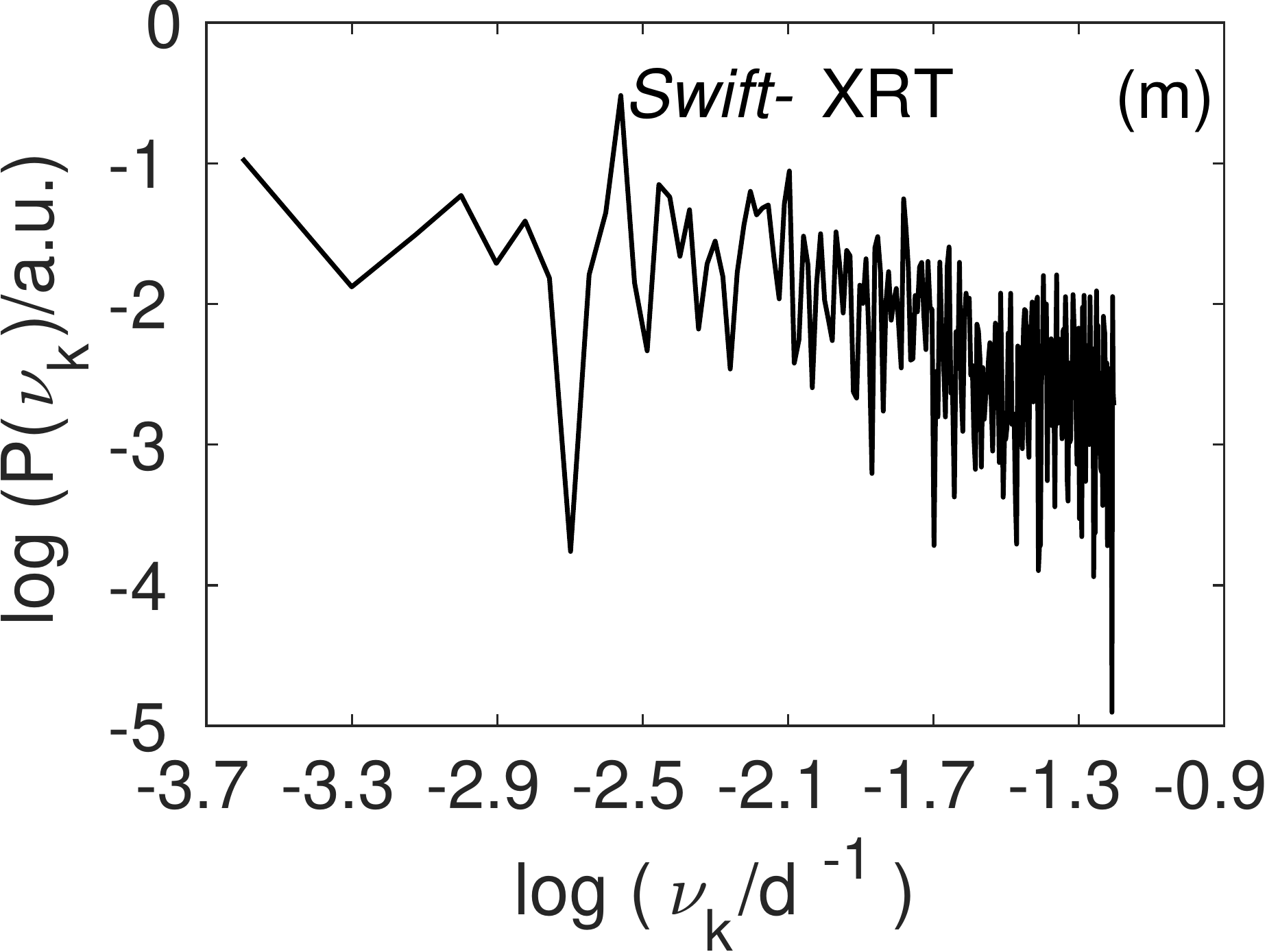}
\hspace*{0.1cm}\includegraphics[width=0.33\textwidth]{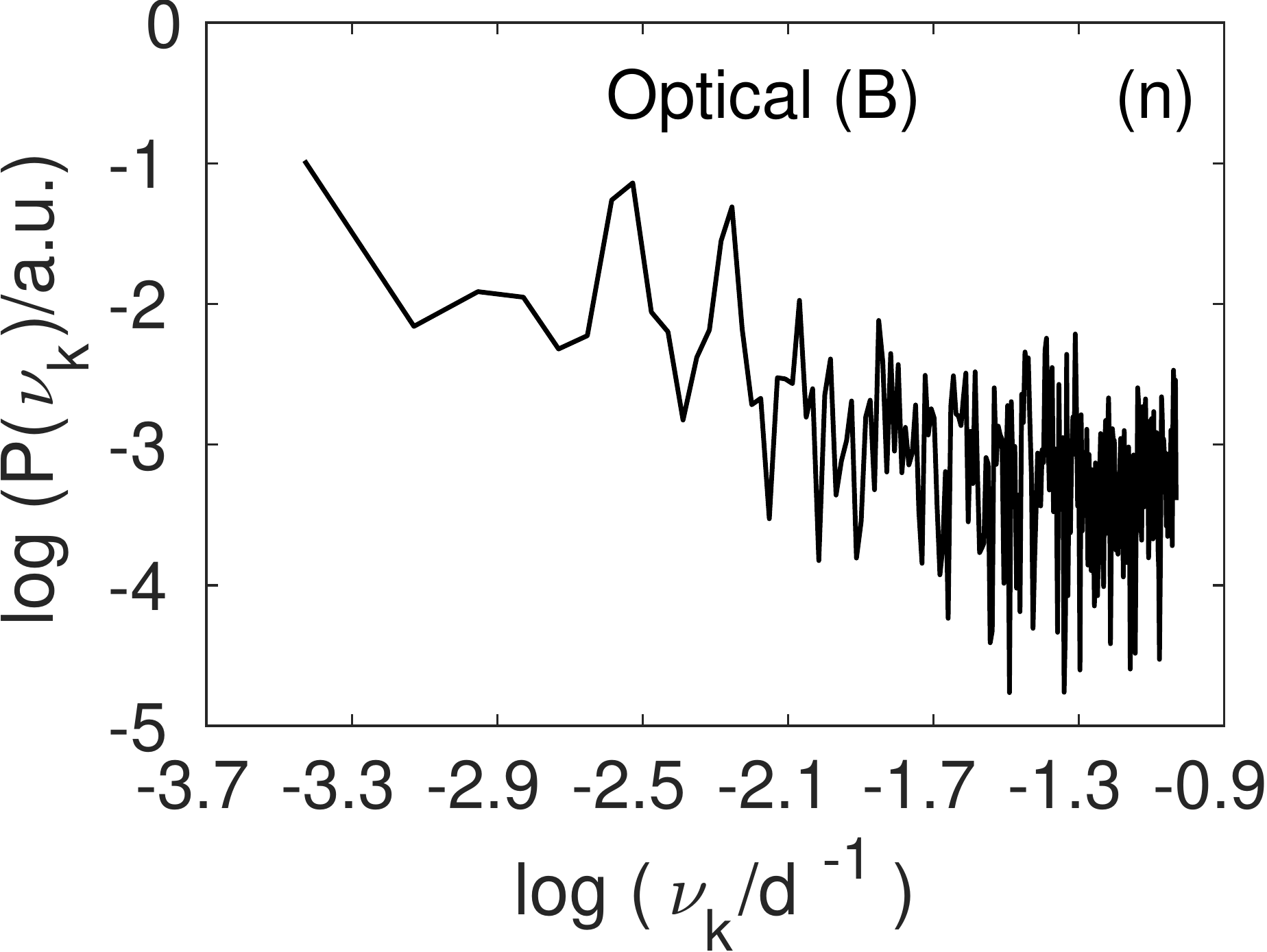}
\hspace*{0.1cm}\includegraphics[width=0.33\textwidth]{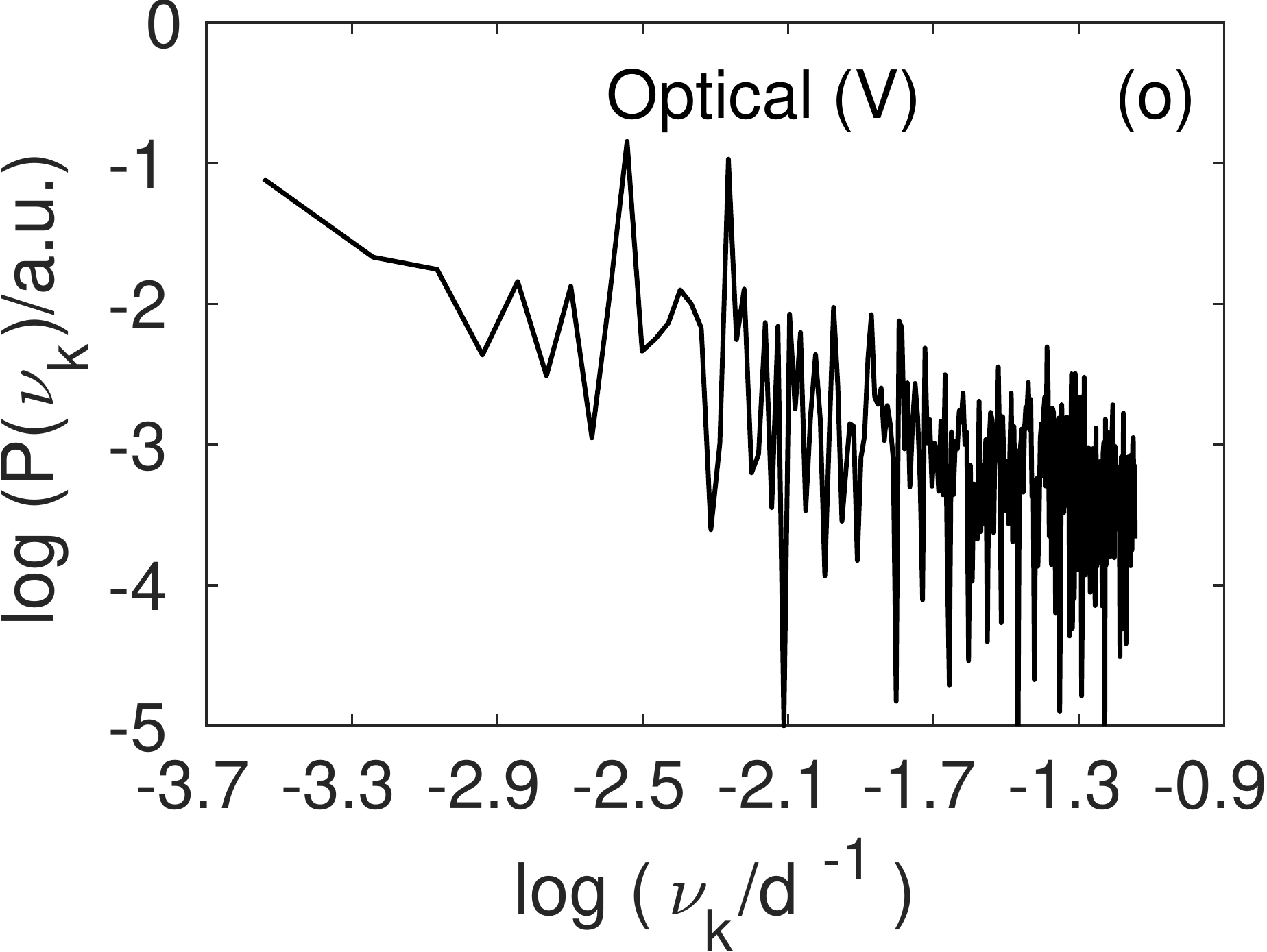}
}
\caption{({\it a-j)}: Distributions of the sampling intervals of the analysed light curves for the blazar PKS\,2155$-$304. For clarity, we have truncated the histograms above the sampling intervals with only two or fewer data points. We note that the sampling interval extends up to 347 days, 21 days, 386 days, 303 days, 325 days, 321 days, 321 days,  321 days, 380 days, and 324 days for the H.E.S.S., {\it Fermi-}LAT, {\it Swift-}XRT, B, V, R, I, J, H, and K-band light curves, respectively (Table~\ref{tab:psd}). ({\it k-t}): Corresponding spectral window functions. }

\label{fig:7}
\end{figure*}

\begin{figure*}
\hbox{
\hspace*{0.1cm}\includegraphics[width=0.33\textwidth]{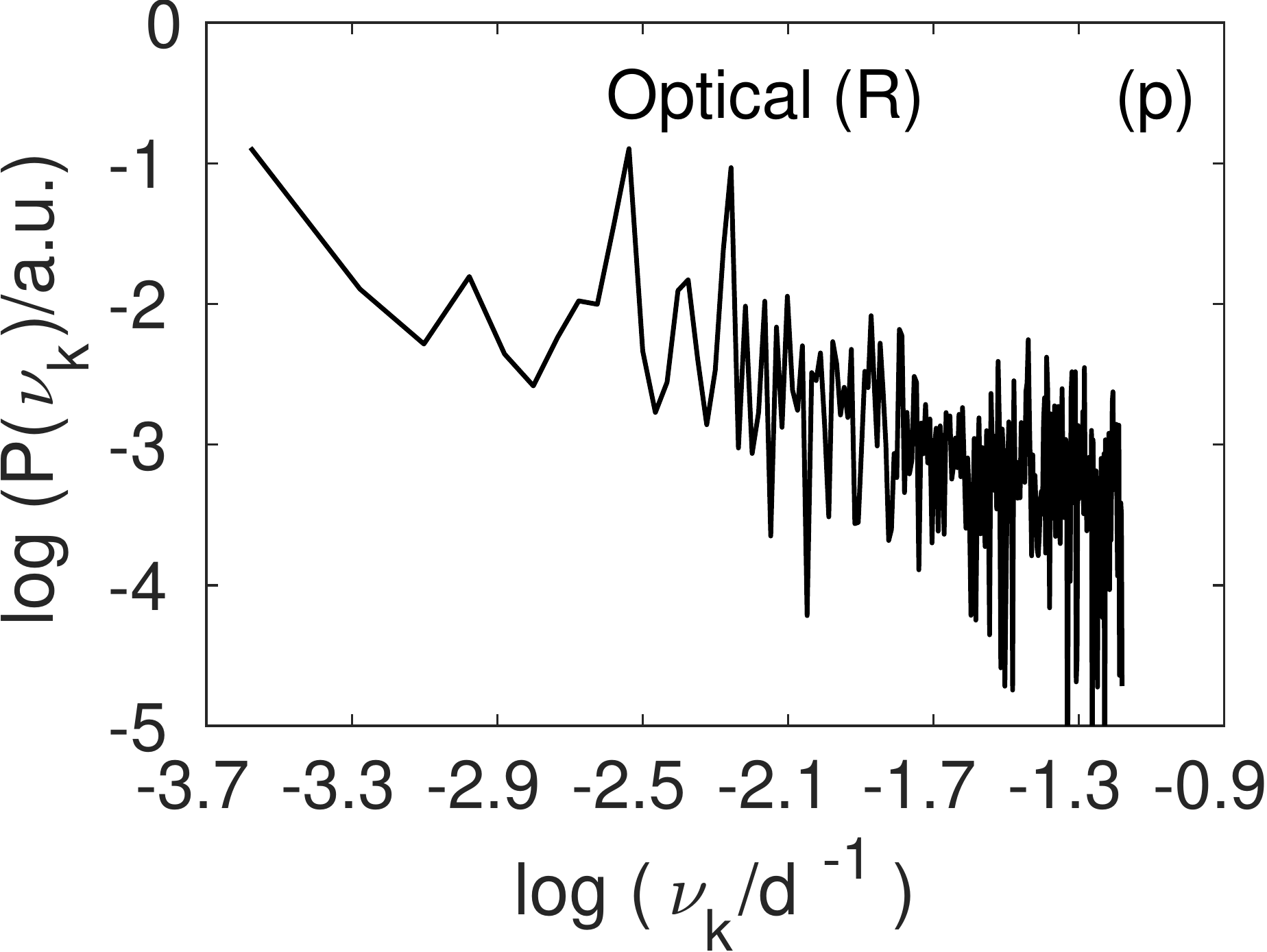}
\hspace*{0.1cm}\includegraphics[width=0.33\textwidth]{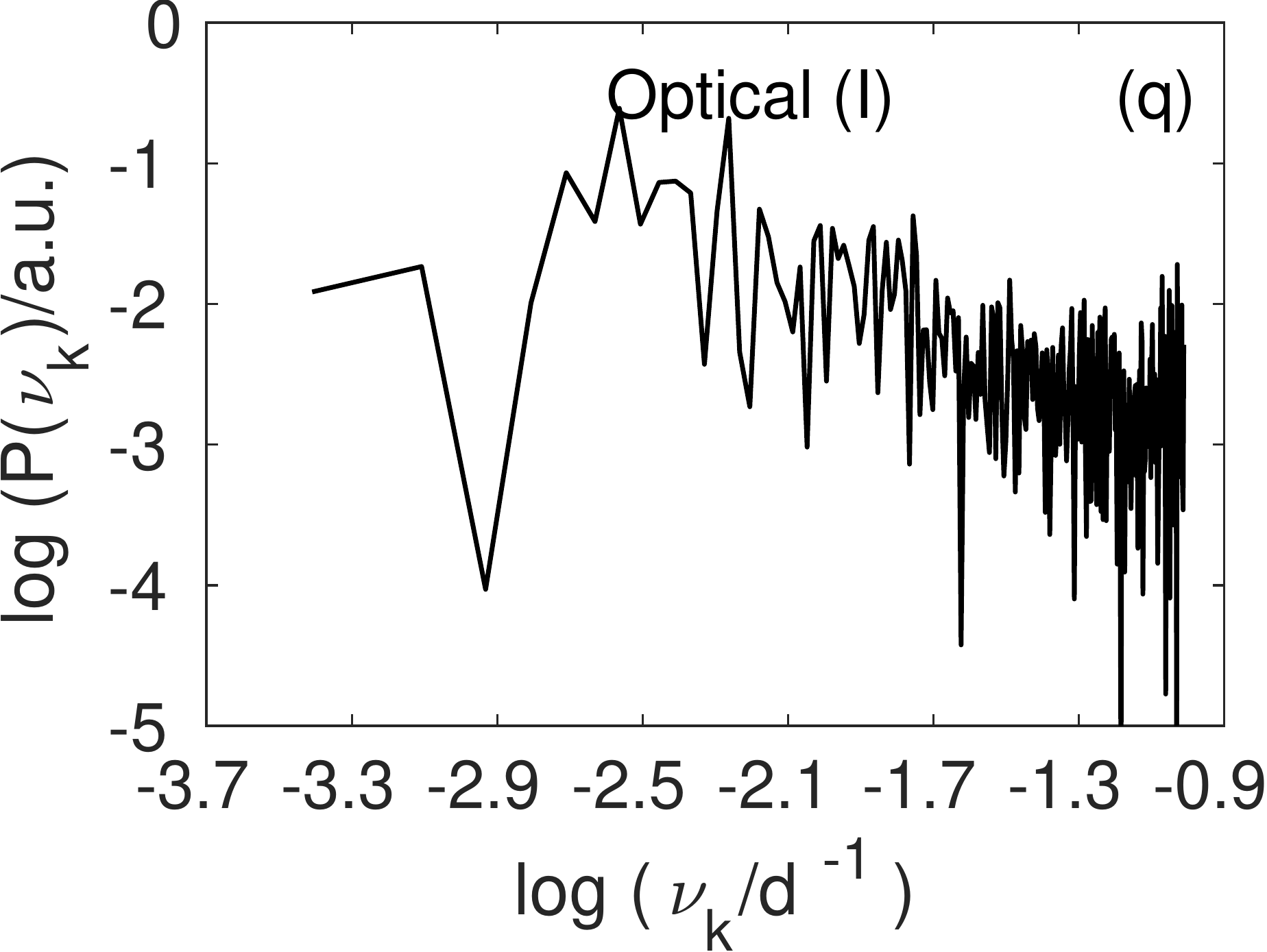}
\hspace*{0.1cm}\includegraphics[width=0.33\textwidth]{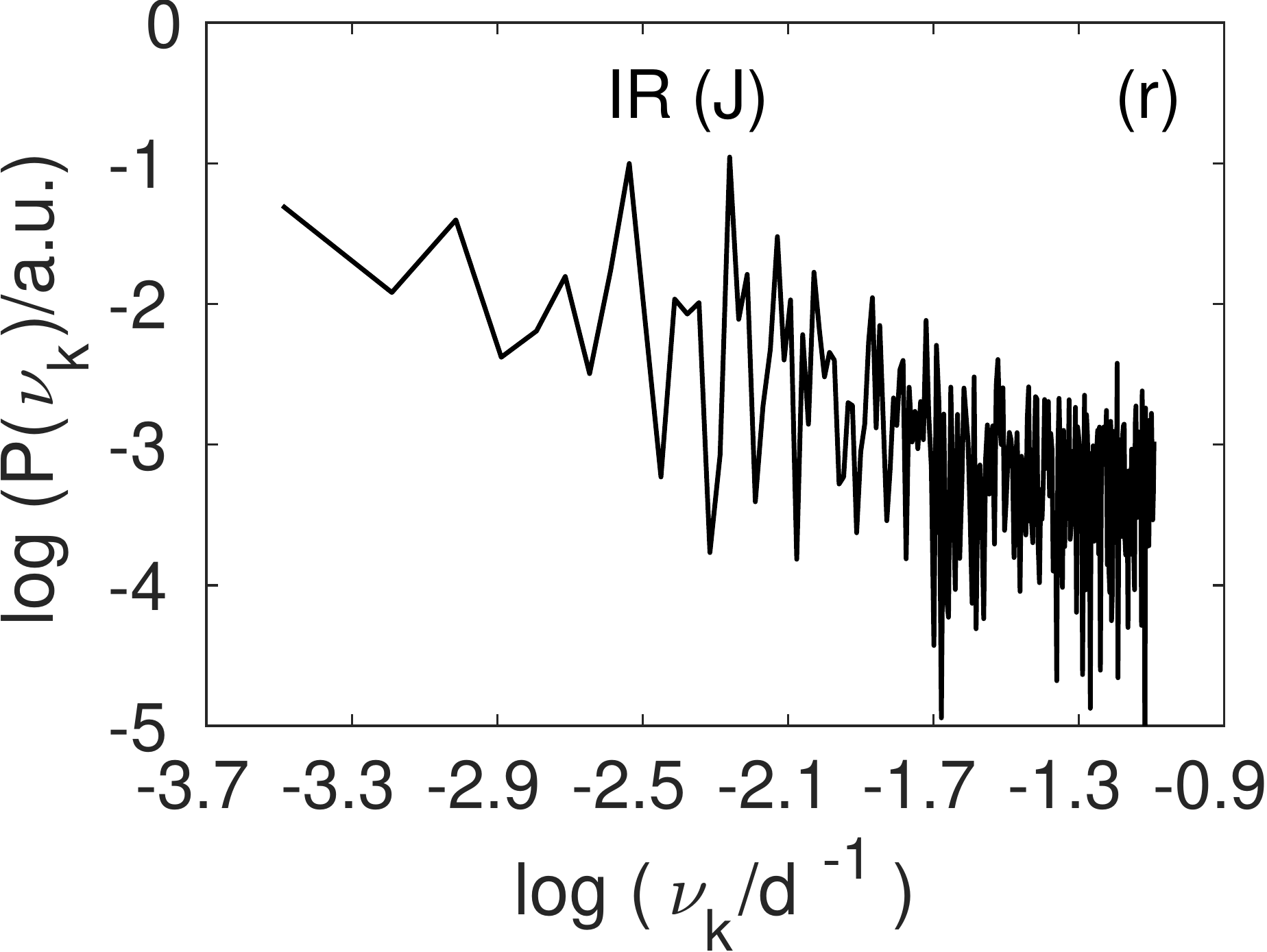}
}
\hspace*{0.1cm}\includegraphics[width=0.33\textwidth]{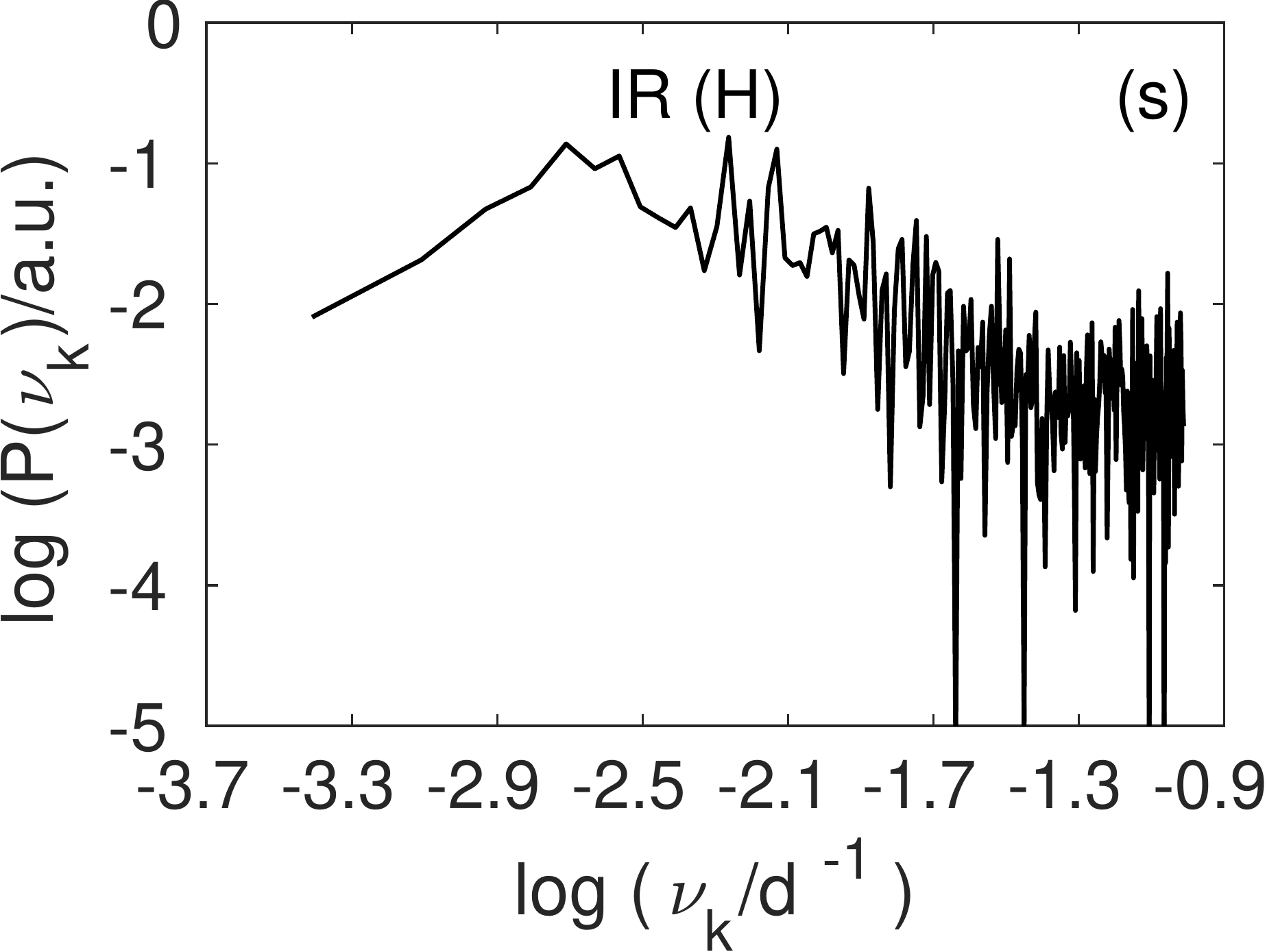}
\hspace*{0.1cm}\includegraphics[width=0.33\textwidth]{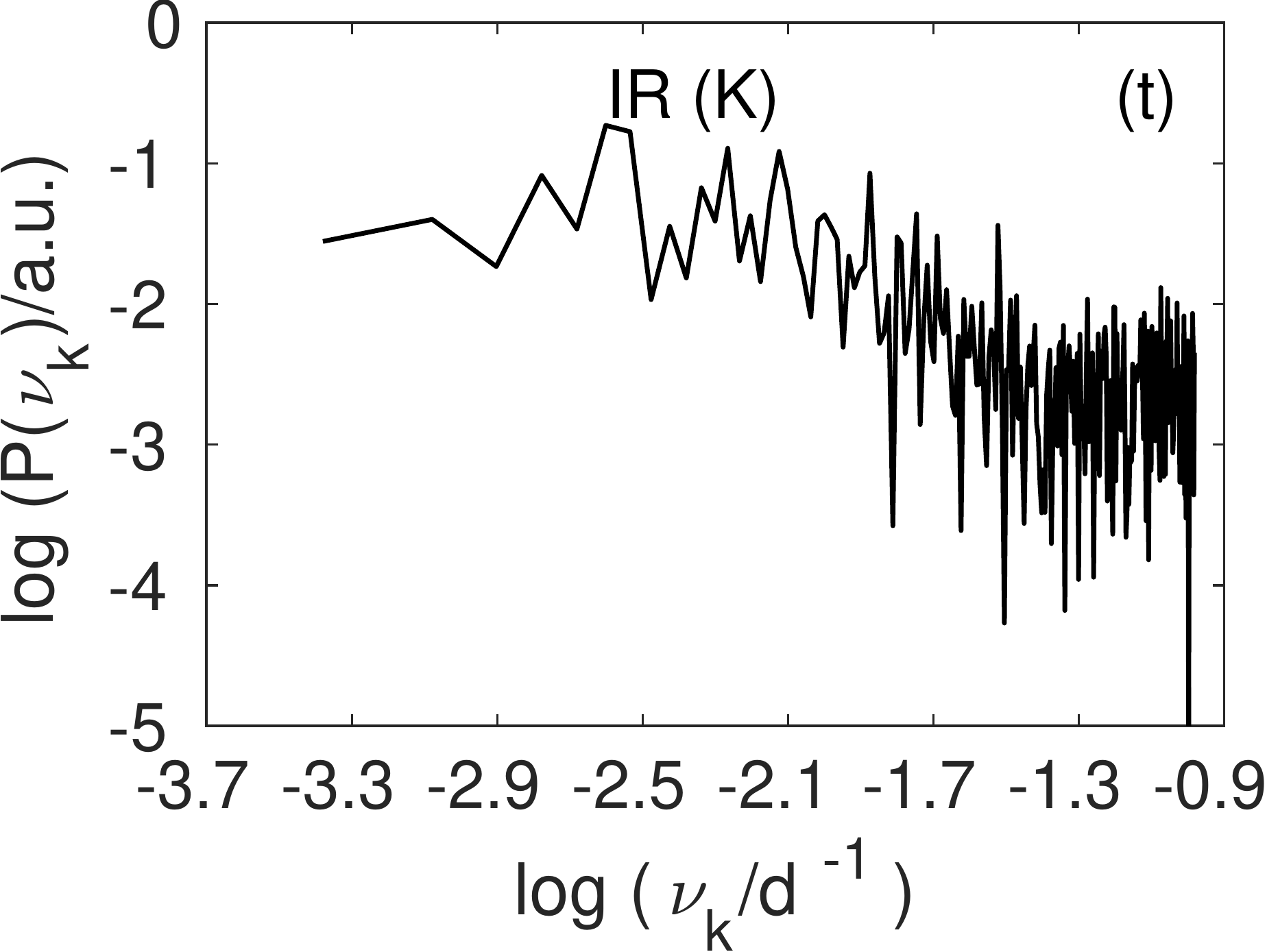}
\contcaption{}
\end{figure*}

\section{Probability distribution curves for the analysed PSDs}
\label{app:B}
The probability that a single power-law fit with a given simulation (input) $\beta$ describe the PSDs within the $\beta$ range 0.2--3.0 is shown in Figs.~\ref{fig:8} and ~\ref{fig:9} for the blazars Mrk\,421 and PKS\,2155$-$304, respectively. We note that $p_\beta$ is higher than 10 percent in all the cases except for OVRO PSD for the blazar Mrk\,421 for which it is 8.4 percent. In general, the high $p_\beta$ values for the periodograms indicate that the single power-law fits describe the PSDs well.   

\begin{figure*}
\hbox{
\hspace*{0.1cm}\includegraphics[width=0.33\textwidth]{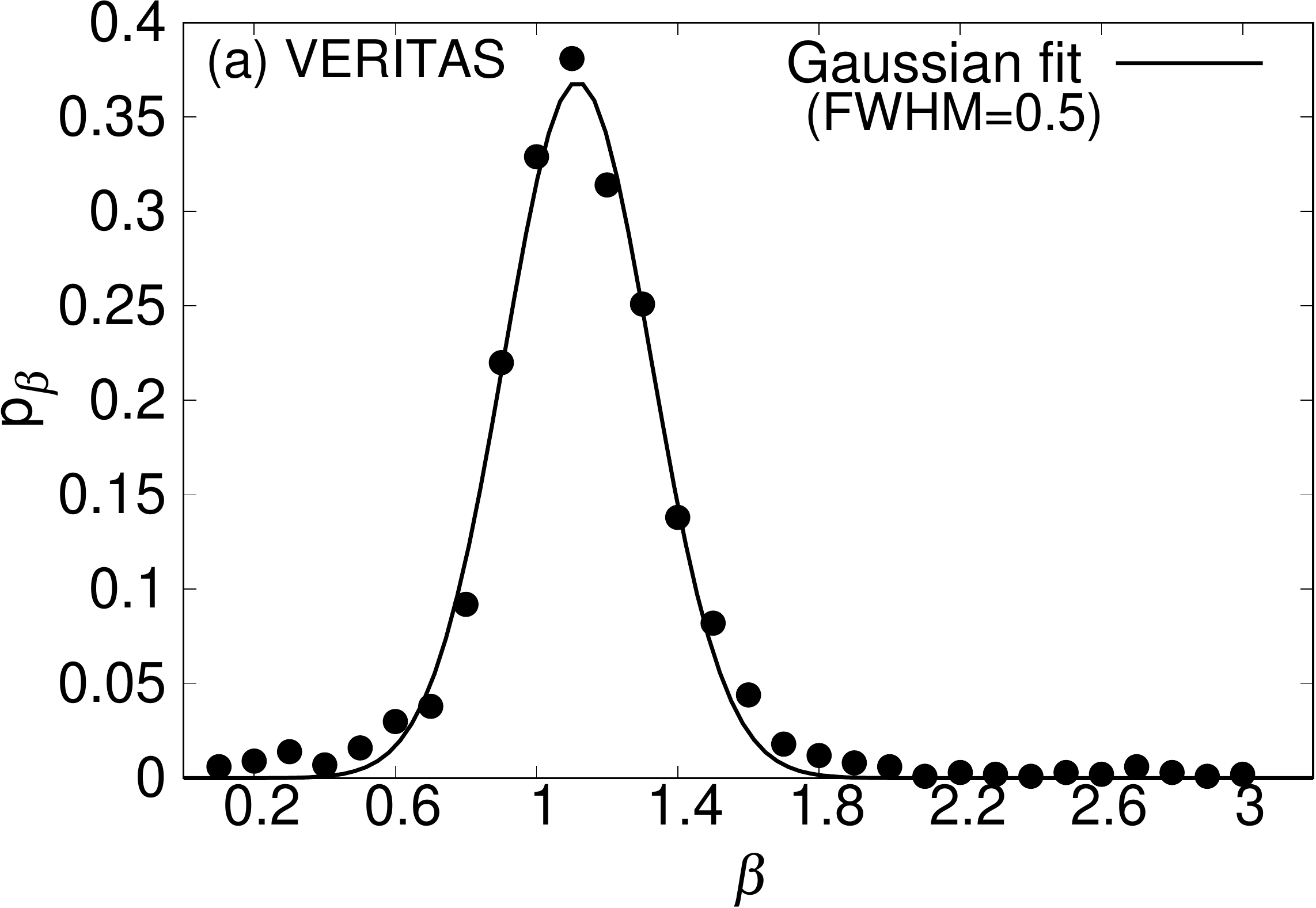}
\hspace*{0.1cm}\includegraphics[width=0.33\textwidth]{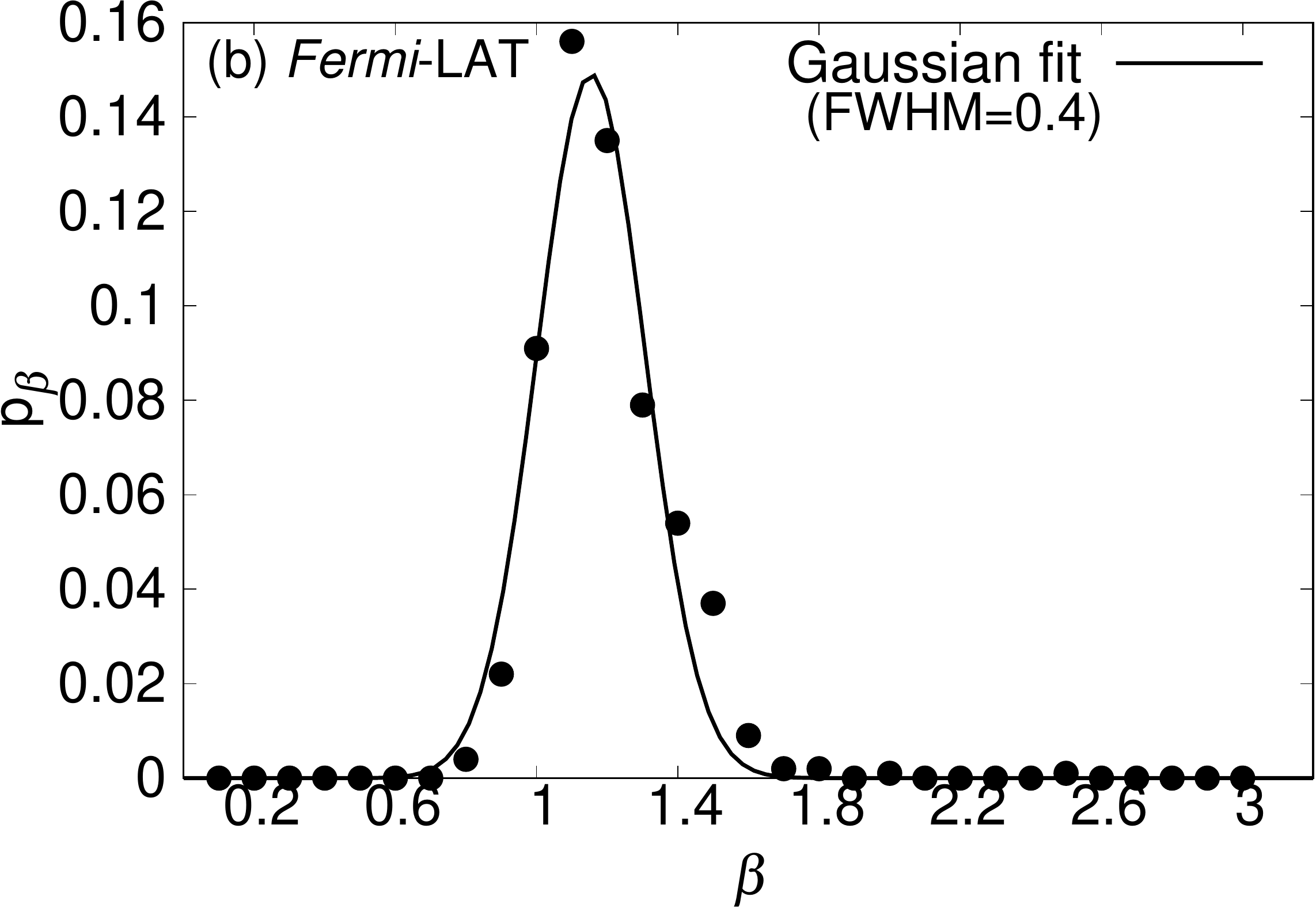}
\hspace*{0.1cm}\includegraphics[width=0.33\textwidth]{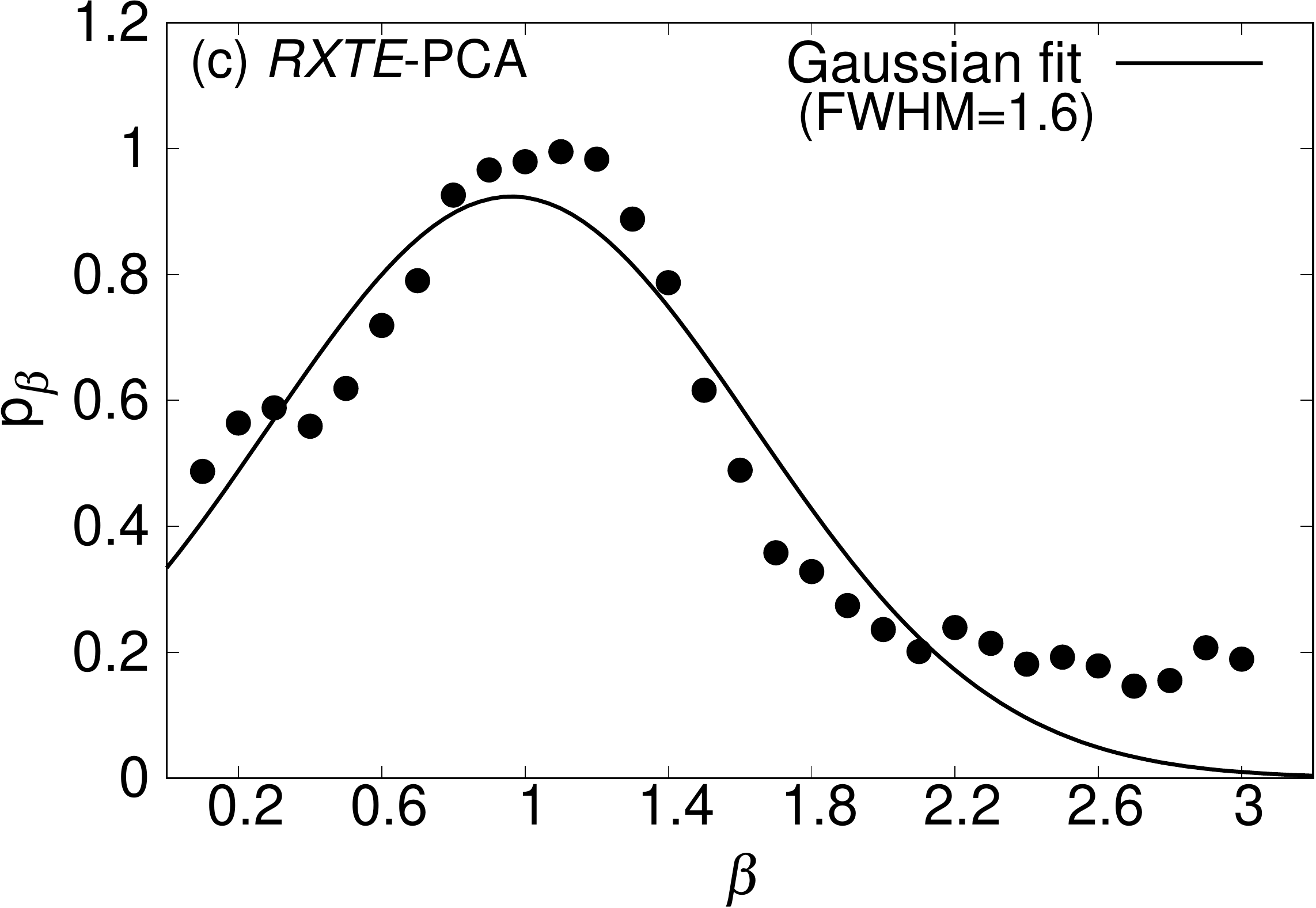}
}
\hbox{
\hspace*{0.1cm}\includegraphics[width=0.33\textwidth]{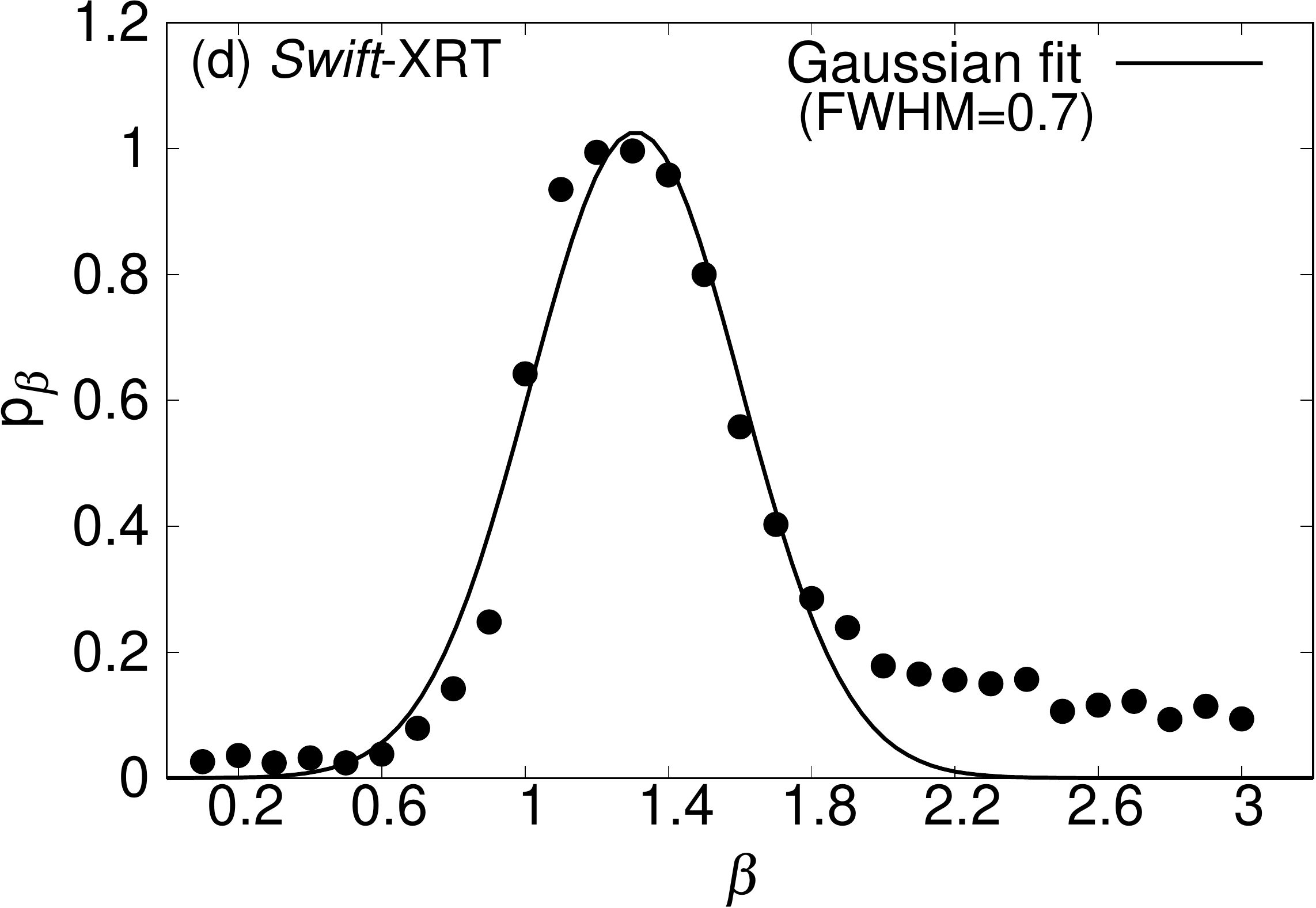}
\hspace*{0.1cm}\includegraphics[width=0.33\textwidth]{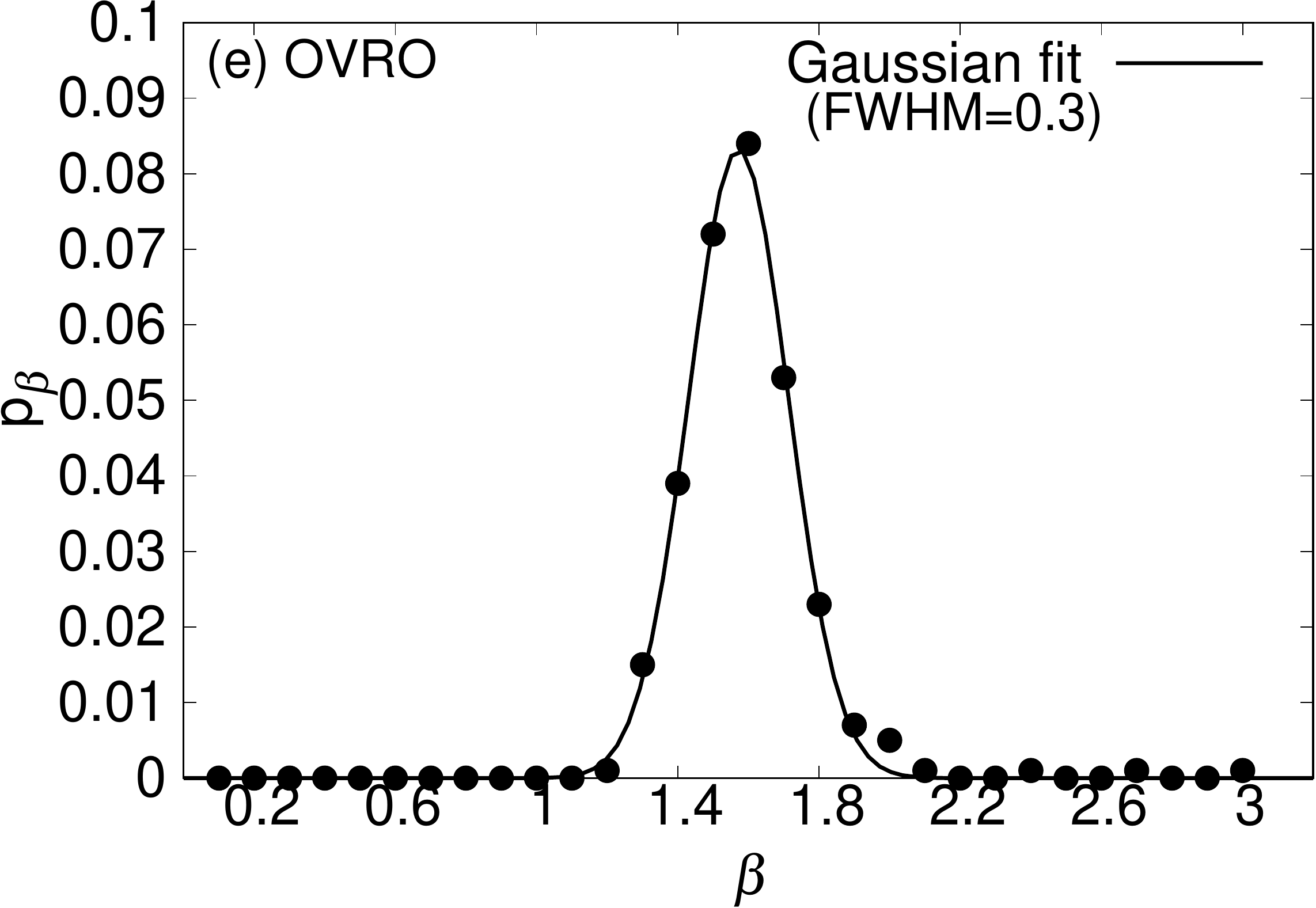}
}

\caption{Probablity, $p_\beta$, that a single power-law slope is acceptable to describe the PSDs for the blazar Mrk\,421. The filled circles show the fraction for which $\chi^2_{\rm dist,i} > \chi^2_{\rm obs}$ was found out of 1,000 simulations for a fixed $\beta$. The solid lines show the Gaussian fits to the distributions.  }
\label{fig:8}
\end{figure*}

\begin{figure*}
\hbox{
\hspace*{0.1cm}\includegraphics[width=0.33\textwidth]{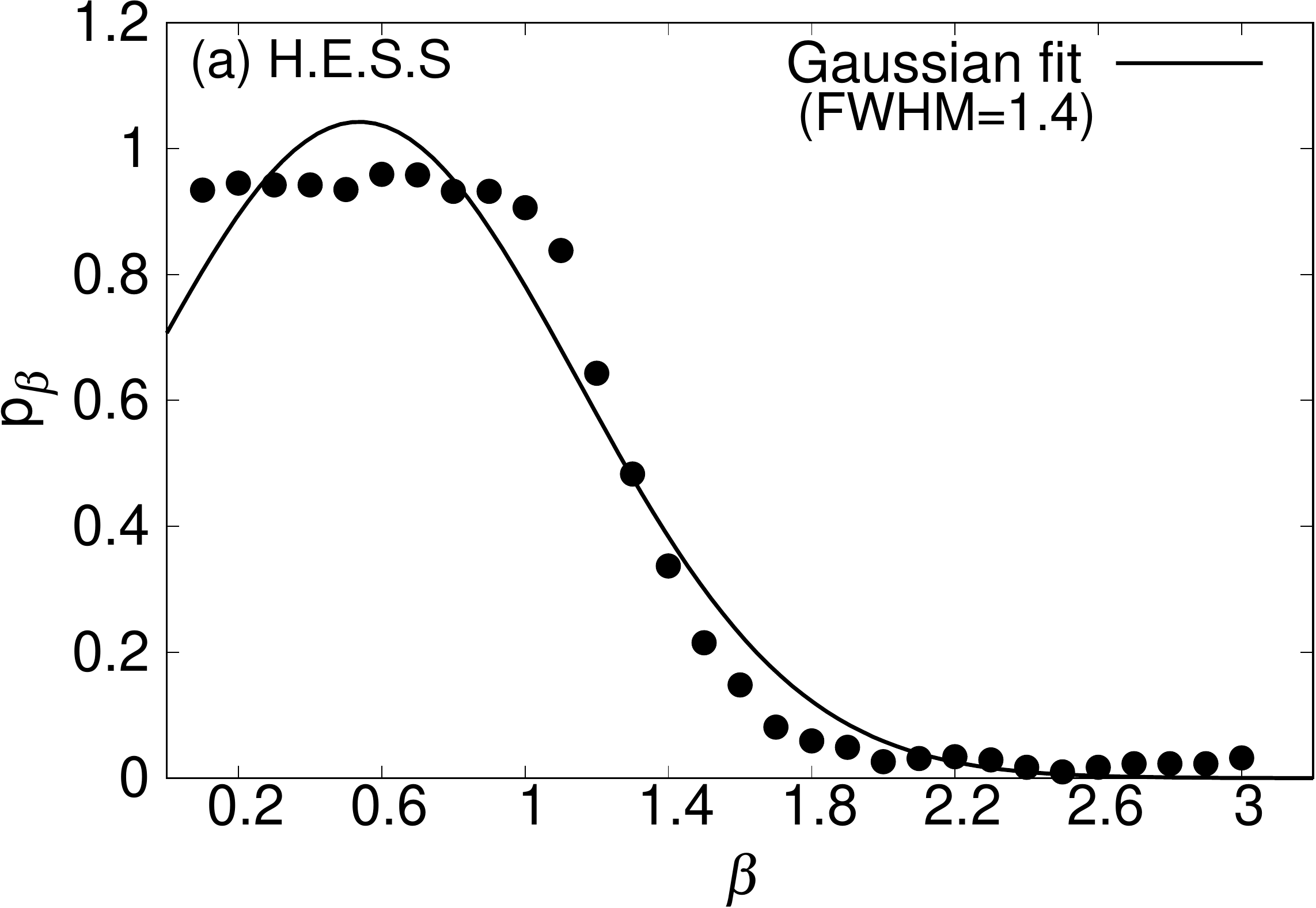}
\hspace*{0.1cm}\includegraphics[width=0.33\textwidth]{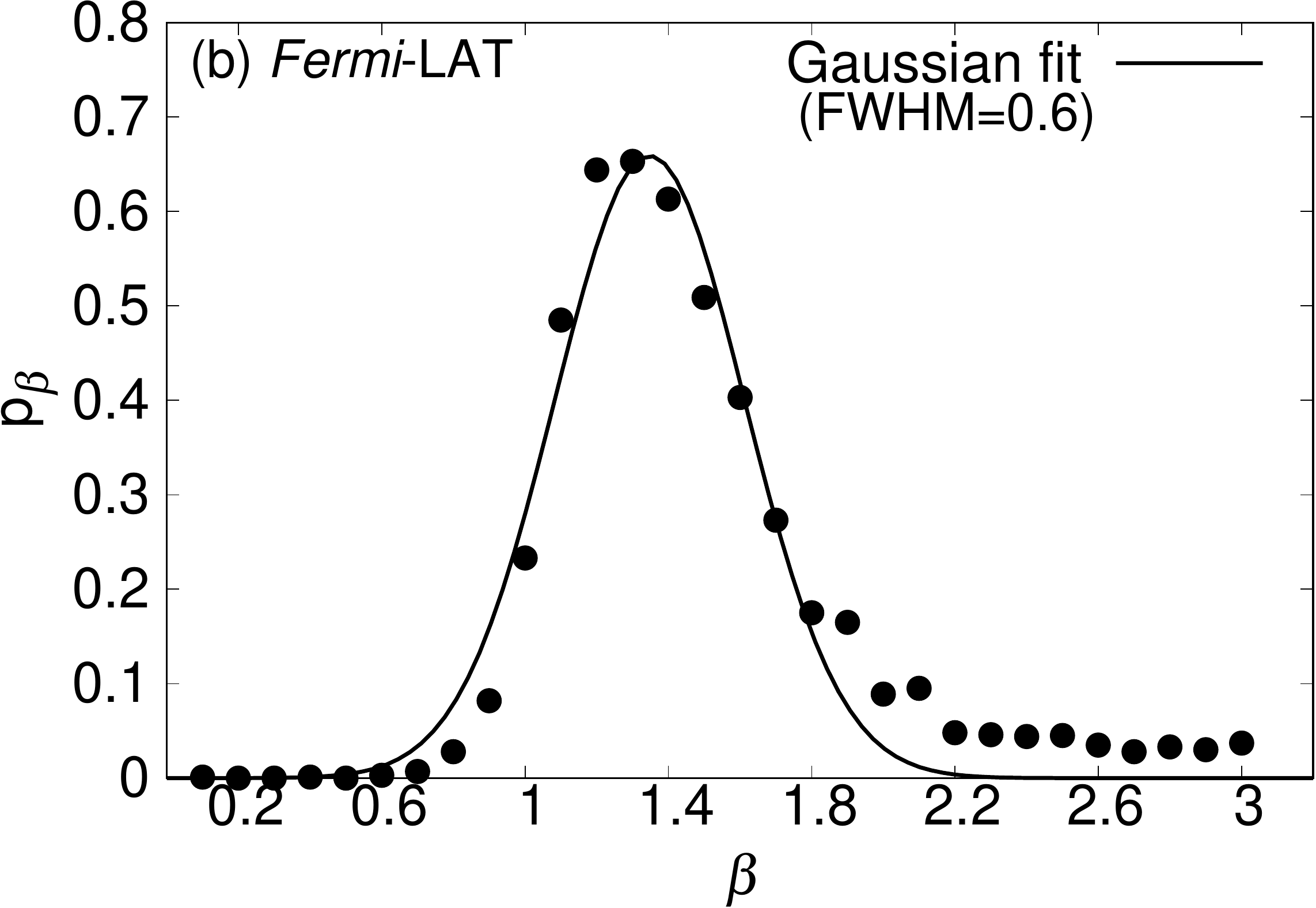}
\hspace*{0.1cm}\includegraphics[width=0.33\textwidth]{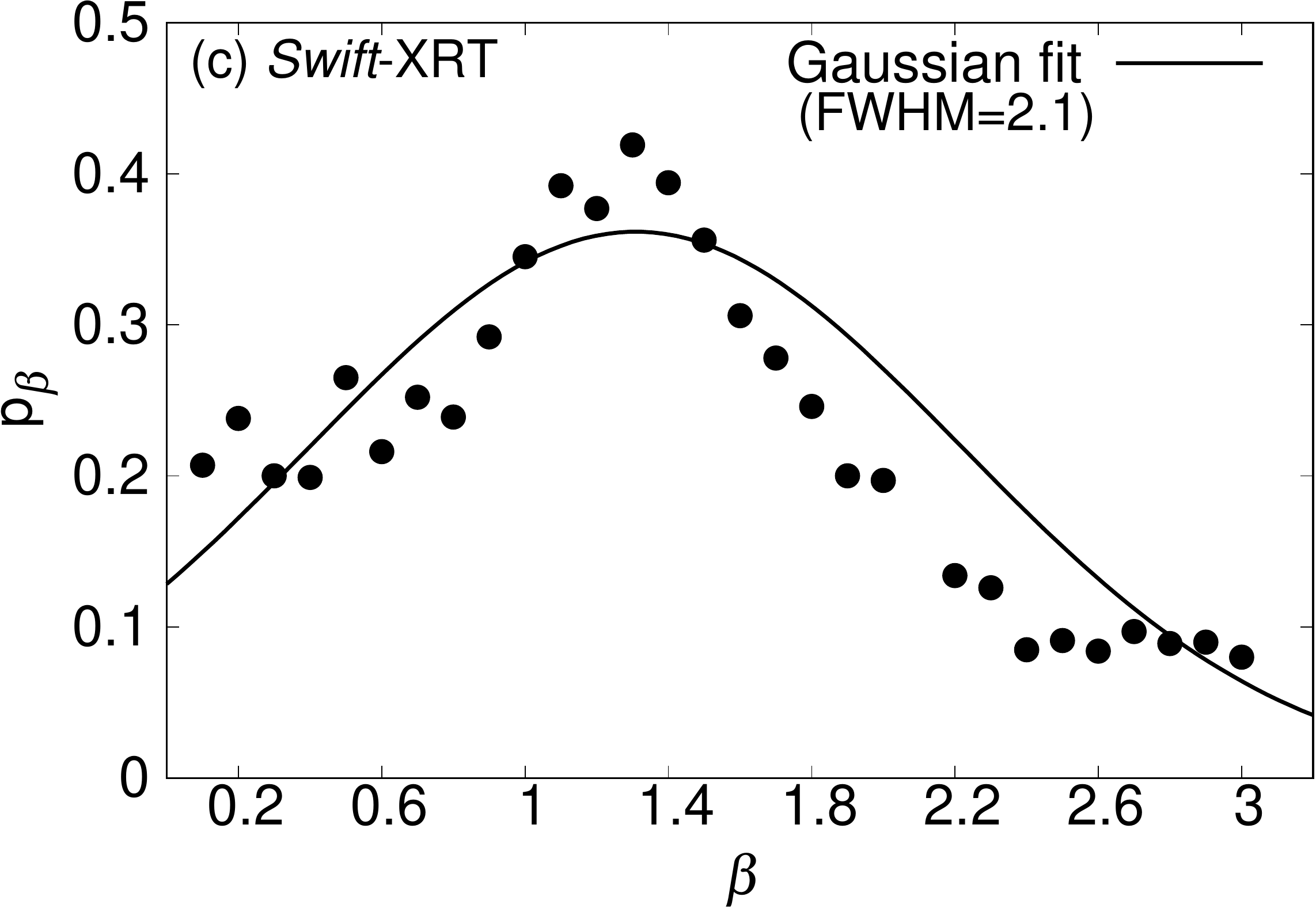}
}
\hbox{
\hspace*{0.1cm}\includegraphics[width=0.33\textwidth]{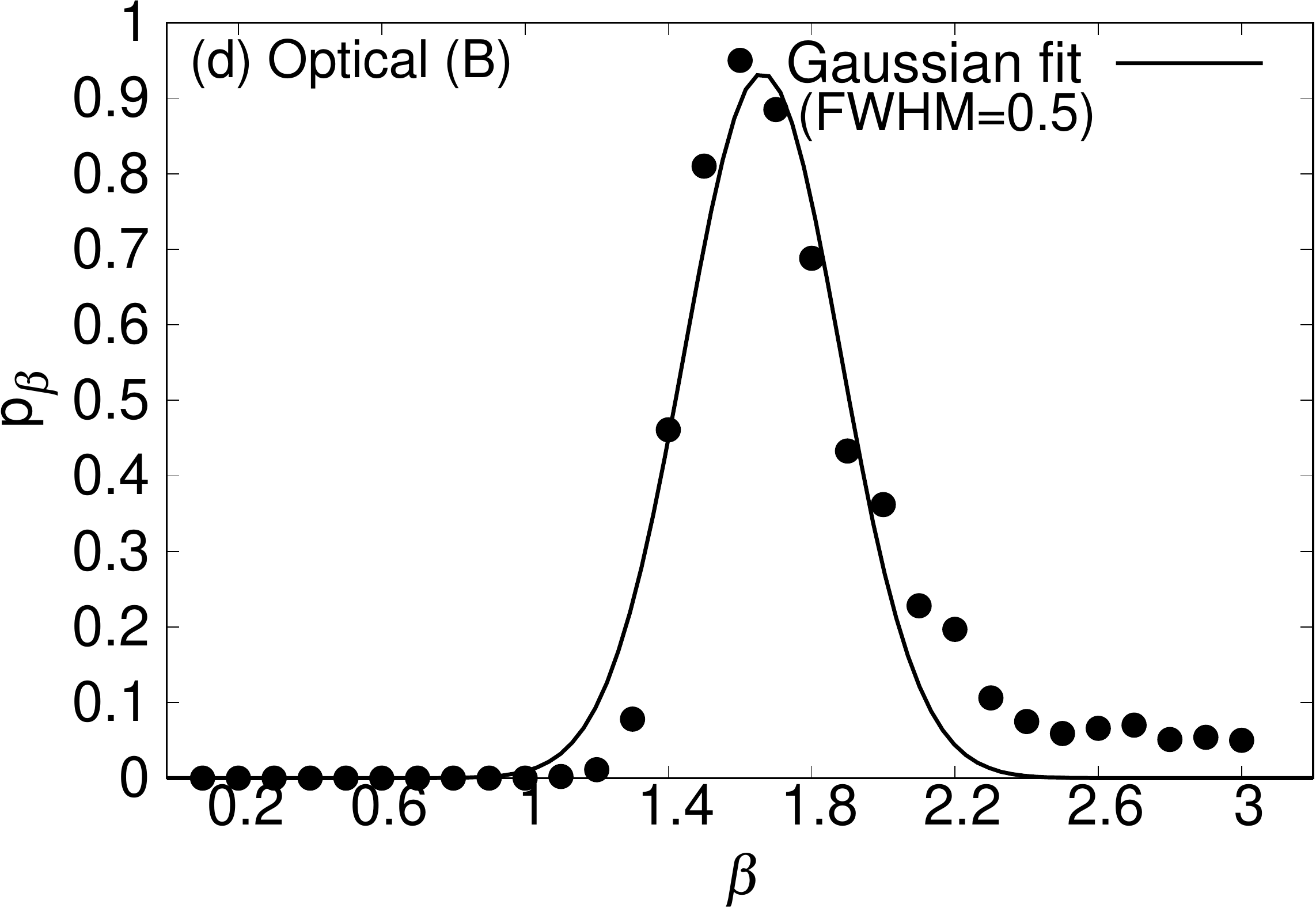}
\hspace*{0.1cm}\includegraphics[width=0.33\textwidth]{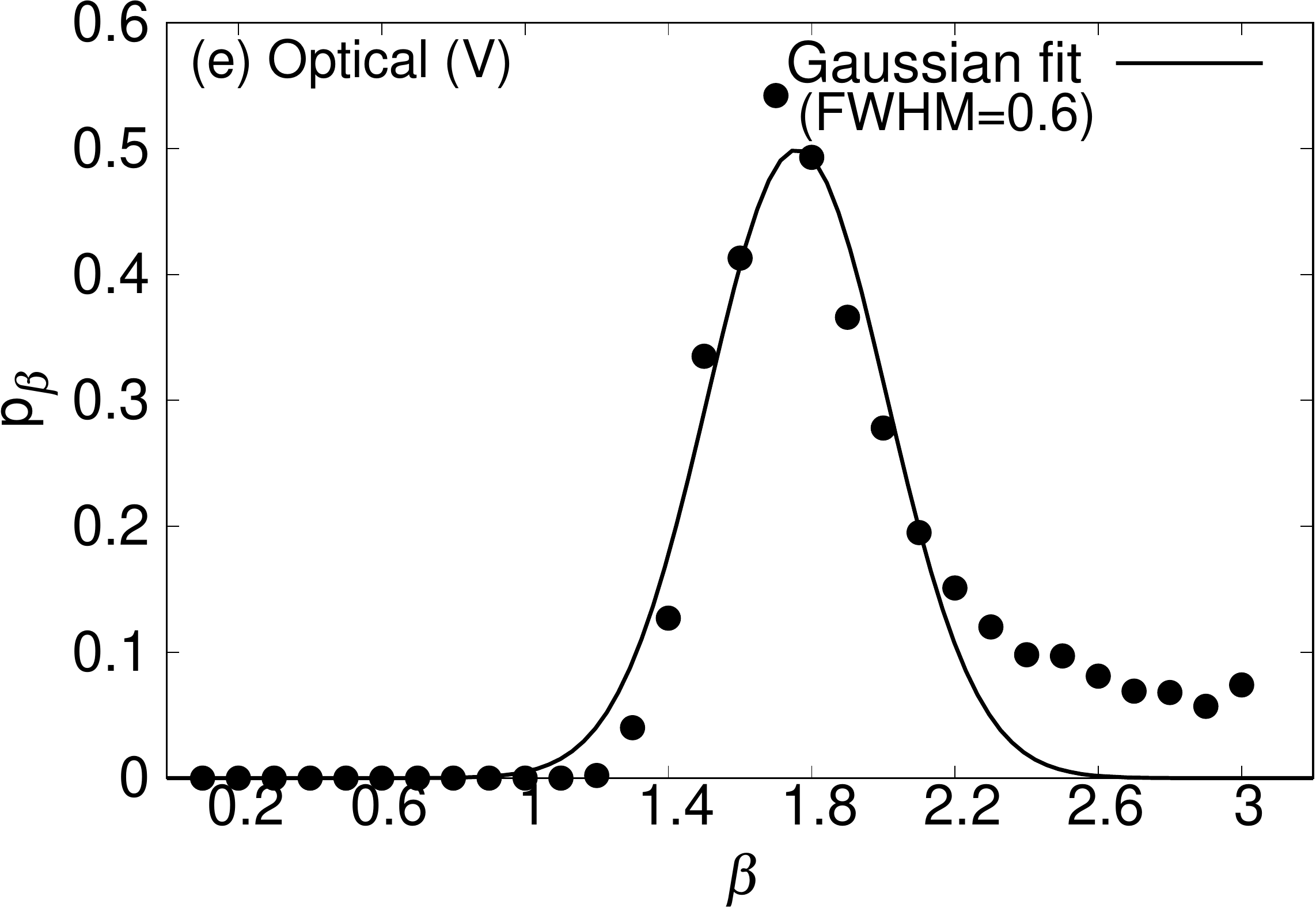}
\hspace*{0.1cm}\includegraphics[width=0.33\textwidth]{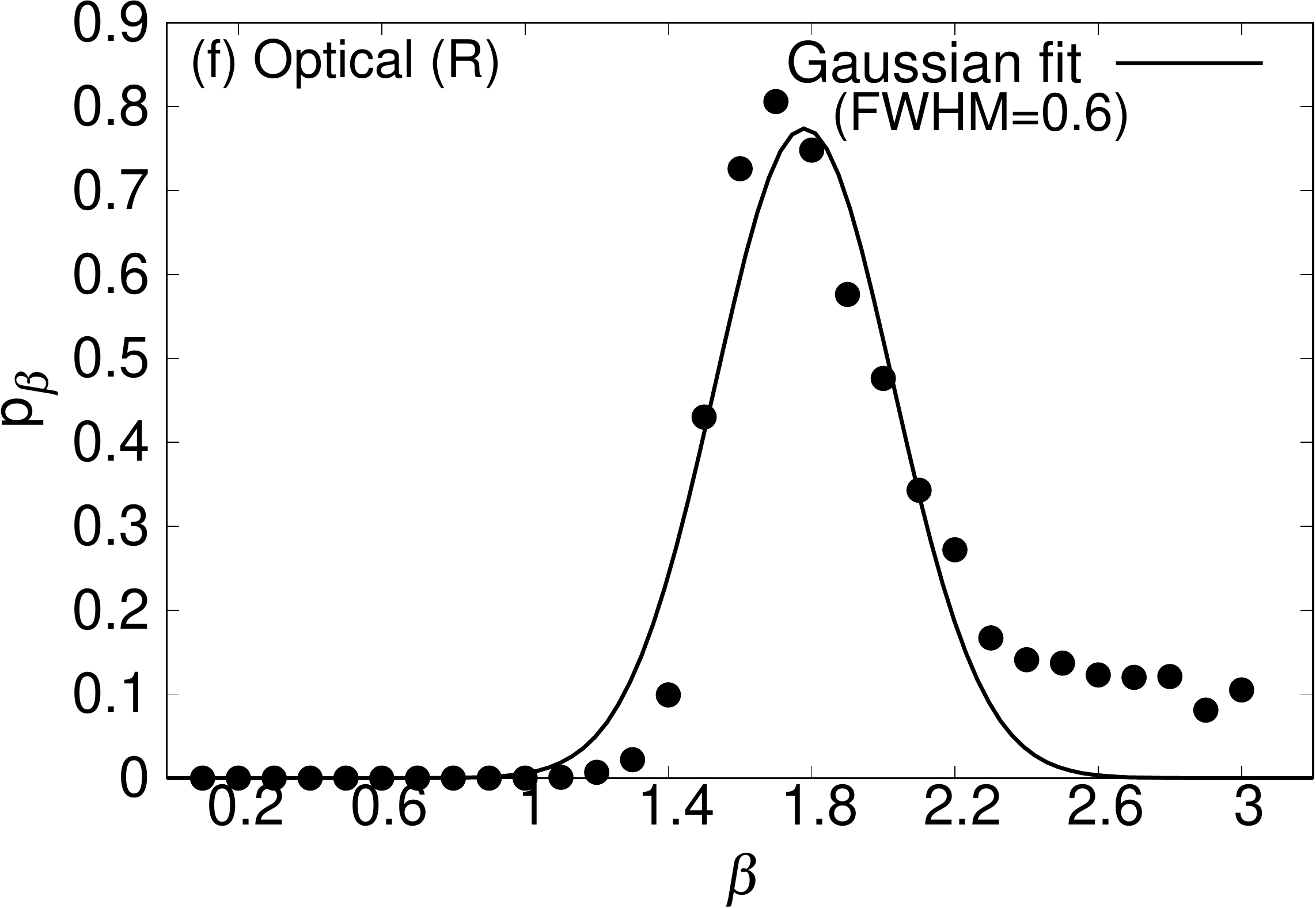}
}
\hbox{
\hspace*{0.1cm}\includegraphics[width=0.33\textwidth]{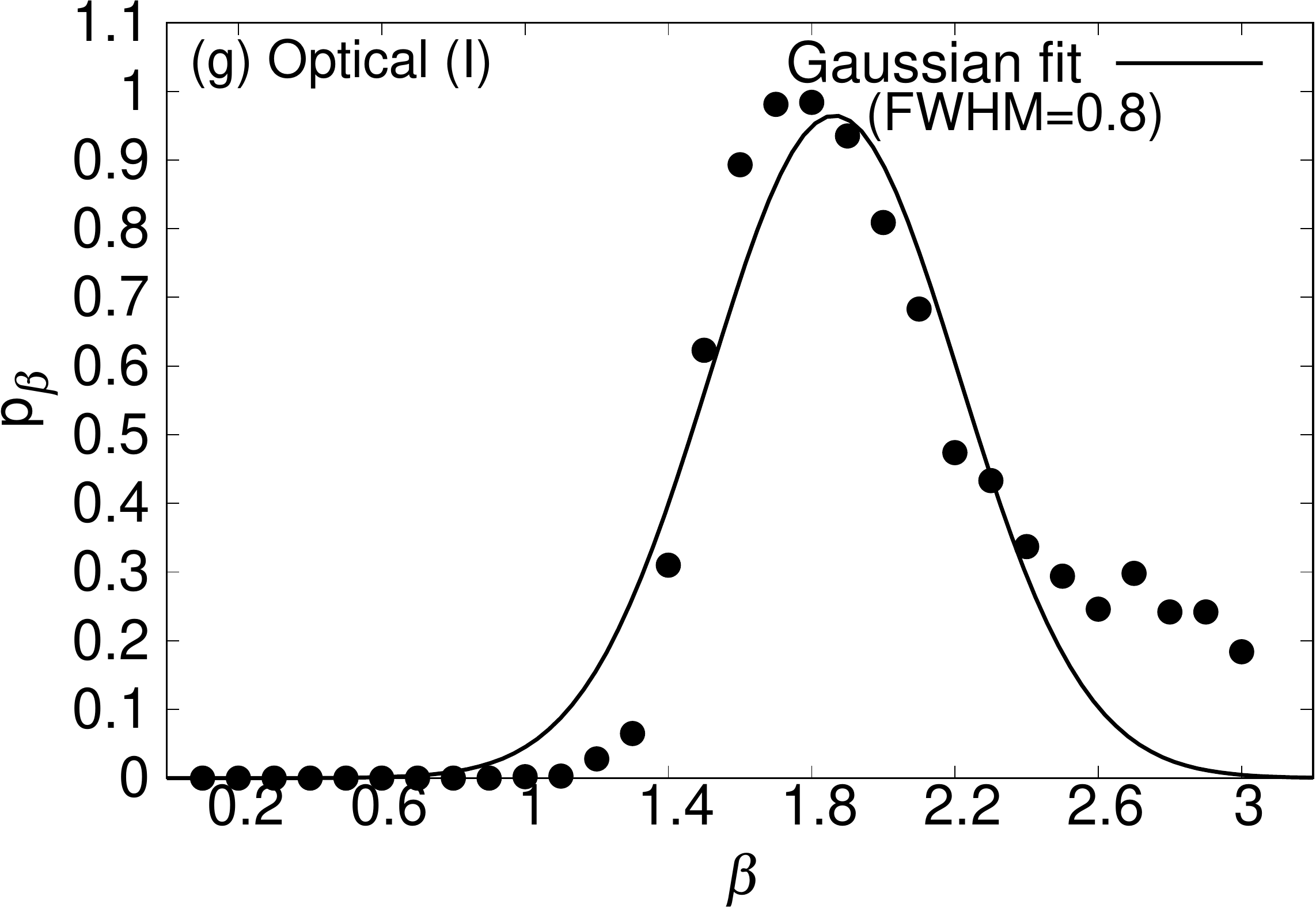}
\hspace*{0.1cm}\includegraphics[width=0.33\textwidth]{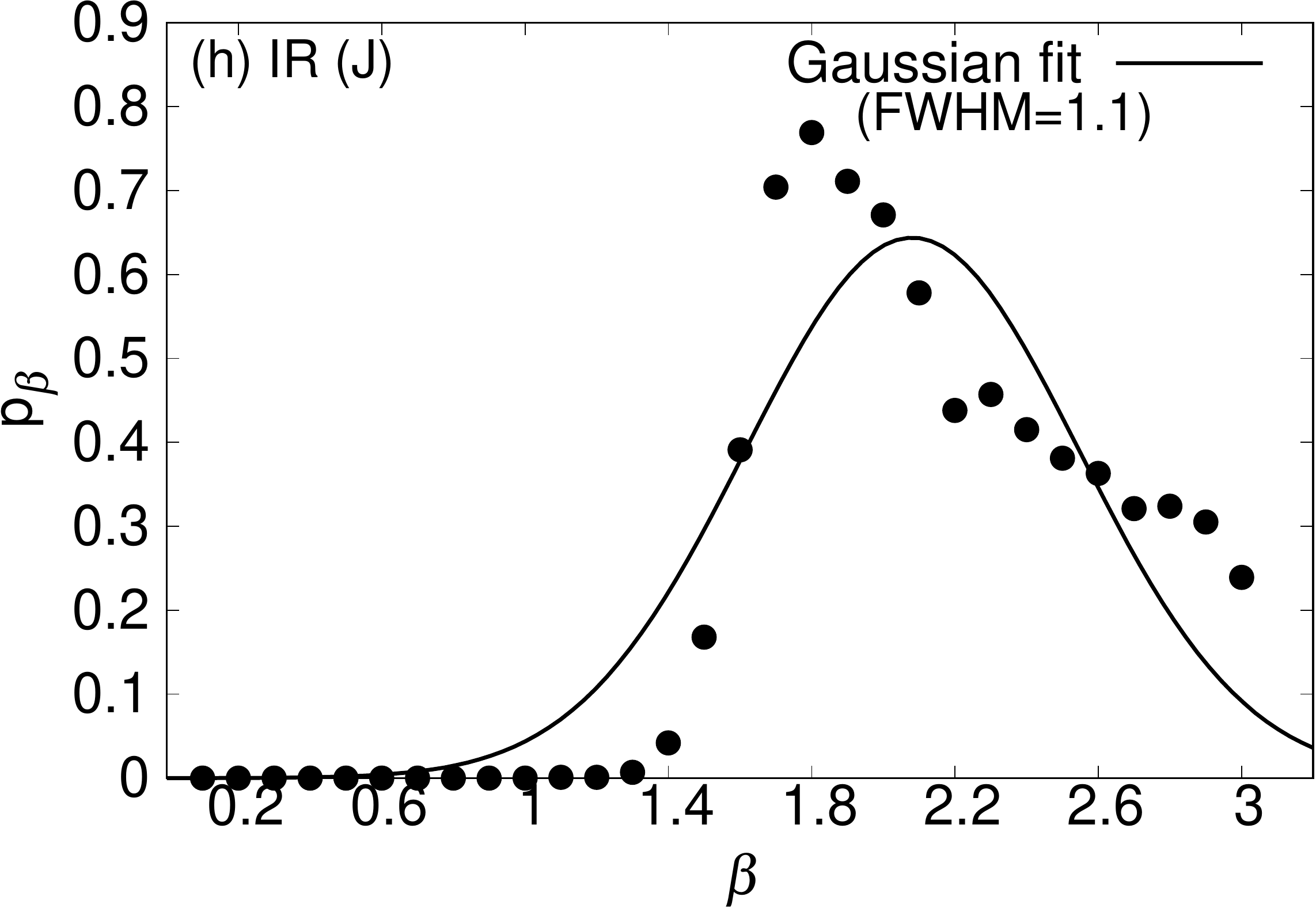}
\hspace*{0.1cm}\includegraphics[width=0.33\textwidth]{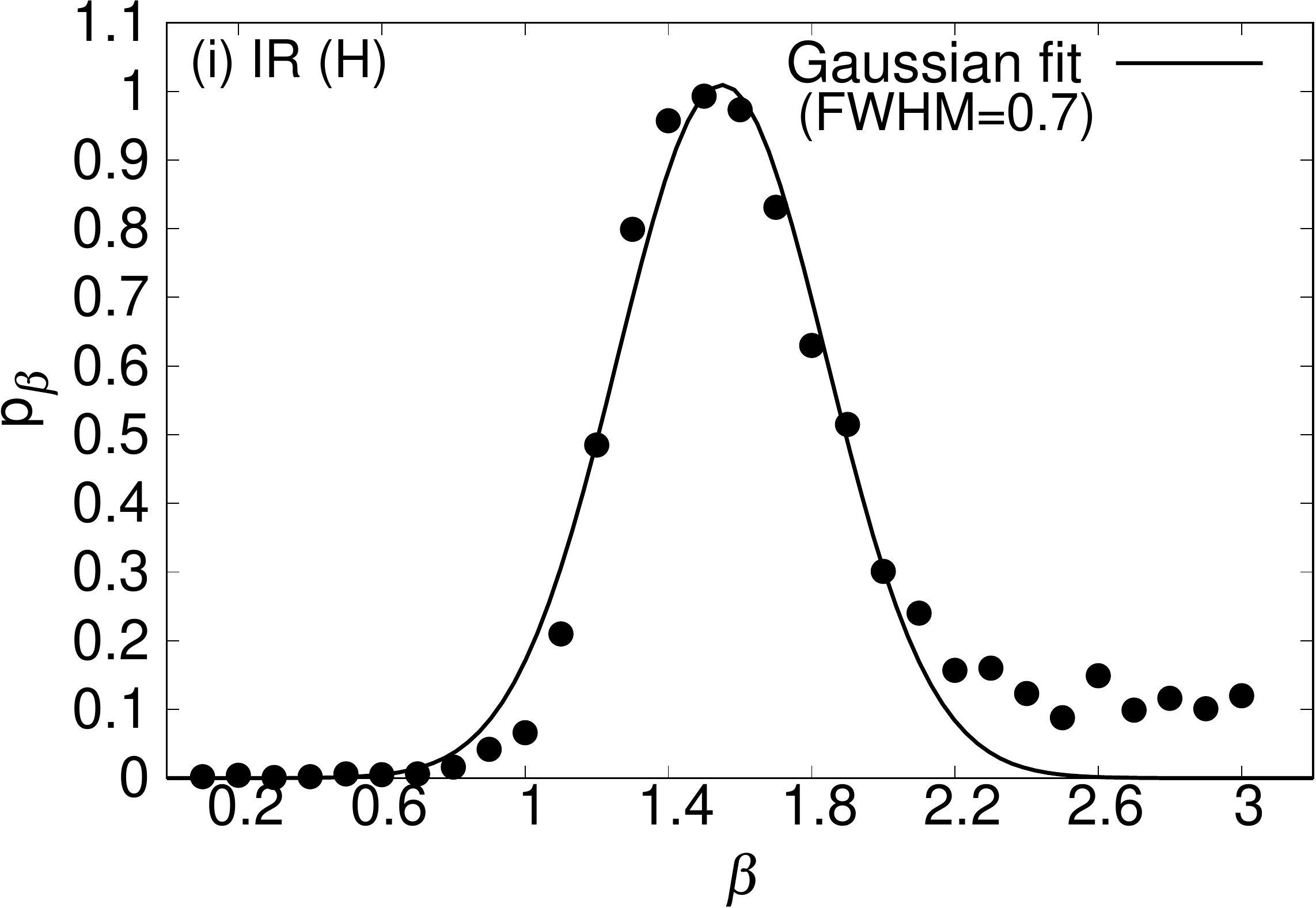}
}
\hbox{
\hspace*{0.1cm}\includegraphics[width=0.33\textwidth]{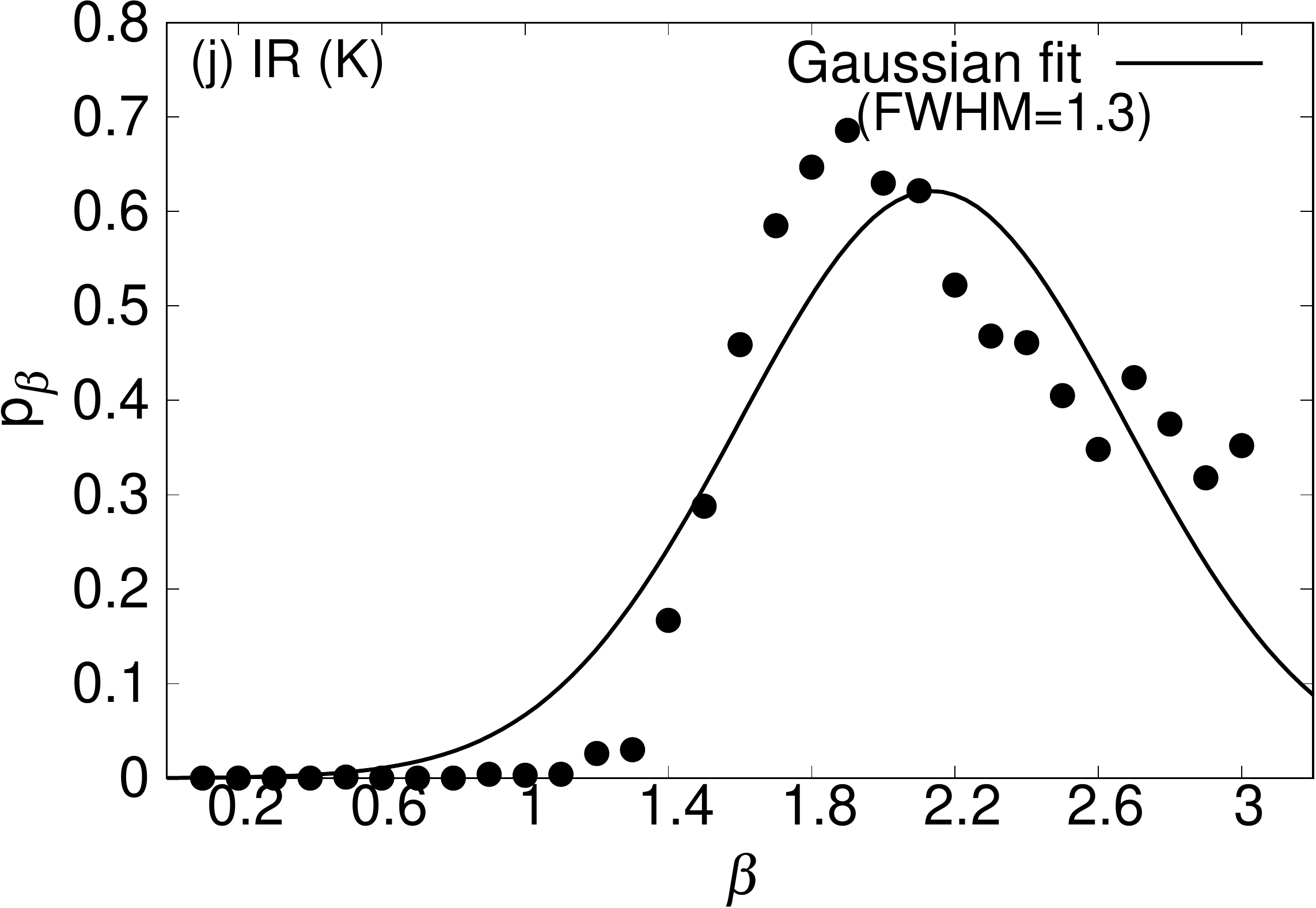}
}

\caption{Probablity, $p_\beta$, that a single power-law slope is acceptable to describe the PSDs for the blazar PKS\,2155$-$304. The filled circles show the fraction for which $\chi^2_{\rm dist,i} > \chi^2_{\rm obs}$ was found out of 1,000 simulations for a fixed $\beta$. The solid lines show the Gaussian fits to the distributions. }
\label{fig:9}
\end{figure*}


\bsp	
\label{lastpage}
\end{document}